\documentclass[aps,prc,twocolumn,showpacs,preprintnumbers,amsmath,amssymb]{revtex4-1}
\usepackage{graphicx}
\usepackage{color}
\definecolor{purple}{rgb}{0.6,0,0.5}
\definecolor{verde}{rgb}{0,0.5,0}
\usepackage[colorlinks=true, pdfstartview=FitV, linkcolor=purple,
citecolor=blue,urlcolor=navyblue]{hyperref}

\begin{document}

\title{Correlations between bulk parameters in relativistic and nonrelativistic hadronic 
mean-field models.} 

\author{B. M. Santos$^1$, M. Dutra$^2$, O. Louren\c{c}o$^3$, and A. Delfino$^1$}

\affiliation{$^1$Instituto de F\'isica, Universidade Federal Fluminense,
24210-150, Niter\'oi, RJ, Brazil  \\
$^2$Departamento de Ci\^encias da Natureza, IHS, Universidade Federal Fluminense, 
28895-532 Rio das Ostras, RJ, Brazil \\
$^3$Departamento de Ci\^encias da Natureza, Matem\'atica e 
Educa\c c\~ao, CCA, Universidade Federal de S\~ao Carlos, 13600-970 Araras, SP,
Brazil}

\date{\today}

\begin{abstract}
In this work, we study the arising of correlations among some isoscalar ($K_o$, $Q_o$, and 
$I_o$) and isovector ($J$, $L_o$, $K_{\mbox{\tiny sym}}^o$, $Q_{\mbox{\tiny sym}}^o$, and
$I_{\mbox{\tiny sym}}^o$) bulk parameters in nonrelativistic and relativistic hadronic
mean-field models. For the former, we investigate correlations in Skyrme and Gogny 
parametrizations, as well as in the nonrelativistic (NR) limit of relativistic 
point-coupling models. We provide analytical correlations among bulk parameters for the NR 
limit, discussing the conditions in which they are linear ones. Based on a recent study 
[B. M. Santos {\it et al}., Phys. Rev. C {\bf 90}, 035203 (2014)], we also show that 
some correlations presented in the NR limit are reproduced for relativistic models 
presenting cubic and quartic self-interactions in the scalar field $\sigma$, mostly 
studied in this work in the context of the relativistic framework. We also discuss how the 
crossing points, observed in the density dependence of some bulk parameters, can be seen 
as a signature of linear correlations between the specific bulk quantity presenting the 
crossing, and its immediately next order parameter.
\end{abstract}

\maketitle  

\section{Introduction} 

In a general way, there are at least two different and competitive approaches in the 
treatment of nuclear matter. One of them is based on nucleon-nucleon interactions, from 
which many-nucleon microscopic relativistic/nonrelativistic Brueckner-Hartree-Fock (BHF)~\cite{bhf} 
calculations are performed to obtain information regarding the entire nuclear system. 
These calculations depend on the chosen nucleon-nucleon potential, classified as phenomenological 
(Reid, Urbana and Argonne interactions, for instance) or theoretical one-boson exchange ones 
(Paris, Bonn and Nijmegen interactions, for instance). For a short review, see Ref.~\cite{rpot}. 
Such nucleon-nucleon interactions reproduce experimental data on phase-shifts and deuteron 
properties and are implemented in complicated many-nucleon BHF codes in order to obtain nuclear 
matter properties. Alternatively, a second (macroscopic) approach does not use nucleon-nucleon 
interaction itself. This is the case of nonrelativistic mean-field models like 
Skyrme~\cite{skyrme} and Gogny~\cite{gog} ones. In relativistic framework, on the 
other hand, the most used models are the relativistic mean-field (RMF) Walecka 
model~\cite{walecka}, and its improved versions~\cite{rmf}.

In both approaches, many of the models and approximations used are intrinsically related 
to the fact that an analytical expression for the nucleon-nucleon potential is unknown. 
Therefore, in a theoretical point of view, correlations between two or more observables 
acquire enormous importance due to the fact that they reduce the set of independent 
relevant quantities to be used in the construction of nuclear models, avoiding redundant 
free-parameter fittings. In an experimental point of view, on the other hand, a 
correlation between two observables $\mathcal{A}$ and $\mathcal{B}$, for instance, allows 
the complete knowledge of $\mathcal{B}$ if $\mathcal{A}$ is experimentally well 
constrained. 

In the few-body nuclear physics, for example, the Tjon line~\cite{tjon}, establishes a 
correlation between the binding energies of $^4\rm He$ and triton, $B_\alpha$ and $B_t$, 
respectively. The parametrization of the numerical results for several two-nucleon 
potentials~\cite{plb} concludes that such a correlation reads roughly $B_\alpha = 4.72(B_t 
- 2.48)$, in MeV units. It means that, if $B_t$ is calculated by using a two 
nucleon-nucleon potential, the value of $B_\alpha$ is predicted, even before any 
four-nucleon calculation. 

Concerning correlations among bulk parameters, there are still a few of them well 
established in the literature. One of them, usually known as the Coester 
line~\cite{coester}, correlates the saturation density $\rho_o$ and the nuclear matter 
binding energy $B_o$. It was analyzed in Ref.~\cite{coester} in a two-nucleon model 
interaction, by varying its tensor force contribution while keeping the deuteron binding 
energy fixed. They showed that the function $B_o\times\rho_o$ roughly follows a line, in fact a band,
also observing its similarity with the curve constructed from $B_o$ and $\rho_o$ 
obtained from distinct two-nucleon potentials. Even modern calculations using 
Brueckner-Hartree-Fock, and including single-particle contribution in the continuum, 
change the results but preserve the Coester line. Another correlation was studied in 
Ref.~\cite{ls-splitting}, and involves the relationship between finite nuclei spin-orbit 
splittings and the ratio $m^*=M^*_o/M$ for a class of finite-range (FR) RMF models, being 
$M^*_o$ the nucleon Dirac effective mass at $\rho=\rho_o$ and $M$ the nucleon rest mass. 
The authors showed that such splittings are experimentally well reproduced if the range for 
$m^*$ is constrained to the following inequality, 
\begin{equation}
0.58\leqslant m^*\leqslant 0.64.
\label{FRS}
\end{equation}  
We will refer the above relation extracted from Ref.~\cite{ls-splitting} as the FRS 
constraint. It is important because it relates bulk infinite nuclear matter calculation 
with the closed shell finite nuclei energy spectrum. Still regarding relationships 
among quantities of finite nuclei and infinite nuclear matter, we also point out the 
correlation between the neutron skin thickness in $^{208}\rm Pb$, and the liquid-to-solid 
transition density in neutron-rich matter, investigated in Ref.~\cite{skin} with RMF 
models.

In a recent paper~\cite{bianca}, we have provided analytical expressions linearly correlating 
the symmetry energy ($J$), its slope ($L_o$) and curvature ($K_{\mbox{\tiny sym}}^o$) at 
the saturation density in the framework of the nonrelativistic (NR) limit of 
nonlinear point-coupling (NLPC) versions of the Boguta-Bodmer models~\cite{boguta} 
(\mbox{FR-RMF} models presenting cubic and quartic self-interactions in the scalar 
field $\sigma$). From such analytical correlations, we were able to predict ranges for 
$L_o$ that Boguta-Bodmer models must present in order furnish good values for finite 
nuclei spin-orbit splittings. In this paper, we further investigate the arising of 
correlations among bulk parameters of infinite nuclear matter at zero temperature in 
nonrelativistic and relativistic hadronic mean-field models. For the former, we choose 
some Skyrme and Gogny parametrizations as well as the NR limit of NLPC models. For the 
latter, we choose \mbox{FR-RMF} models, specially the Boguta-Bodmer ones, and show the 
conditions to be satisfied in order to their bulk parameters present correlations. In 
particular, we show, in a general way, how linear correlations are connected with crossing 
points in the density dependence of bulk parameters. For this purpose, we generalize the 
procedure used in Ref.~\cite{margueron}, where the authors associated the crossing point 
in the incompressibility function for different nonrelativistic Skyrme models, but not 
for \mbox{FR-RMF} ones, with the linear correlation between $K_o$ and $Q_o$ 
(incompressibility and skewness coefficient at $\rho_o$, respectively). We show here that 
there are other crossing densities different from $\rho_c\simeq 0.7\rho_o$, value found 
in Refs.~\cite{margueron,prl}, for different bulk parameters in nonrelativistic models, 
as well as in relativistic ones. In the following, we will see that such crossing 
densities may be seen as signatures of linear correlations between higher order 
derivatives of the specific bulk parameter presenting the crossing.

The paper is organized as follows. In Sec.~\ref{lincorr} we show how linear correlations 
are connected with crossing densities in bulk parameters of infinite nuclear matter. In 
Sec.~\ref{nonrel}, we use some results of our previous study of the NR 
limit~\cite{bianca}, in order to apply the calculations of the previous section. In 
Sec.~\ref{rel}, we use the predictions of the NR limit to study the conditions that 
establish linear correlations in \mbox{FR-RMF} models. In Sec.~\ref{twoicc}, we 
study relativistic and nonrelativistic models presenting two isovector coupling 
constants in the context of the correlation between the symmetry energy and its slope.
Finally, we present our main conclusions in Sec.~\ref{sum-con}.

\section{Connection between linear correlations and crossing densities} 
\label{lincorr}

In the literature, it is verified that crossing points can occur in the density 
dependence of the symmetry energy~\cite{19}, in pure neutron matter equation of 
state~\cite{20}, and in pairing gap of nuclear matter~\cite{22}, for instance. Recently, 
in Refs.~\cite{prl,margueron}, the authors found a specific crossing point in the 
incompressibility of nuclear matter for different nonrelativistic Skyrme 
models, and not confirmed in \mbox{FR-RMF} ones. They showed that this crossing density 
($\rho_c\simeq 0.7\rho_o\simeq 0.11$~fm$^{-3}$) when used in the calculation of the 
derivative of the incompressibility, makes this quantity better correlated to the centroid 
energy of the isoscalar giant monopole resonance than $K_o$. The authors also pointed out 
that this crossing density $\rho_c$ is closer to the average density in the $^{208}{\rm 
Pb}$ nucleus, $\langle\rho\rangle=0.12$~fm$^{-3}$, than the saturation density itself, 
$\rho_o\simeq 0.16$~fm$^{-3}$, with $\langle\rho\rangle$ also obtained by using the 
Skyrme model. The question we pose here is what properly mean, or indicate, such crossing 
points. In the following, we try to answer this question.

Actually, crossing points in different bulk parameters of infinity nuclear matter can be 
viewed as a signature of linear correlations between higher order derivatives of that 
particular bulk parameter presenting the crossing in its density dependence. In order to 
make it clear, we proceed to generalize the calculation performed in 
Ref.~\cite{margueron}, where the authors associated a crossing in the $K(\rho)$ function 
with the linear correlation between $K_o$ and $Q_o$ for some Skyrme parametrizations. 
Firstly, let us define a function of the density, $\mathcal{F}(\rho)$, expanded in terms 
of the dimensionless variable $x=\frac{\rho-\rho_o}{3\rho_o}$, and around the saturation 
density as,
\begin{multline}
\mathcal{F}(\rho) = \mathcal{F}(\rho_o) 
+ \mathcal{F}^\prime(\rho_o)x 
+ \frac{\mathcal{F}^{\prime\prime}(\rho_o)}{2!}x^2 
+ \frac{\mathcal{F}^{\prime\prime\prime}(\rho_o)}{3!}x^3 
+ \cdots
\label{f}
\end{multline}
with the derivatives of $\mathcal{F}(\rho)$ given by
\begin{eqnarray}
\mathcal{F}^{(m)}(\rho)=(3\rho)^{m} \frac{\partial}{\partial \rho} \left[ 
\frac{\mathcal{F}^{(m-1)}(\rho)}{(3\rho)^{m-1}}  \right],
\label{derf}
\end{eqnarray}
for $m=1,2,3\,...$

For the values $m=1,2,3$, for instance, and by noting that $\frac{\rho}{\rho_o}=3x+1$ 
and $\frac{\partial}{\partial \rho}= \frac{\partial x}{\partial 
\rho}\frac{\partial}{\partial x}=\frac{1}{3\rho_o}\frac{\partial}{\partial x}
$, we see that Eq.~(\ref{derf}) leads to
\begin{eqnarray}
\mathcal{F}^{\prime}(\rho)&=&3\rho \frac{\partial \mathcal{F}}{\partial \rho 
}=\frac{3\rho}{3\rho_o}\frac{\partial \mathcal{F}}{\partial x}=(3x+1)\frac{\partial 
\mathcal{F}}{\partial x},
\end{eqnarray}
\begin{eqnarray}
\mathcal{F}^{\prime\prime}(\rho)&=&(3\rho)^2 \frac{\partial}{\partial \rho} \left[ 
\frac{\mathcal{F}'(\rho)}{3 \rho } \right]=(3\rho)^2 \frac{\partial}{\partial \rho} \left[ 
\frac{\partial \mathcal{F}}{\partial \rho } \right] \nonumber \\
&=& \frac{(3\rho)^{2}}{(3\rho_o)^{2}} \frac{\partial^{2} \mathcal{F}}{\partial x^{2} }
=(3x+1)^{2}\frac{\partial^{2} \mathcal{F}}{\partial x^{2}},
\label{fpp}
\end{eqnarray}
and
\begin{eqnarray}
\mathcal{F}^{\prime\prime\prime}(\rho)&=&(3\rho)^3 \frac{\partial}{\partial \rho} \left[ 
\frac{\mathcal{F}''(\rho)}{(3 \rho)^{2} } \right]
=(3\rho)^3 \frac{\partial}{\partial \rho} \left[ \frac{\partial^{2} \mathcal{F}}{\partial 
\rho^{2} } \right] \nonumber \\
&=& \frac{(3\rho)^{3}}{(3\rho_o)^{3}} \frac{\partial^{3} \mathcal{F}}{\partial x^{3} }
=(3x+1)^{3}\frac{\partial^{3} \mathcal{F}}{\partial x^{3}}.
\label{f3p}
\end{eqnarray}
The pattern verified in Eqs.~(\ref{f})-(\ref{f3p}) allows us to write $\mathcal{F}(\rho)$ 
and its derivatives in a compact form as,
\begin{eqnarray}
\mathcal{F}^{(m)}(\rho)&=& (3x+1)^{m} \left\lbrace  \mathcal{F}^{(m)}  (\rho_o)  + 
\mathcal{F}^{(m+1)}  (\rho_o)  x\right.  \nonumber \\
&+&\left.\frac{ \mathcal{F}^{(m+2)}  (\rho_o)}{2!} 
x^{2}   
+\frac{ \mathcal{F}^{(m+3)}  (\rho_o)}{3!} x^{3}  +...  \right\rbrace,
\label{fgeral}
\end{eqnarray}
for $m=0,1,2,3\,...$ and with $\mathcal{F}^{(m)}(\rho_o)$, $\mathcal{F}^{(m+1)}(\rho_o)$, 
$\mathcal{F}^{(m+2)}  (\rho_o)\,...$ being the bulk parameters evaluated at the saturation 
density. 

In order to make our analysis simpler, we consider that $\mathcal{F}^{(m)}/(3x+1)^{m}$ is 
well described by its expansion until order $x^3$. Now, let us assume that the bulk 
parameters of the higher order derivatives of $\mathcal{F}^{(m)}$, namely, 
$\mathcal{F}^{(m+1)}(\rho_o)$, $\mathcal{F}^{(m+2)}(\rho_o)$ and 
$\mathcal{F}^{(m+3)}(\rho_o)$, are correlated with $\mathcal{F}^{(m)}(\rho_o)$, more 
specifically in a linear way, as given below,
\begin{eqnarray}
\mathcal{F}^{(m+1)}(\rho_o)&=&b_1 + a_1 \mathcal{F}^{(m)}(\rho_o),
\label{fm1} \\
\mathcal{F}^{(m+2)}(\rho_o)&=&b_2 + a_2 \mathcal{F}^{(m)}(\rho_o),
\label{fm2}\\
\mathcal{F}^{(m+3)}(\rho_o)&=&b_3 + a_3 \mathcal{F}^{(m)}(\rho_o),
\label{fm3}
\end{eqnarray}
with $a_i$ and $b_i$ independent of any other parameters of the model. If such linear 
correlations hold, then, Eq.~(\ref{fgeral}) can be rewritten as
\begin{multline}
\mathcal{F}^{(m)}(\rho) \simeq (3x+1)^{m}\times\\
\times\left\lbrace f(x)\mathcal{F}^{(m)}(\rho_o) + b_1 x + \frac{b_2}{2!}x^{2} 
+ \frac{b_3}{3!}x^3 \right\rbrace,
\label{fm}
\end{multline}
where
\begin{eqnarray}
f(x)= 1 + a_1 x +\frac{  a_2 }{2!} x^2 +\frac{a_3 }{3!} x^3.
\label{fx}
\end{eqnarray}
For a specific value (or values) of the density, named as $\rho_c$, that makes 
$f(x_c)=0$, with $x_c=\frac{\rho_c-\rho_o}{3\rho_o}$, $\mathcal{F}^{(m)}(\rho_c)$ will be 
exactly the same for any hadronic model, according to Eq.~(\ref{fm}) since the $b_i$ 
parameters are model independent. Therefore, the functions $\mathcal{F}^{(m)}(\rho)$ of 
any model will cross each other exactly at $\rho=\rho_c$. This is the crossing point 
presented in the $\mathcal{F}^{(m)}(\rho)$ function. Thus, one can see this 
crossing as a signature of the linear correlations given by Eqs.~(\ref{fm1})-(\ref{fm3}). 
Naturally, such a signature only holds if we can express the function 
$\mathcal{F}^{(m)}(\rho)$ in terms of the expansion presented in Eq.~(\ref{fgeral}).

In the case of $\mathcal{F}^{(m)}/(3x+1)^{m}$ is expanded until order $x^3$, 
Eq.~(\ref{fx}) generates a cubic equation, when one imposes $f(x_c)=0$, in order to 
localize the values of $\rho_c$. In the general case, in which is needed to expand 
$\mathcal{F}^{(m)}/(3x+1)^{m}$ until order $x^N$, then
\begin{eqnarray}
f(x_c)=1+\sum_{i=1}^{N}\frac{a_ix_c^i}{i!}=0,
\label{fxgeral}
\end{eqnarray}
produces an equation of order $N$ to be solved in order to determine the possible values 
of $x_c$, and, consequently, the crossing densities $\rho_c$ of the function 
$\mathcal{F}^{(m)}(\rho)$. 

The determination of linear correlations from the searching of crossing points is 
suitable, for instance, for the isovector bulk properties of hadronic models, in which the 
quantities and its derivatives can be expanded around the saturation density exactly as in 
Eq.~(\ref{fgeral}). The energy per particle of a system of proton fraction given by 
$y=\rho_p/\rho$ ($\rho_p$ is the proton density) can be expanded in terms of the isospin 
asymmetry parameter $\beta=1-2y$ as
\begin{eqnarray}
E(\rho,\beta)= E(\rho) + \mathcal{S}(\rho)\beta^2 + \mathcal{S}_{4}(\rho)\beta^4 + 
\cdots
\end{eqnarray}
where $E(\rho)$ is the energy per particle related to the symmetric nuclear matter 
($\beta=0$). The other two coefficients of the expansion are, respectively, the symmetry 
energy and the fourth-order symmetry energy, with the density dependence, associated with 
the isovector sector, expanded in terms of the density as
\begin{eqnarray}
\mathcal{S}(\rho)&=& J + L_ox + \frac{K_{\mbox{\tiny sym}}^o}{2!}x^2 
+ \frac{Q_{\mbox{\tiny sym}}^o}{3!}x^3 + \cdots 
\label{s}
\end{eqnarray}
and
\begin{eqnarray}
\mathcal{S}_4(\rho)&=& J_{4} + L_4^ox + \frac{K_{\mbox{\tiny sym,4}}^o}{2!}x^2  
+ \frac{Q_{\mbox{\tiny sym,4}}^o}{3!}x^3 + \cdots
\label{s4}
\end{eqnarray}
The derivatives of $\mathcal{S}(\rho)$ are defined exactly as in Eq.~(\ref{derf}), 
\mbox{i. e.}, $L=3\rho\frac{\partial\mathcal{S}}{\partial\rho}$, 
$K_{\mbox{\tiny sym}}=(3\rho)^2\frac{\partial^2\mathcal{S}}{\partial\rho^2}$, 
$Q_{\mbox{\tiny sym}}=(3\rho)^3\frac{\partial^2\mathcal{S}}{\partial\rho^3}$, and so on. 
The isovector bulk parameters, $L_o$, $K_{\mbox{\tiny sym}}^o$ and $Q_{\mbox{\tiny 
sym}}^o$ are the derivatives evaluated at $\rho=\rho_o$, and $J$ is given by 
$J=\mathcal{S}(\rho_o)$. Analogous quantities are also defined from $\mathcal{S}_4(\rho)$. 
Based on this structure, the search of linear correlations from the location of crossing 
densities can be naturally performed. This will be done for nonrelativistic and 
relativistic models in the next two sections.

As a last remark, we mention here that the procedure described above was first used in 
Ref.~\cite{margueron} specifically to justify the crossing point in the 
$K(\rho)\times\rho$ curve, and not as a route to find linear correlations as we are doing 
in the present work. Moreover, we are also generalizing this method to any bulk parameter.

\section{Correlations in nonrelativistic models} 
\label{nonrel}

\subsection{Theoretical framework of the NR limit}

In the background of nonrelativistic mean-field models, we analyse some parametrizations 
of Skyrme and Gogny models, and also those from the NR limit of NLPC versions of the 
Boguta-Bodmer model. It is important to mention that as the FR-RMF models, relativistic 
point-coupling ones also describe very well the infinite nuclear matter bulk parameters 
and finite nuclei properties~\mbox{\cite{nlpc2,nlpc3,nlpc4,newrefpc1,newrefpc2,nlpc1}}. 
In Ref.~\cite{nlpc1}, for instance, the authors were able to obtain, by using a NLPC 
model, ground state binding energies, spin-orbit splittings, and rms charge radii of a 
large set of closed shell nuclei, as well as, of nuclei outside the valley of beta 
stability, clearly showing the success of these kind of model. Their nonrelativistic 
versions, besides follows this same pattern, at least concerning the infinite nuclear 
matter that is the scope of our work, are also useful in the sense that they can be used 
to predict correlations also exhibited in \mbox{FR-RMF} models, as observed in our 
previous study of Ref.~\cite{bianca}.

The relativistic NLPC versions of the Boguta-Bodmer models are described by the following 
Lagrangian density
\begin{eqnarray}
\mathcal{L}_{\mbox{\tiny NLPC}}&=&\bar{\psi}(i\gamma^{\mu}\partial_{\mu}-M)\psi
-\frac{1}{2}G^{2}_{\mbox{\tiny
V}}(\bar{\psi}\gamma^{\mu}\psi)^{2}+\frac{1}{2}G^{2}_{\mbox{\tiny S}}(\bar{
\psi } \psi)^{2} \nonumber \\
&+&\frac{A}{3}(\bar{\psi}\psi)^{3}+\frac{B}{4}(\bar{\psi}\psi)^{4}-
\frac{1}{2}G^{2}_{\mbox{\tiny TV}}(\bar{\psi}\gamma^{\mu}\vec{\tau}\psi)^{2},
\label{lagrangeana}
\end{eqnarray}
that mimics the two-, three- and four-body pointlike interactions. In this equation, the 
last term is included in order to take into account the asymmetry of the system (different 
number of protons and neutrons). In the NR limit of the NLPC model, and by using the 
mean-field approximation, the energy density functional at zero temperature for asymmetric 
nuclear matter is written as
\begin{eqnarray}
\varepsilon^{\mbox{\tiny (NR)}}(\rho,y)&=&(G^{2}_{\mbox{\tiny V}}-G^{2}_{\mbox{\tiny
S}})\rho^{2} -A\rho^{3} -B\rho^{4} 
\nonumber \\
&+& G^{2}_{\mbox{\tiny TV}}\rho^{2}(2y-1)^{2} +
\frac{3}{10M^*(\rho,y)}\lambda\rho^{\frac{5}{3}},
\label{enr}
\end{eqnarray}
where the effective mass is
\begin{eqnarray}
M^*(\rho,y)=\frac{M^2}{(M+G^2_{\mbox{\tiny S}}\rho +
2A\rho^2 +3B\rho^3)H_{\frac{5}{3}}},
\end{eqnarray}
with $H_{\frac{5}{3}} =2^{\frac{2}{3}}[y^{\frac{5}{3}}+(1-y)^{\frac{5}{3}}]$, and
$\lambda=(3\pi^2/2)^{\frac{2}{3}}$, see Ref.~\cite{bianca}.

From the energy density in Eq.~(\ref{enr}), it is possible to obtain pressure, 
incompressibility and the symmetry energy of the model, since 
$P=\rho^2\frac{\partial(\mathcal{E}/\rho)}{\partial\rho}$, $K=9\frac{\partial 
P}{\partial\rho}$, and 
$\mathcal{S}=\frac{1}{8}\left[\frac{\partial^2(\varepsilon/\rho)}{\partial 
y^2}\right]_{y=\frac{1}{2}}$. These expressions are, respectively, given by
\begin{eqnarray}
P^{\mbox{\tiny (NR)}}(\rho,y)&=& (G^{2}_{\mbox{\tiny V}}-G^{2}_{\mbox{\tiny S}})\rho^{2} 
-2A\rho^{3} -3B\rho^{4}\nonumber \\
&+& G^{2}_{\mbox{\tiny TV}}\rho^{2}(2y-1)^{2} + \frac{\lambda H_{5/3}}{5M^{2}}\times 
\nonumber \\
&\times&\left(M +\frac{5}{2}G^{2}_{\mbox{\tiny S}}\rho +8A\rho^{2} 
+\frac{33}{2}B\rho^{3}\right)\rho^{\frac{5}{3}},\quad
\label{pnr}
\end{eqnarray}
\begin{eqnarray}
K^{\mbox{\tiny (NR)}}(\rho, y)&=& 18(G^{2}_{\mbox{\tiny V}}-G^{2}_{\mbox{\tiny S}})\rho 
-54A\rho^{2} -108B\rho^{3} \nonumber \\
&+&18G^{2}_{\mbox{\tiny TV}}\rho(2y-1)^{2} + \frac{3\lambda H_{5/3}}{M^2}\times 
\nonumber \\
&\times&\left(M +4G^{2}_{\mbox{\tiny S}}\rho +\frac{88}{5}A\rho^{2} 
+\frac{231}{5}B\rho^{3}\right)\rho^{\frac{2}{3}},\qquad
\label{knr}
\end{eqnarray}
and
\begin{eqnarray}
\mathcal{S}^{\mbox{\tiny (NR)}}(\rho)= G^{2}_{\mbox{\tiny TV}}\rho + 
\frac{\lambda\rho^{\frac{2}{3}}}{6M^*(\rho,1/2)}.
\label{snr}
\end{eqnarray}
The symmetry energy $\mathcal{S}^{\mbox{\tiny (NR)}}$ is used in order to obtain its 
slope, curvature and skewness. The results are
\begin{eqnarray}
L^{\mbox{\tiny (NR)}}(\rho)&=&\frac{\lambda\rho^{\frac{2}{3}}}{3M^{2}} 
\left(M+ \frac{5}{2}G^2_{\mbox{\tiny S}}\rho + 8A\rho^{2} + 
\frac{33}{2}B\rho^3\right)
\nonumber\\
&+& 3G^2_{\mbox{\tiny TV}}\rho,
\label{lrhonr}
\end{eqnarray}
\begin{eqnarray}
K_{\mbox{\tiny sym}}^{\mbox{\tiny (NR)}}(\rho)&=&\frac{\lambda\rho^{\frac{2}{3}}}{3M^{2}} 
\left(-M +5 G^2_{\mbox{\tiny S}}\rho+40 A\rho^2 +132B\rho^3\right),\quad\,\,\,\,\,
\label{ksymrhonr}
\end{eqnarray}
and
\begin{equation}
Q_{\mbox{\tiny sym}}^{\mbox{\tiny (NR)}}(\rho)=\frac{4\lambda\rho^{\frac{2}{3}}}{3M^2} 
\left(M-\frac{5}{4} G^2_{\mbox{\tiny S}}\rho +20A\rho^{2} +165 B\rho^{3}\right),
\label{qsymrhonr}
\end{equation}
respectively.

The coupling constants of the model are $G^{2}_{\mbox{\tiny S}}$, $G^{2}_{\mbox{\tiny 
V}}$, $A$, $B$ and $G^{2}_{\mbox{\tiny TV}}$. The first four of them are adjusted in
order to fix $\rho_o$, $B_o$, $K_o$ and $M^*_o$. This is done by solving a system of four 
equations, namely, $\varepsilon^{\mbox{\tiny (NR)}}(\rho_o,1/2)/\rho_o=-B_o$, 
$K^{\mbox{\tiny (NR)}}(\rho_o,1/2)=K_o$, $P^{\mbox{\tiny (NR)}}(\rho_o,1/2)=0$ (nuclear 
matter saturation), and $M^*(\rho_o,1/2)=M^*_o$. The last coupling constant, 
$G^2_{\mbox{\tiny TV}}$, is obtained by imposing upon the model the requirement of 
presenting a particular value for \mbox{$J=\mathcal{S}^{\mbox{\tiny (NR)}}(\rho_o)$}. The 
explicit forms of the constants $G^{2}_{\mbox{\tiny S}}$, $G^{2}_{\mbox{\tiny V}}$, $A$, 
$B$ and $G^{2}_{\mbox{\tiny TV}}$ in terms of $m^*$, $\rho_o$, $B_o$, $K_o$ and $J$, are 
given in the Appendix.

\subsection{Results from the isovector sector}

In Ref.~\cite{bianca}, we rewritten the coupling constants of the model in terms of the 
bulk parameters $m^*$, $\rho_o$, $B_o$, and $K_o$ (an analogous procedure is done in the 
context of the Skyrme models in Ref.~\cite{sam}). This method allowed us to explicitly 
write the slope of the symmetry energy at the saturation density also as a function of 
$m^*$, $\rho_o$, $B_o$, and $K_o$, and thus, find the following correlation between $J$ 
and $L_o$, namely,
\begin{eqnarray}
L_o=3J + b(m^{*},\rho_{o},B_{o},K_{o}),
\label{lj}
\end{eqnarray}
with
\begin{multline}
b(m^*,\rho_o,B_o,K_o)=\frac{1}{\left( 3M^{2} - 19E_{\mbox{\tiny F}}^oM  + 18E_{\mbox{\tiny F}}^{o2}\right)}\times \\
\times\bigg\{ \frac{10E_{\mbox{\tiny F}}^{o}}{ 9m^{*} }(3M^{2} - 14ME_{\mbox{\tiny F}}^o) 
 - 5E_{\mbox{\tiny F}}^{o}(M^{2} - 5 M E_{\mbox{\tiny F}}^{o} )\\
+ 30B_{o}E_{\mbox{\tiny F}}^{o2}  - \frac{5K_{o}}{ 9} E_{\mbox{\tiny F}}^{o}M  
\bigg\},
\label{functionb}
\end{multline}
where $E_{\mbox{\tiny F}}^o=3\lambda\rho_{o}^{\frac{2}{3}}/10M$.

The expression in Eq.~(\ref{ksymrhonr}) along with the correlation in Eq.~(\ref{lj}),
and the definitions of the coupling constants in terms of the bulk parameters are used to 
find
\begin{eqnarray} 
K_{\mbox{\tiny sym}}^o=(L_o-3J)p(\rho_o) +q(\rho_o,B_o,K_o)
\label{ksymo}
\end{eqnarray}
where
\begin{eqnarray}
p(\rho_o)=\frac{3\left( 5M^{2} - 18E_{\mbox{\tiny F}}^oM  \right)}{3M^{2} - 
14ME_{\mbox{\tiny F}}^o},
\end{eqnarray}
and
\begin{eqnarray}
q(\rho_o,B_o,K_o)=\frac{5M^{2}E_{\mbox{\tiny F}}^{o} + 
90B_{o} ME_{\mbox{\tiny F}}^{o} - \frac{70}{9}K_oE_{\mbox{\tiny F}}^{o}M}{3M^{2} - 
14ME_{\mbox{\tiny F}}^o},\quad
\,\,\,\,
\end{eqnarray}
As pointed out in Ref.~\cite{bianca}, if $J$ and $K_o$ are kept fixed in 
Eq.~(\ref{ksymo}), $K_{\mbox{\tiny sym}}^o$ will present a linear correlation with $L_o$, 
since, the binding energy and the saturation density are well established closely around 
the values of $B_o=16$~MeV and $\rho_o=0.15$~fm$^{-3}$, and do not vary too much for each 
parametrizations of any nuclear mean-field model. For this reason, we consider $B_o$ and 
$\rho_o$ as constants hereafter. The same conditions also occur with the skewness of 
$\mathcal{S}$ of the NR limit, due to its analytical structure in Eq.~(\ref{qsymrhonr}). 
This quantity, obtained through $Q_{\mbox{\tiny sym}}^o=Q_{\mbox{\tiny sym}}^{\mbox{\tiny 
(NR)}}(\rho_o)$, can also be written in the following form,
\begin{eqnarray}
Q_{\mbox{\tiny sym}}^o= (L_o-3J)u(\rho_o) + v(\rho_o,B_o,K_o)
\label{qsymo}
\end{eqnarray}
with
\begin{eqnarray}
u(\rho_o)=-\frac{5\left( 3M^{2} - 22E_{\mbox{\tiny F}}^{o}M -144E_{\mbox{\tiny F}}^{o2} 
\right)}{3M^2 - 14ME_{\mbox{\tiny F}}^o},
\end{eqnarray}
and
\begin{multline}
v(\rho_o,B_o,K_o)= \frac{1}{3M^2 - 14ME_{\mbox{\tiny F}}^o}\times \\
\times\bigg\{ 5E_{\mbox{\tiny F}}^{o}M\left( M +52E_{\mbox{\tiny F}}^{o}  \right)- 
\frac{850}{9} K_{o} E_{\mbox{\tiny F}}^{o}M \\
 + 150B_{o}E_{\mbox{\tiny F}}^{o}\left(9M-8 E_{\mbox{\tiny F}}^{o}\right)\bigg\}.
\end{multline}

\begin{figure}[!htb]
\centering
\includegraphics[scale=0.33]{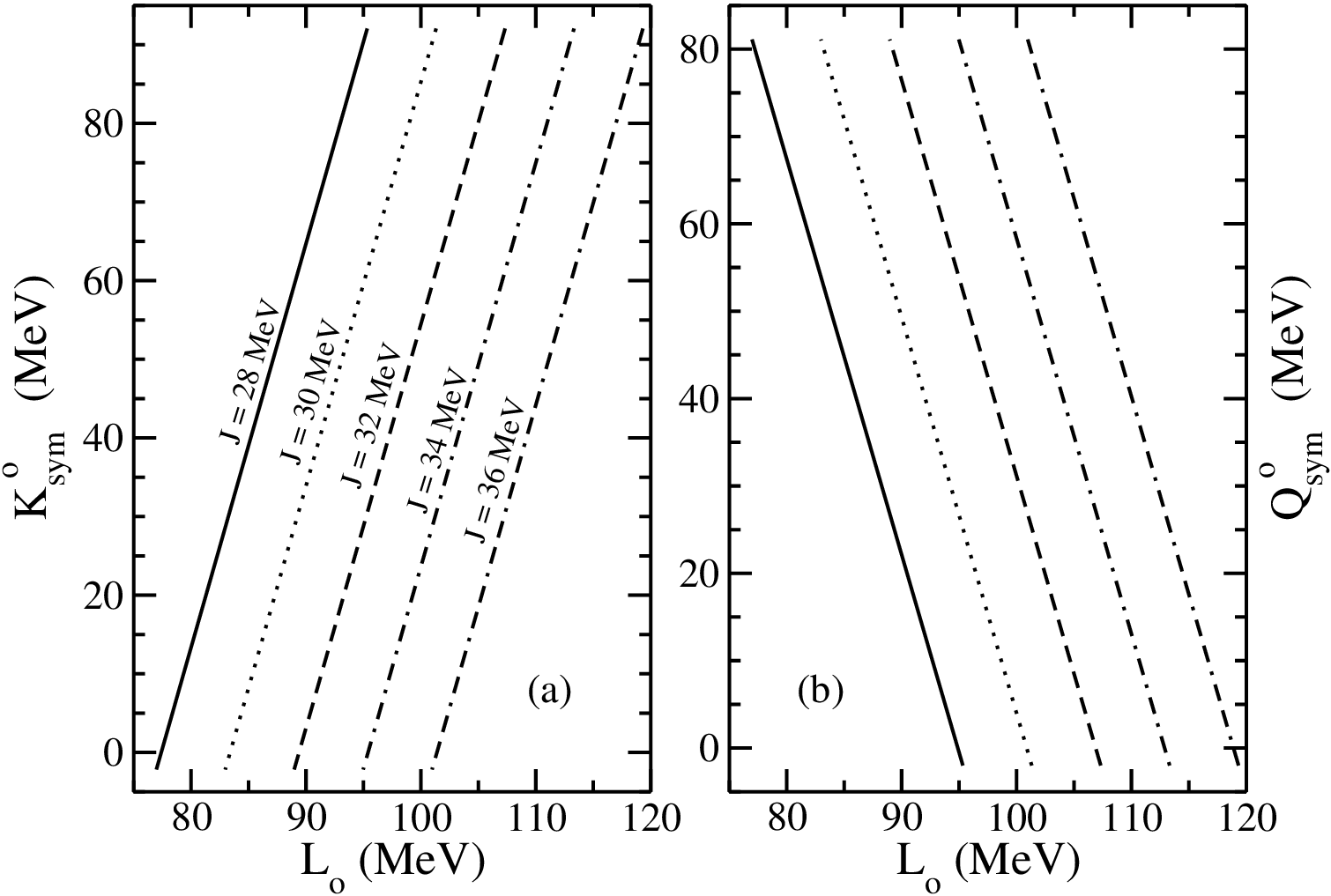}
\vspace{-0.2cm}
\caption{(a) $K_{\mbox{\tiny sym}}^o$ and (b) $Q_{\mbox{\tiny sym}}^o$ as a function 
of $L_o$ for the NR limit for different values of $J$. The effective mass varies in the 
range of $0.50\leqslant m^*\leqslant 0.80$. For each panel, $\rho_o=0.15$~fm$^{-3}$, 
$B_o=16$~MeV and $K_o=270$~MeV.} 
\label{ksymo-qsymo-nr}
\end{figure}

The linear dependence of $K_{\mbox{\tiny sym}}^o$ and $Q_{\mbox{\tiny sym}}^o$ as a 
function of $L_o$ for fixed $J$ and $K_o$ is depicted in Fig.~\ref{ksymo-qsymo-nr}. 

We remark to the reader that it was possible to investigate how $K_{\mbox{\tiny sym}}^o$ 
and $Q_{\mbox{\tiny sym}}^o$ depend on $L_o$, by using parametrizations presenting $J$ 
fixed and different values of the effective mass, specifically in the range of 
$0.50\leqslant m^*\leqslant 0.80$. In this way, it was possible to keep $J$ fixed and 
still be able to vary $L_o$ from the variation of $b(m^*,\rho_o,B_o,K_o)$ in 
Eq.~(\ref{lj}).

\begin{figure}[!htb]
\centering
\includegraphics[scale=0.33]{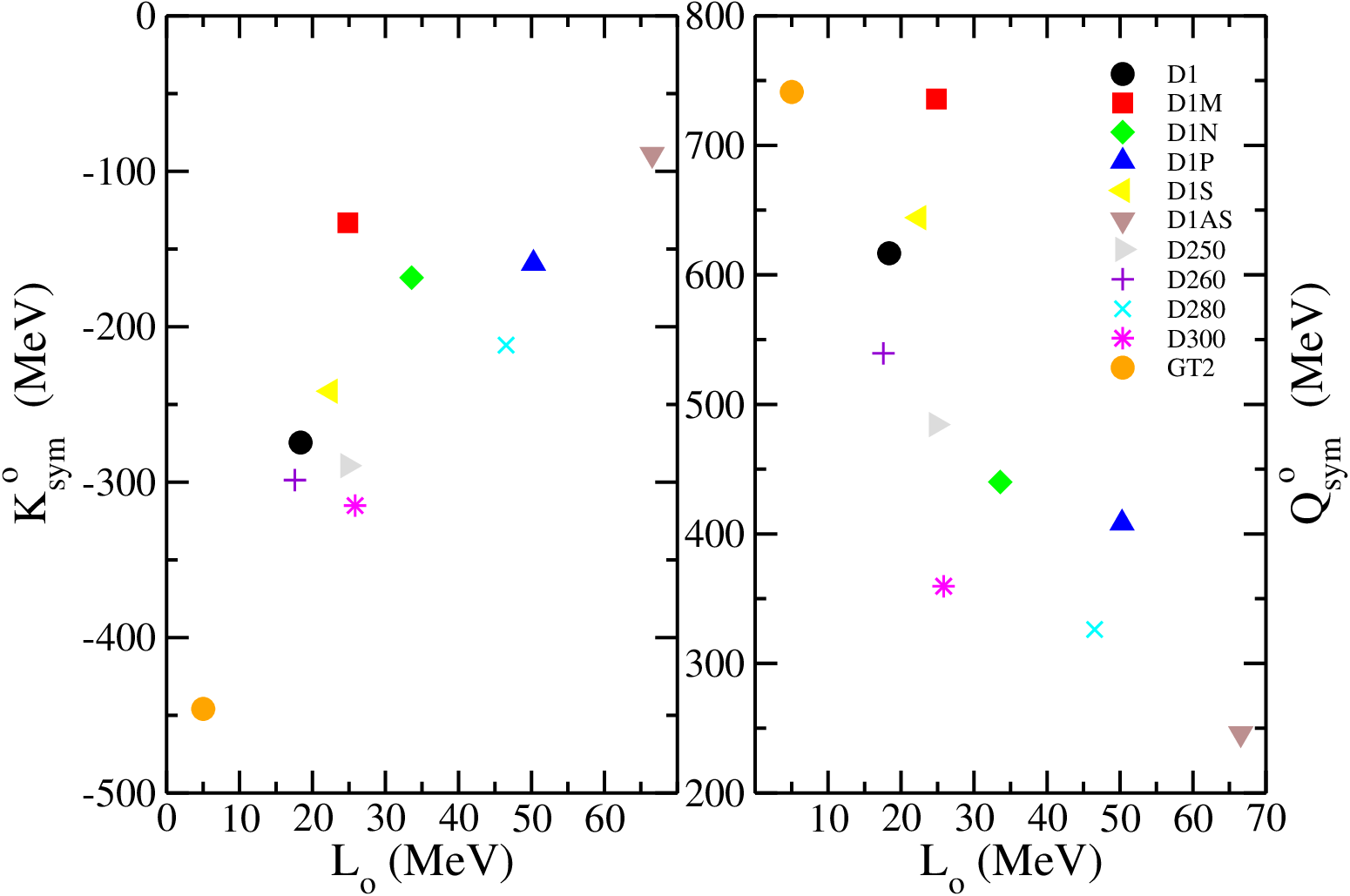}
\vspace{-0.2cm}
\caption{(Color online) (a) $K_{\mbox{\tiny sym}}^o$ and (b) $Q_{\mbox{\tiny sym}}^o$ as 
a function of $L_o$ for Gogny parametrizations studied in Ref.~\cite{gogny}.} 
\label{ksymo-qsymo-gogny}
\end{figure}

The increasing of $K_{\mbox{\tiny sym}}^o$, and decreasing of $Q_{\mbox{\tiny sym}}^o$ 
as a function of $L_o$, are also verified for Gogny interactions presented in 
Ref.~\cite{gogny}. In this work, the authors analysed the isovector properties of this 
specific nonrelativistic model, providing analytical expressions for symmetric and 
asymmetric nuclear matter, see Fig.~\ref{ksymo-qsymo-gogny}.
Notice that although some parametrizations present a linear dependence, this 
behavior is not verified for all of them.

\begin{figure}[!htb]
\centering
\includegraphics[scale=0.33]{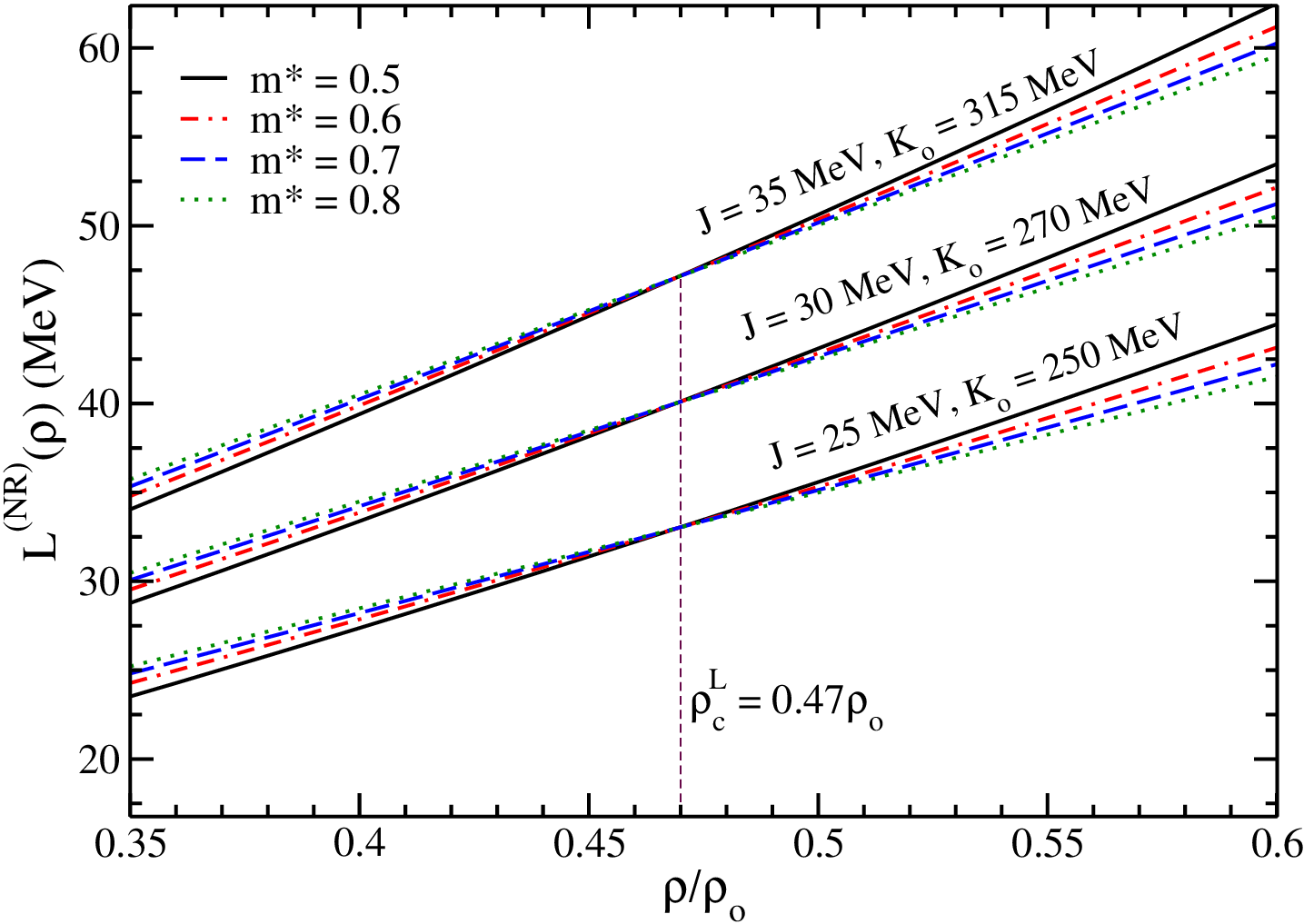}
\vspace{-0.2cm}
\caption{(Color online) Symmetry energy slope as a function of $\rho/\rho_o$ for the NR 
limit, Eq.~(\ref{lrhonr}). In this figure, $\rho_o=0.15$~fm$^{-3}$, and $B_o=16$~MeV.} 
\label{crossing-l-nr}
\end{figure}

According to the discussion of Sec.~\ref{lincorr}, the searching of such linear 
correlations could also have been done if we had looked for their possible signature, 
in this case in the density dependence of the symmetry energy slope. If the function 
$L^{\mbox{\tiny (NR)}}(\rho)$ can be expanded, and if its density dependence presents a 
crossing point, then one can ensure at least the linear behavior presented in 
Fig.~\ref{ksymo-qsymo-nr}{\color{purple}a}. In fact, this crossing around 
$\rho_c^{\mbox{\tiny L}}/\rho_0=0.47$ is verified in Fig.~\ref{crossing-l-nr} in the NR 
limit for parametrizations in which the values of $J$ and $K_o$ are kept fixed.

It is worthwhile to note that the crossing density displayed in 
Fig.~\ref{crossing-l-nr}, namely, $\rho_c^{\mbox{\tiny L}}/\rho_0=0.47$,  is exactly the 
same for all curves. In this figure, we group three different sets of parametrizations, 
each one presenting three distinct values for the pair ($J$, $K_o$). The values were 
chosen inside the ranges of \mbox{$25\leqslant J \leqslant 35$~MeV}, and 
\mbox{$250\leqslant K_o \leqslant 315$~MeV}. One can still notice that the quantity 
$L^{\mbox{\tiny (NR)}}(\rho_c^{\mbox{\tiny L}})$ is different for each set. However, the 
values of $L^{\mbox{\tiny (NR)}}(\rho_c^{\mbox{\tiny L}})$ of these parametrizations 
present an overlap of $48\%$ with the constraint established in Ref.~\cite{slope}, 
namely, $L(\rho_c^{\mbox{\tiny L}})=47.3\pm7.8$~MeV. We still point out that the range of 
$25\leqslant J \leqslant 35$~MeV is showed to be totally compatible~\cite{rmf} with 
experimental values from analyses of different terrestrial nuclear experiments and 
astrophysical observations~\cite{bali}, and the range of $250\leqslant K_o \leqslant 
315$~MeV was based on the recent reanalysis of data on isoscalar giant monopole resonance 
energies~\cite{stone}.

If we consider the expansion of $L^{\mbox{\tiny (NR)}}/(3x+1)$ until order $x$, namely, 
$L(\rho)\simeq(3x+1)(L_o + K_{\mbox{\tiny sym}}^ox)$, it is possible to explain the 
crossing point in Fig.~\ref{crossing-l-nr} if $K_{\mbox{\tiny sym}}^o$ is linearly 
correlated with $L_o$, i. e., if we can write $K_{\mbox{\tiny sym}}^o=b_1+a_1L_o$, and if 
there exist a real root for the equation $f(x_c)=a_1x_c+1=0$, since in this case one 
writes $L(\rho)=(3x+1)\left\lbrace L_o\left(a_1x +  1 \right)  + b_1x\right\rbrace$. As we 
can see in Eq.~(\ref{ksymo}) and in Fig.~\ref{ksymo-qsymo-nr}{\color{purple}a}, the 
linear correlation between $K_{\mbox{\tiny sym}}^o$ and $L_o$ holds in fact for $J$ and 
$K_o$ fixed. However, if we solve the equation $f(x_c)=0$ we find $\rho_c^{\mbox{\tiny 
L}}/\rho_0=0.41$ (since $a_1=p(\rho_o)=5.13$ and $\rho_c^{\mbox{\tiny 
L}}/\rho_0=3x_c^{\mbox{\tiny L}}+1$), not exactly the same crossing density presented in 
Fig.~\ref{crossing-l-nr}. This suggests that $f(x_c)$ should be modified in order to 
produces a more exact root, what means that $L^{\mbox{\tiny (NR)}}/(3x+1)$ actually needs
to be expanded in higher orders in $x$, at least in the density region around the crossing 
density $\rho_c^{\mbox{\tiny L}}/\rho_0=0.47$. If we now take the expansion of 
$L^{\mbox{\tiny (NR)}}/(3x+1)$ until order $x^2$, the crossing point will exist if 
$K_{\mbox{\tiny sym}}^o$ and $Q_{\mbox{\tiny sym}}^o$ are linearly correlated with $L_o$. 
The latter correlation is verified in Eq.~(\ref{qsymo}) and 
Fig.~\ref{ksymo-qsymo-nr}{\color{purple}b} as we already discussed. For this case we 
will have
\begin{multline}
L(\rho)\simeq(3x+1)\left\lbrace L_o + K_{\mbox{\tiny sym}}^ox + \frac{Q_{\mbox{\tiny 
sym}}^o}{2}x^2 \right\rbrace \\
=(3x+1)\left\lbrace L_o\left(\frac{a_2}{2} x^2 + a_1x +  1 \right)  
+ \left( \frac{b_2}{2} x + b_1\right)x\right\rbrace,
\label{lrhonrexp}
\end{multline}
with $b_1=q-3Jp$, $a_1=p$, $b_2=v-3Ju$, and $a_2=u$. The crossing is explained if 
$f(x_c^{\mbox{\tiny L}})= 1 + px_c^{\mbox{\tiny L}} + \frac{u}{2}{x_c^{\mbox{\tiny 
L}}}^2=0$, satisfied for
\begin{multline}
x_c^{\mbox{\tiny L}}=\frac{1}{5\left(3M^{2} - 22E_{\mbox{\tiny F}}^{o}M 
-144E_{\mbox{\tiny F}}^{o2} \right)}\bigg\{15M^2 - 54E_{\mbox{\tiny F}}^oM - \\
(315M^4 - 2700E_{\mbox{\tiny F}}^oM^3 + 1676E_{\mbox{\tiny F}}^{o2}M^2 + 
20160E_{\mbox{\tiny F}}^{o3}M)^{\frac{1}{2}}\bigg\},
\end{multline}
that produces $\rho_c^{\mbox{\tiny L}}/\rho_0=0.46$, a value much more close to the 
crossing density than $\rho_c^{\mbox{\tiny L}}/\rho_0=0.41$, found previously.

Therefore, it becomes clear that a crossing point in a density dependence of a bulk 
parameters indicates a route for the searching of linear correlations in its higher order 
derivatives. Nevertheless, we point out to the reader that a crossing point itself does 
not ensure linear correlations in all higher order bulk parameters. For the previous 
analysis, for example, one can not affirm that $L_o$ will be correlated with 
$I_{\mbox{\tiny sym}}^o=(3\rho_o)^4\left(\frac{\partial^4{\cal 
S}}{\partial\rho^4}\right)_{\rho=\rho_o}$, simply by the fact that the expansion of 
$L^{\mbox{\tiny (NR)}}/(3x+1)$ until order $x^3$ is better than those until order $x^2$. 
It is needed to check whether such linear correlation really holds. Actually, the crossing 
ensure at least the linear correlation between $L_o$ and the immediately next order bulk 
parameter $K_{\mbox{\tiny sym}}^o$. As a matter of fact, we investigate if $L_o$ 
correlates with $I_{\mbox{\tiny sym}}^o$ by obtaining the expression of the fourth order 
derivative of $\mathcal{S}^{\mbox{\tiny (NR)}}$, namely,
\begin{align}
I_{\mbox{\tiny sym}}^{\mbox{\tiny (NR)}}(\rho)=\frac{28\lambda\rho^{\frac{2}{3}}}{3M^2} 
\left(-M +\frac{5}{7} G^2_{\mbox{\tiny S}}\rho -\frac{20}{7}A\rho^{2} +\frac{330}{7} 
B\rho^{3}\right).
\label{isymrho-nr}
\end{align}

From Eq.~(\ref{isymrho-nr}), is possible to write $I_{\mbox{\tiny sym}}^o=I_{\mbox{\tiny 
sym}}^{\mbox{\tiny (NR)}}(\rho_o)$ in terms of $L_o$, $J$, $\rho_o$, $B_o$ and $K_o$ in 
the following way,
\begin{eqnarray}
I_{\mbox{\tiny sym}}^o= (L_o-3J)r(\rho_o) + s(\rho_o,B_o,K_o)
\label{isymo}
\end{eqnarray}
with
\begin{eqnarray}
r(\rho_o)=\frac{20\left(3M^2 -56ME_{\mbox{\tiny F}}^{o} +324E_{\mbox{\tiny F}}^{o2} 
\right)}{3M^{2} - 14ME_{\mbox{\tiny F}}^o},
\end{eqnarray}
and
\begin{multline}
s(\rho_o,B_o,K_o) = \frac{5}{3M^{2} - 14ME_{\mbox{\tiny F}}^o}\times \\
\Bigg\{1080B_{o}E_{\mbox{\tiny F}}^{o}\left(M -2E_{\mbox{\tiny F}}^{o}\right)\\
-12 E_{\mbox{\tiny F}}^{o}M \left(M -18E_{\mbox{\tiny F}}^{o}\right)
 -\frac{160 }{3} K_{o}E_{\mbox{\tiny F}}^{o}M
 \Bigg\}.
\end{multline}
Thus, it is verified that $I_{\mbox{\tiny sym}}^o$ also linearly correlates with $L_o$, 
under the same conditions that makes $L_o$ also correlated with $K_{\mbox{\tiny sym}}^o$ 
and $Q_{\mbox{\tiny sym}}^o$, namely, fixed values for $J$ and $K_o$. Furthermore, a new 
expansion of $L^{\mbox{\tiny (NR)}}/(3x+1)$ until order $x^3$ 
generates the cubic equation $f(x_c^{\mbox{\tiny L}})= 1 + px_c^{\mbox{\tiny L}} + 
\frac{u}{2!}{x_c^{\mbox{\tiny L}}}^2 + \frac{r}{3!}{x_c^{\mbox{\tiny L}}}^3=0$, 
presenting a root corresponding to $\rho_c^{\mbox{\tiny L}}/\rho_0=0.47$ (since 
$r(0.15)=14.17$), the exact value for the crossing density.

\begin{figure}[!htb]
\centering
\includegraphics[scale=0.33]{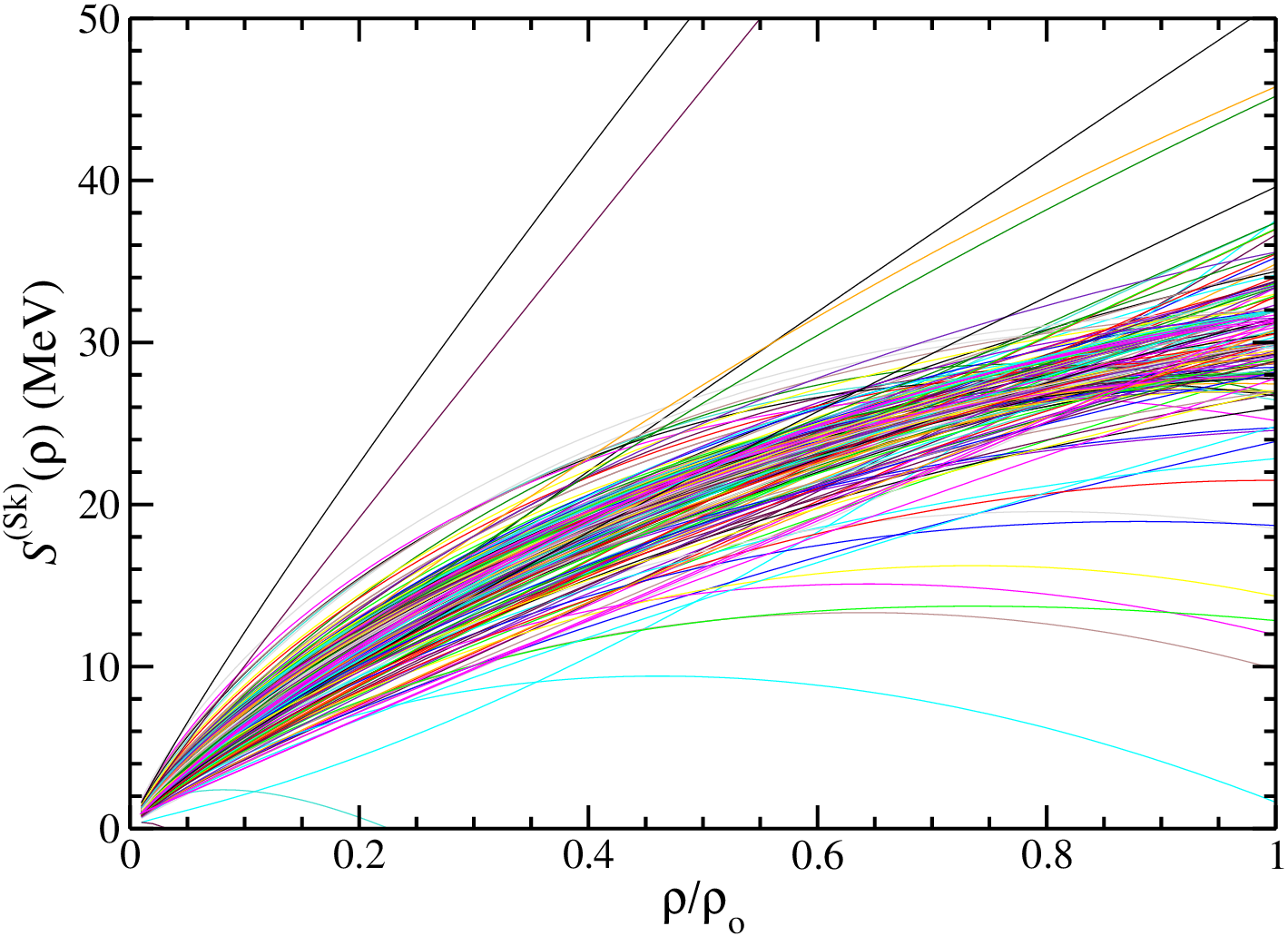}
\vspace{-0.2cm}
\caption{(Color online) Symmetry energy as a function of $\rho/\rho_o$ for the $240$ 
Skyrme parametrizations of Ref.~\cite{skyrme}.} 
\label{ssk240}
\end{figure}

\begin{figure}[!htb]
\centering
\includegraphics[scale=0.33]{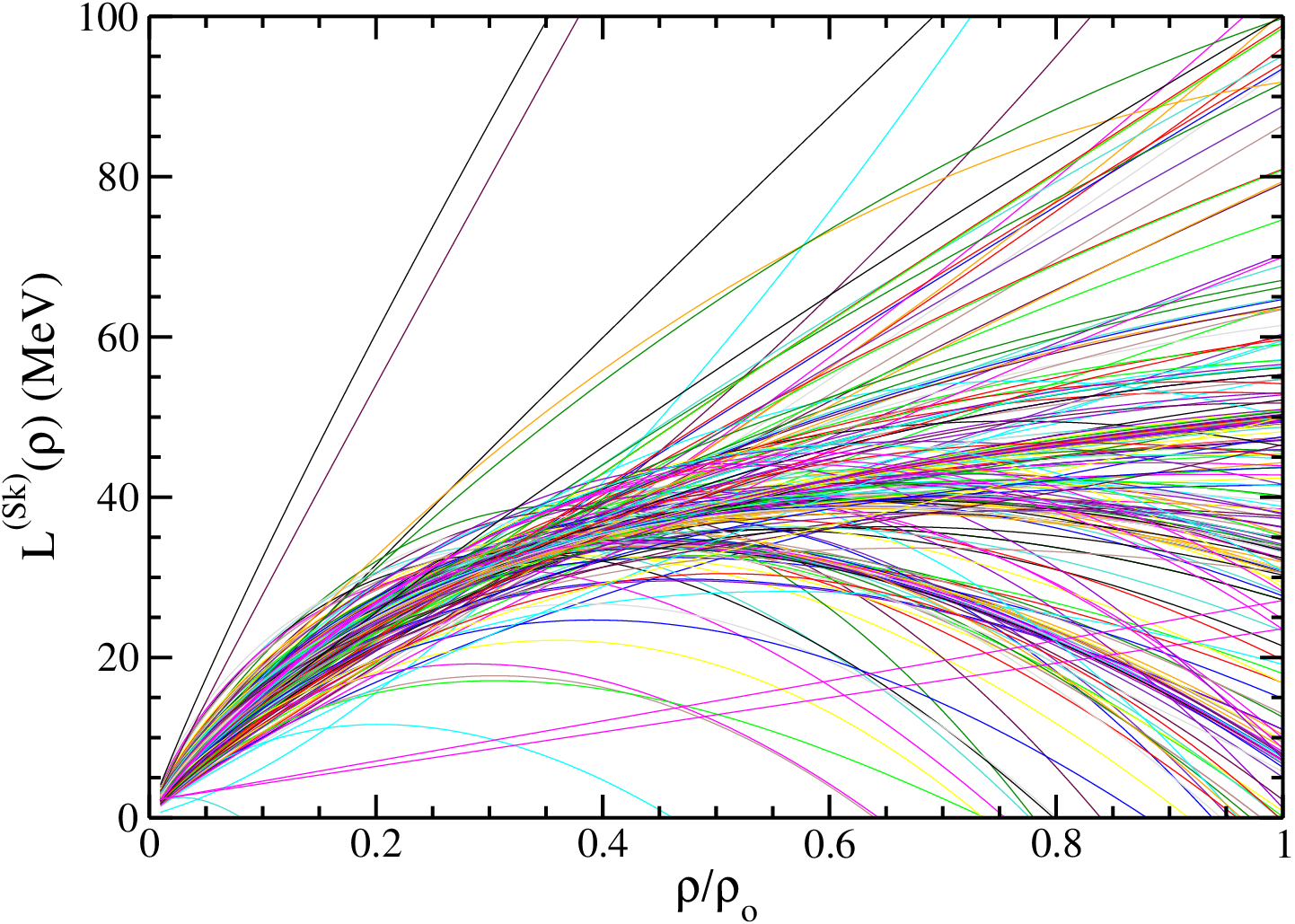}
\vspace{-0.2cm}
\caption{(Color online) Symmetry energy slope as a function of $\rho/\rho_o$ for the 
$240$ Skyrme parametrizations of Ref.~\cite{skyrme}.} 
\label{lsk240}
\end{figure}

Following these same ideas, we search for signatures of linear correlations in the Skyrme 
model. At this point, we remind the reader there is no unique crossing point at 
the density dependence of the symmetry energy and its slope for the Skyrme model, as we 
can see in Figs.~\ref{ssk240} and \ref{lsk240}, respectively, where we display the $240$ 
parametrizations of Ref.~\cite{skyrme}.

\noindent
This lack of a unique crossing in the density dependence of $\mathcal{S}(\rho)$ and 
$L(\rho)$ functions can also be seen in Fig.~2 (left) of Ref.~\cite{corecrust}, where the 
authors studied $21$ Skyrme parametrizations.

\begin{figure}[!htb]
\centering
\includegraphics[scale=0.33]{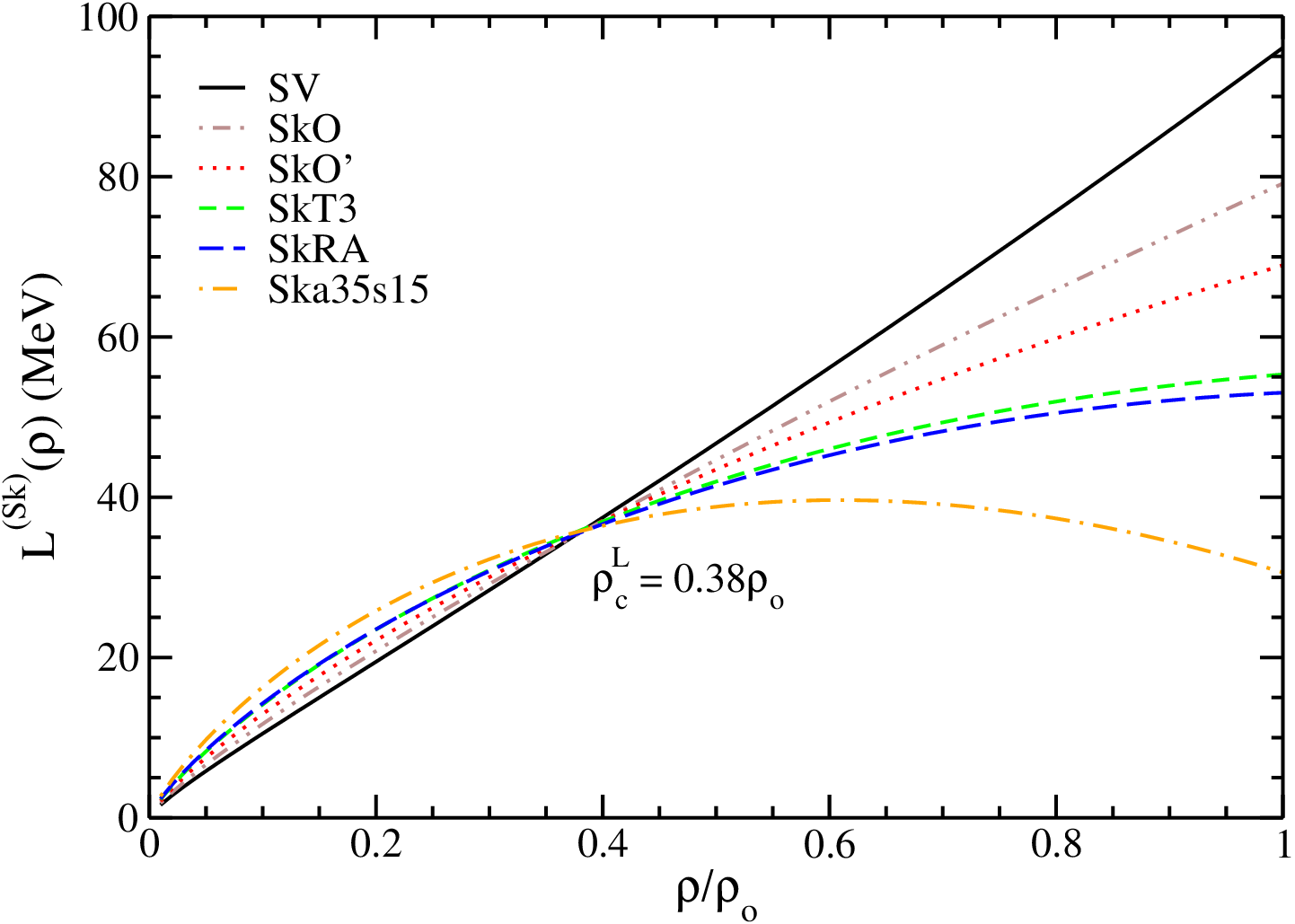}
\vspace{-0.2cm}
\caption{(Color online) Symmetry energy slope as a function of $\rho/\rho_o$ for some 
Skyrme parametrizations.} 
\label{crossing-l-skyrme}
\end{figure}

Specifically for the density dependence of the symmetry energy slope, we found a 
crossing density for the SV~\cite{sv}, SkO~\cite{sko}, SkO'~\cite{sko}, SkT3~\cite{skt3}, 
SkRA~\cite{skra} and Ska35s15~\cite{skyrme} parametrizations at $\rho_c^{\mbox{\tiny 
L}}/\rho_0=0.38$. It is displayed in Fig.~\ref{crossing-l-skyrme}.

\begin{figure}[!htb]
\centering
\includegraphics[scale=0.33]{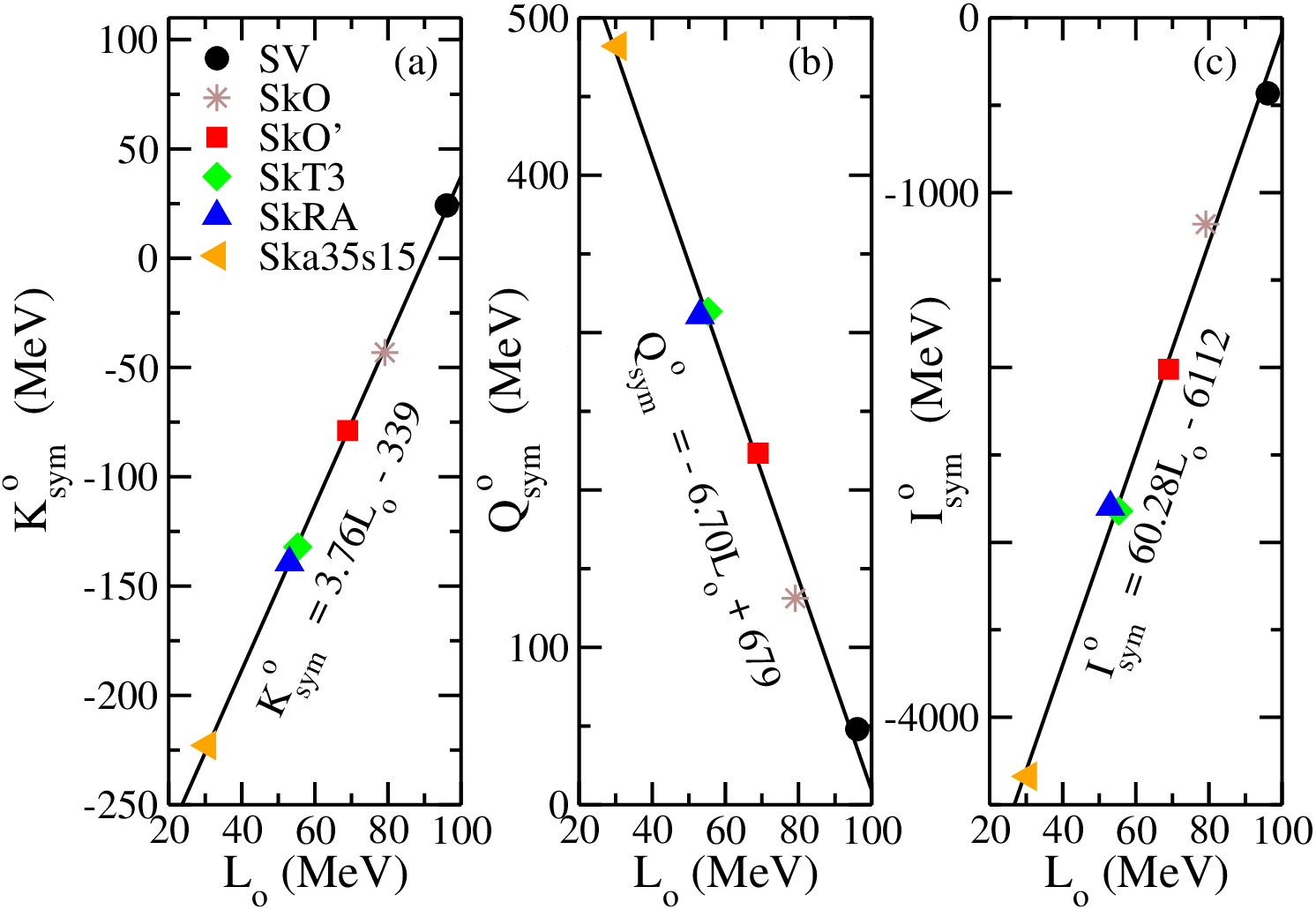}
\vspace{-0.2cm}
\caption{(Color online) (a) $K_{\mbox{\tiny sym}}^o$ (b) $Q_{\mbox{\tiny sym}}^o$, and 
(c) $I_{\mbox{\tiny sym}}^o$ as a function of $L_o$ for the Skyrme parametrizations of 
Fig.~\ref{crossing-l-skyrme}. The linear fits are indicated in each panel.} 
\label{ksymo-qsymo-isymo-skyrme}
\end{figure}

According to the discussed so far, these Skyrme parametrizations will present linear 
correlation at least regarding $K_{\mbox{\tiny sym}}^o$ and $L_o$. This is confirmed in 
Fig.~\ref{ksymo-qsymo-isymo-skyrme}{\color{purple}a}. Moreover, in 
Figs.~\ref{ksymo-qsymo-isymo-skyrme}{\color{purple}b} and 
~\ref{ksymo-qsymo-isymo-skyrme}{\color{purple}c} it is also checked the linear 
correlations of $L_o$ with $Q_{\mbox{\tiny sym}}^o$ and $I_{\mbox{\tiny sym}}^o$, 
respectively, like in the case of the NR limit.
For the sake of completeness, we have checked that the expansion of $L(\rho)/(3x+1)$ 
that approaches to the exact function around $\rho/\rho_o=0.38$ is taken until order 
$x^3$ for these Skyrme parametrizations. Therefore, the angular coefficients found in 
Fig.~\ref{ksymo-qsymo-isymo-skyrme} are used to define the function $f(x_c^{\mbox{\tiny 
L}})$ that needs to be null. This will provide the cubic equation $1 + 
3.76x_c^{\mbox{\tiny L}} - \frac{6.70}{2!}{x_c^{\mbox{\tiny L}}}^2 + 
\frac{60.28}{3!}{x_c^{\mbox{\tiny L}}}^3= 0$, that has one of the roots given by 
$\rho_c^{\mbox{\tiny L}}/\rho_0=0.38$, precisely the crossing density verified in 
Fig.~\ref{crossing-l-skyrme}.

\begin{figure}[!htb]
\centering
\includegraphics[scale=0.33]{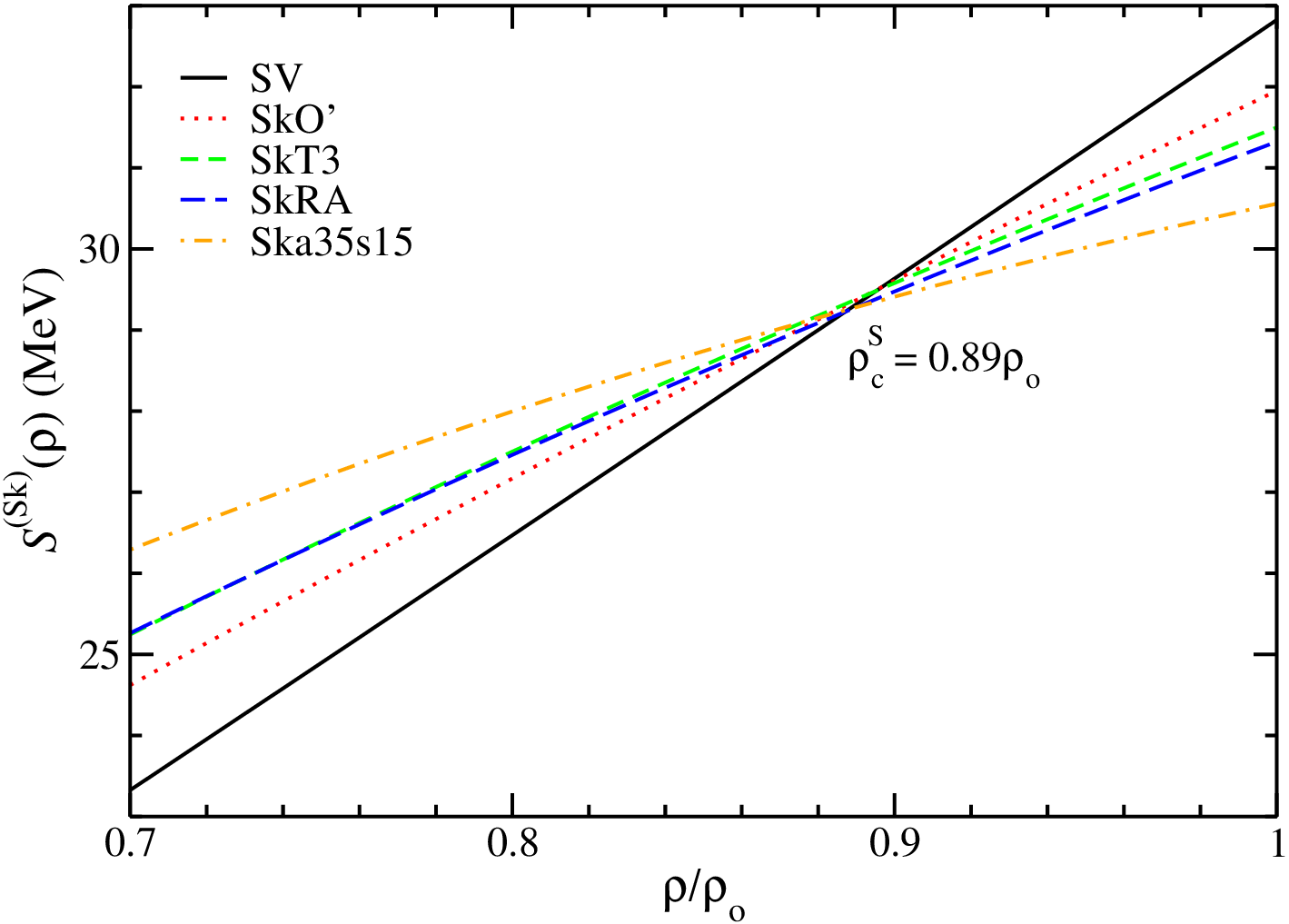}
\vspace{-0.2cm}
\caption{(Color online) Symmetry energy as a function of $\rho/\rho_o$ for some Skyrme 
parametrizations.} 
\label{crossing-s-skyrme}
\end{figure}

\begin{figure}[!htb]
\centering
\includegraphics[scale=0.33]{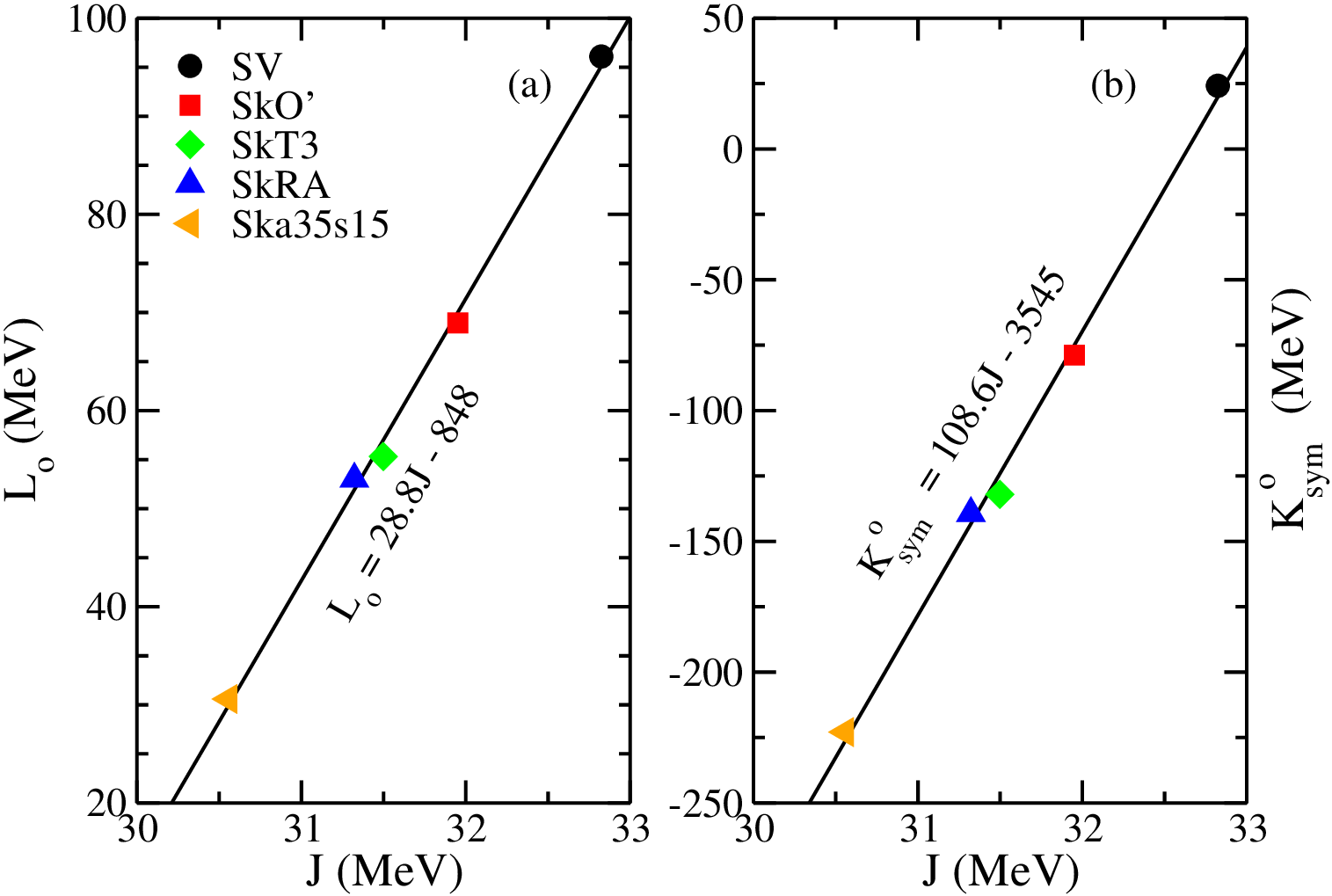}
\vspace{-0.2cm}
\caption{(Color online) (a) $L_o$ and (b) $K_{\mbox{\tiny sym}}^o$ as a function of $J$ 
for the Skyrme parametrizations of Fig.~\ref{crossing-s-skyrme}. The linear fits are 
indicated in each panel.} 
\label{lo-ksymo-skyrme}
\end{figure}

\begin{figure}[!htb]
\centering
\includegraphics[scale=0.33]{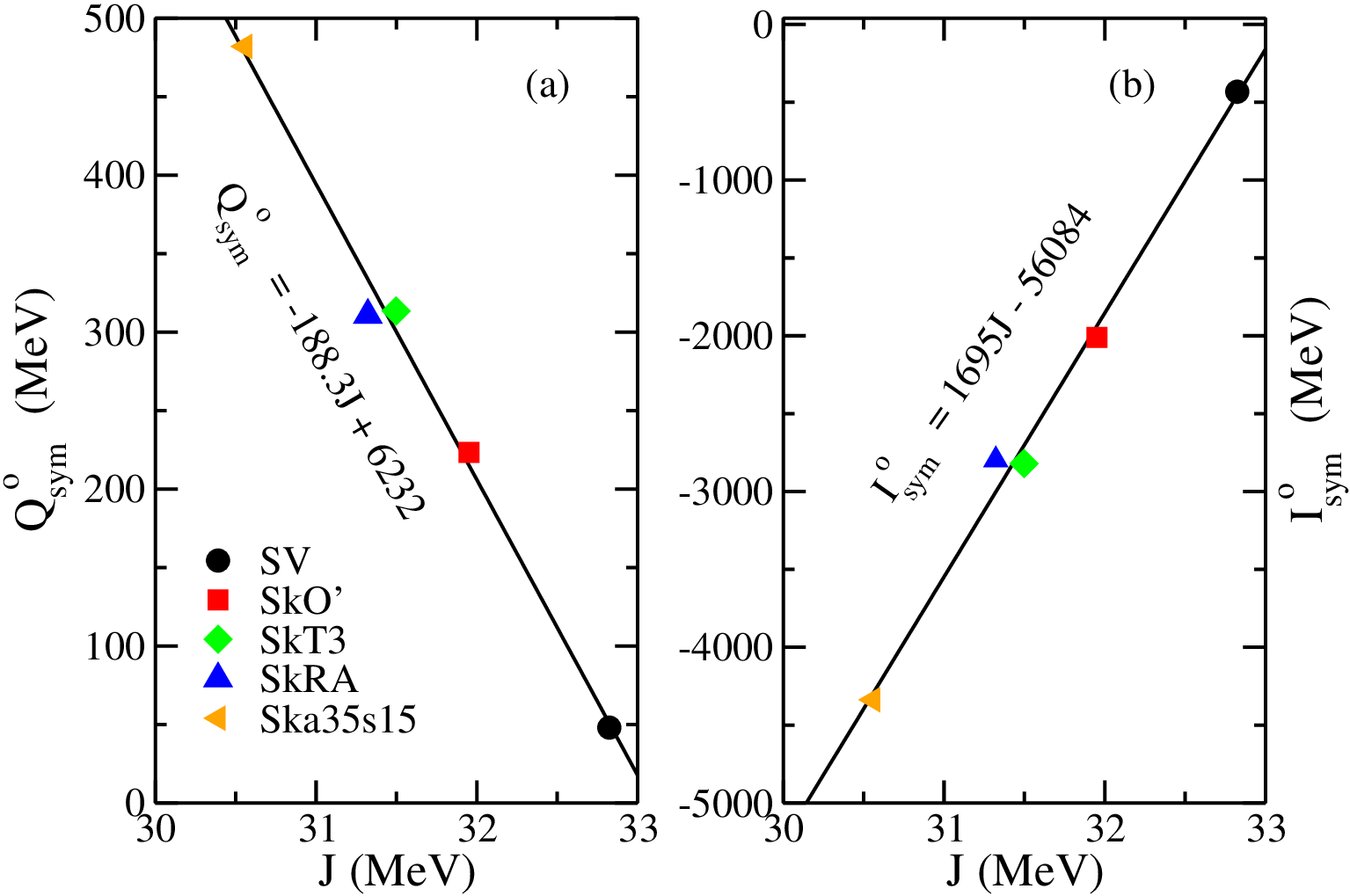}
\vspace{-0.2cm}
\caption{(Color online) (a) $Q_{\mbox{\tiny sym}}^o$ and (b) $I_{\mbox{\tiny sym}}^o$ as 
a function of $J$ for the Skyrme parametrizations of Fig.~\ref{crossing-s-skyrme}. The 
linear fits are indicated in each panel.} 
\label{qsymo-isymo-skyrme}
\end{figure}

Still at the framework of the Skyrme parametrizations, another crossing point is observed 
in the isovector sector, specifically in the density dependence of the symmetry energy 
itself. As pointed out in Fig.~\ref{crossing-s-skyrme}, such a crossing occurs at 
$\rho_c^{\mbox{\tiny S}}/\rho_0=0.89$.
We have noticed that for these parametrizations, $J$ is linearly correlated with $L_o$, 
$K_{\mbox{\tiny sym}}^o$, $Q_{\mbox{\tiny sym}}^o$ and $I_{\mbox{\tiny sym}}^o$ as one 
can see in Figs.~\ref{lo-ksymo-skyrme} and \ref{qsymo-isymo-skyrme}. The angular 
coefficients of these lines are used to define the quartic equation to be solved in order 
to determine the value of the crossing density, namely, $1 + 28.8x_c^{\mbox{\tiny S}} + 
\frac{108.6}{2!}{x_c^{\mbox{\tiny S}}}^2 - \frac{188.3}{3!}{x_c^{\mbox{\tiny S}}}^3 + 
\frac{1695}{4!}{x_c^{\mbox{\tiny S}}}^4 = 0$. One root of this equation provides the 
value $\rho_c^{\mbox{\tiny S}}/\rho_0=0.89$, observed in Fig.~\ref{crossing-s-skyrme}.

\begin{figure}[!htb]
\centering
\includegraphics[scale=0.33]{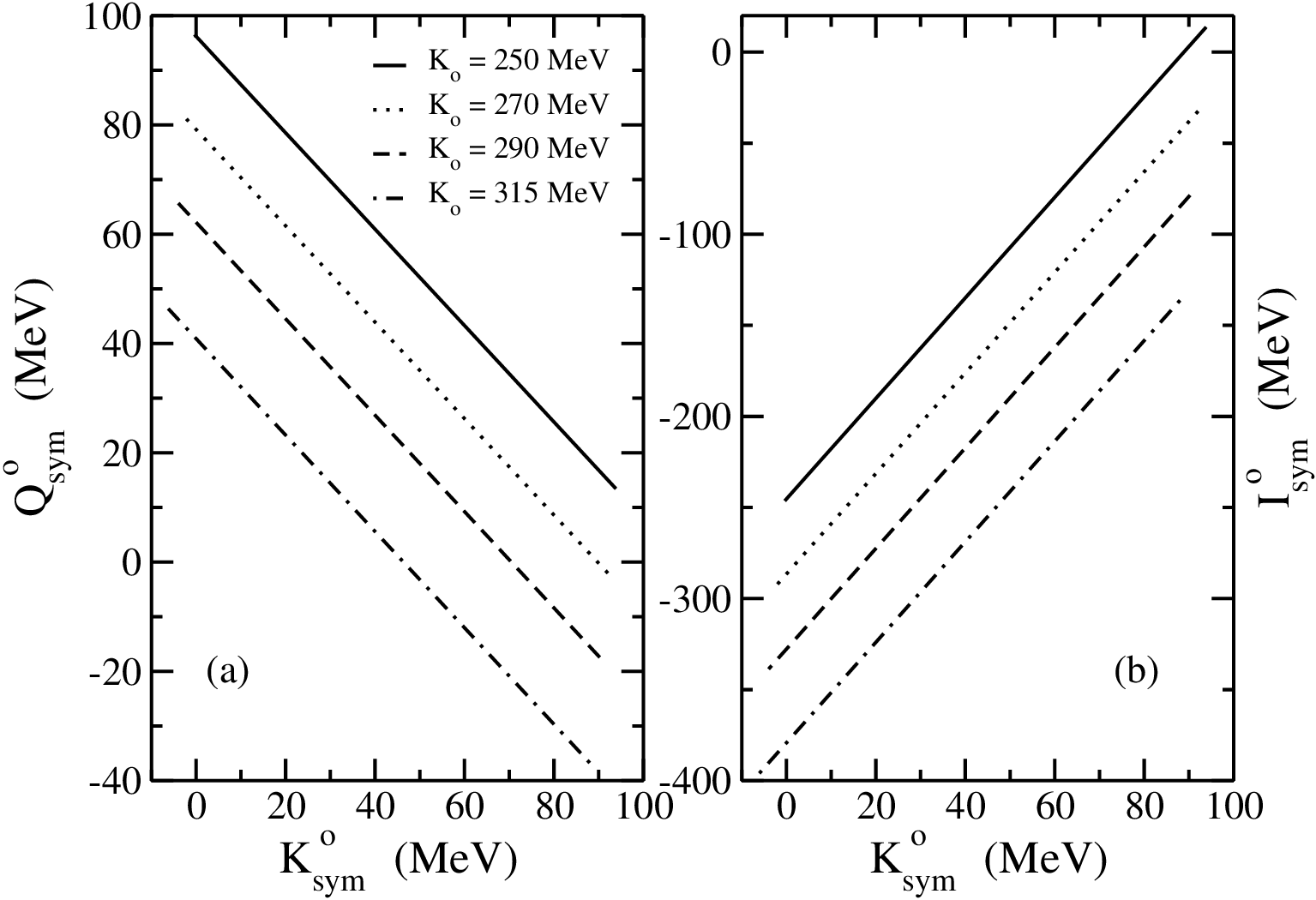}
\vspace{-0.2cm}
\caption{(a) $Q_{\mbox{\tiny sym}}^o$ and (b) $I_{\mbox{\tiny sym}}^o$ as a function 
of $K_{\mbox{\tiny sym}}^o$ for the NR limit. The effective mass varies in the range of 
$0.50\leqslant m^*\leqslant 0.80$. For each panel, $\rho_o=0.15$~fm$^{-3}$ and 
$B_o=16$~MeV.} 
\label{qsymo-isymo-nr}
\end{figure}

As a remark, it is worthwhile to note that a linear correlation itself between two bulk 
parameters is not a sufficient condition to guarantee a crossing point in the density 
dependence of the immediately preceding bulk parameter. As an example of this statement, 
we focus on the analytical structure of the NR limit to find other two specific linear 
correlations. From Eqs.~(\ref{ksymo}) and (\ref{qsymo}) it is straightforward to obtain
\begin{eqnarray}
Q_{\mbox{\tiny sym}}^o = \frac{u}{p}K_{\mbox{\tiny sym}}^o + v - \frac{uq}{p}.
\end{eqnarray}
From Eqs.~(\ref{ksymo}) and (\ref{isymo}), an analogous expression can be written relating 
$I_{\mbox{\tiny sym}}^o$ and $K_{\mbox{\tiny sym}}^o$, namely,
\begin{eqnarray}
I_{\mbox{\tiny sym}}^o = \frac{r}{p}K_{\mbox{\tiny sym}}^o + s - \frac{rq}{p}.
\end{eqnarray}
Therefore, one see for a fixed value of $K_o$, that the bulk parameter 
$K_{\mbox{\tiny sym}}^o$ is linearly correlated with $Q_{\mbox{\tiny sym}}^o$ and 
$I_{\mbox{\tiny sym}}^o$, see Fig.~\ref{qsymo-isymo-nr}.

\noindent
Here, we highlight that one can vary $K_{\mbox{\tiny sym}}^o$ without any information 
regarding $L_o$. Let us remind that $K_{\mbox{\tiny sym}}^o$ can also be written only 
as a function of the isoscalar parameters, according to Eq.~(12) of Ref.~\cite{bianca}. 
Thus, for fixed values of $K_o$, it is possible to investigate how $Q_{\mbox{\tiny 
sym}}^o$ and $I_{\mbox{\tiny sym}}^o$ depend on $K_{\mbox{\tiny sym}}^o$ only from the 
variation of $m^*$.

\begin{figure}[!htb]
\centering
\includegraphics[scale=0.33]{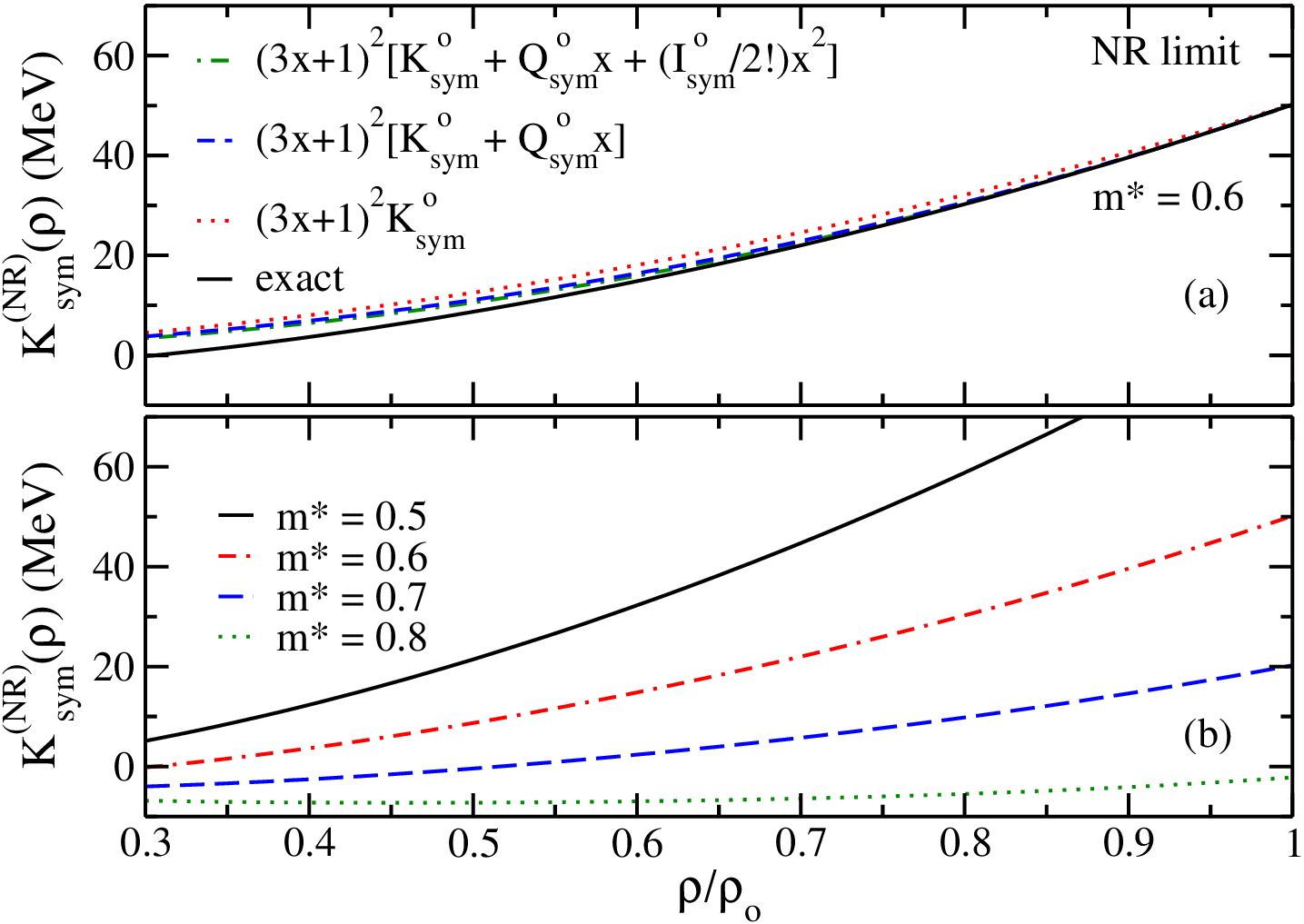}
\vspace{-0.2cm}
\caption{(Color online) $K_{\mbox{\tiny sym}}^{\mbox{\tiny (NR)}}$ as a function of 
$\rho/\rho_o$. (a) Exact function for $m^*=0.6$, Eq.~(\ref{ksymrhonr}), compared with its 
expansion. (b) Exact function for different effective mass values. The bulk parameters 
fixed in the two panels are $K_o=270$~MeV, $\rho_o=0.15$~fm$^{-3}$ and $B_o=16$~MeV.} 
\label{ksymrhoexp-crossing-nr}
\end{figure}

Notice that, the expansion of $K_{\mbox{\tiny sym}}^{\mbox{\tiny (NR)}}/(3x+1)^2$ until order 
$x^2$ can describe the exact function in Eq.~(\ref{ksymrhonr}) in a density region of 
subsaturation densities, see Fig.~\ref{ksymrhoexp-crossing-nr}{\color{purple}a}. In spite 
of this, the quadratic equation constructed from the angular coefficients extracted 
from Fig.~\ref{qsymo-isymo-nr}, namely, $1 - 0.88x_c^{{\mbox{\tiny K}}_{\mbox{\tiny sym}}} 
+ \frac{2.76}{2!}{x_c^{{\mbox{\tiny K}}_{\mbox{\tiny sym}}}}^2 = 0$, presents no real 
roots, indicating no crossing points in the $K_{\mbox{\tiny sym}}^{\mbox{\tiny 
(NR)}}(\rho)$ function. This finding is confirmed in 
Fig.~\ref{ksymrhoexp-crossing-nr}{\color{purple}b}.

\begin{figure}[!htb]
\centering
\includegraphics[scale=0.33]{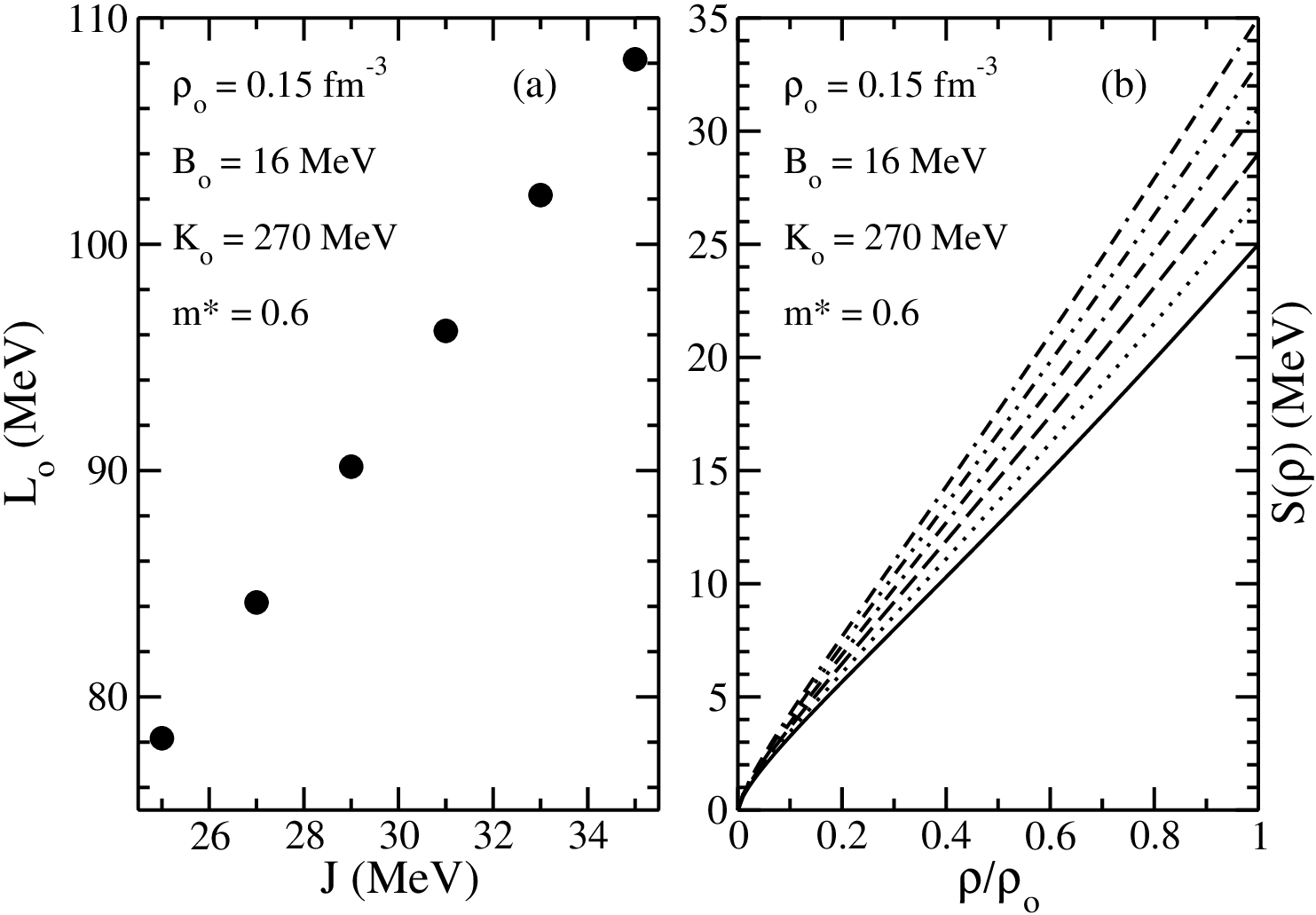}
\vspace{-0.2cm}
\caption{(a) $L_o\times J$ correlation obtained from Eq.~(\ref{lj}). (b) density 
dependence of symmetry energy, Eq.~(\ref{snr}).}
\label{jl-srho-nrl}
\end{figure}

Another example of linear correlations between bulk parameters and the lack of 
crossing points in the density dependence in one of them, is the case of the quantities 
$L_o$ and $J$. According to Eq.~(\ref{lj}), a linear correlation between $J$ and $L_o$ is 
established if the function $b(m^*,\rho_o,B_o,K_o)$ is kept fixed, i. e., if the analyzed 
parametrizations have exactly the same isoscalar bulk parameters. Thus, parametrizations 
with different $J$ values but same isoscalar parameters, present the behavior depicted in 
Fig.~\ref{jl-srho-nrl}{\color{purple}a}.
 \\
As one can see in Fig.~\ref{jl-srho-nrl}{\color{purple}b}, these parametrizations 
do not generate any crossing point in the density dependence of $\mathcal{S}(\rho)$. From 
the point of view of the discussion of Sec.~\ref{lincorr}, one can understand the lack of 
crossing points from the equation $1 + 3x_c^{\mbox{\tiny S}} = 0$, satisfied only 
for $x_c^{\mbox{\tiny S}} = -\frac{1}{3}$, i.~e., for $\rho_c^{\mbox{\tiny 
S}}/\rho_0=0$ (since we have $\rho/\rho_0=3x+1$).

In summary, one can associate linear correlations as signatures of crossing points only if 
the exact function studied can be approximated by its expansion, and simultaneously, if 
the equation $f(x_c)=0$ present nonzero real roots in the analysed range of densities. In 
the case of the NR limit, the latter condition is not satisfied in the study of the 
$K_{\mbox{\tiny sym}}^{\mbox{\tiny (NR)}}(\rho)$ function, implying in no crossing points 
in its density dependence in the range of subsaturation densities.

\subsection{Results from the isoscalar sector}

Regarding the quantities related to the isoscalar sector of the nonrelativistic hadronic 
models, we underline here the relationship between $K_o$, $Q_o$ and $I_o$. For the former 
two quantities, a correlation was firstly found in Refs.~\cite{prl,margueron} for some 
Skyrme parametrizations. In particular, they found it as a linear one. Here we proceed to 
find, in the NR limit framework, the conditions that the parametrizations must satisfy in 
order to gives rise to the same relationship. For this purpose, we first need to obtain 
$Q(\rho,y)$ from the energy density, Eq.~(\ref{enr}). This is done by calculating 
$Q=(3\rho)^3\left[\frac{\partial^3(\mathcal{E}/\rho)}{\partial\rho^3}\right]$, with the 
full expression given by
\begin{multline}
Q^{\mbox{\tiny (NR)}}(\rho,y)=-162B\rho^3 + \frac{12\lambda 
H_{5/3}}{5M^2}\rho^{\frac{2}{3}}\times \\
\times\left(M -\frac{5}{4} G^2_{\mbox{\tiny S}}\rho + 20A\rho^{2} + 165B\rho^{3}\right),
\end{multline}
where $Q_o$ is defined by $Q_o=Q^{\mbox{\tiny (NR)}}(\rho_o,1/2)$.

\begin{figure}[!htb]
\centering
\includegraphics[scale=0.33]{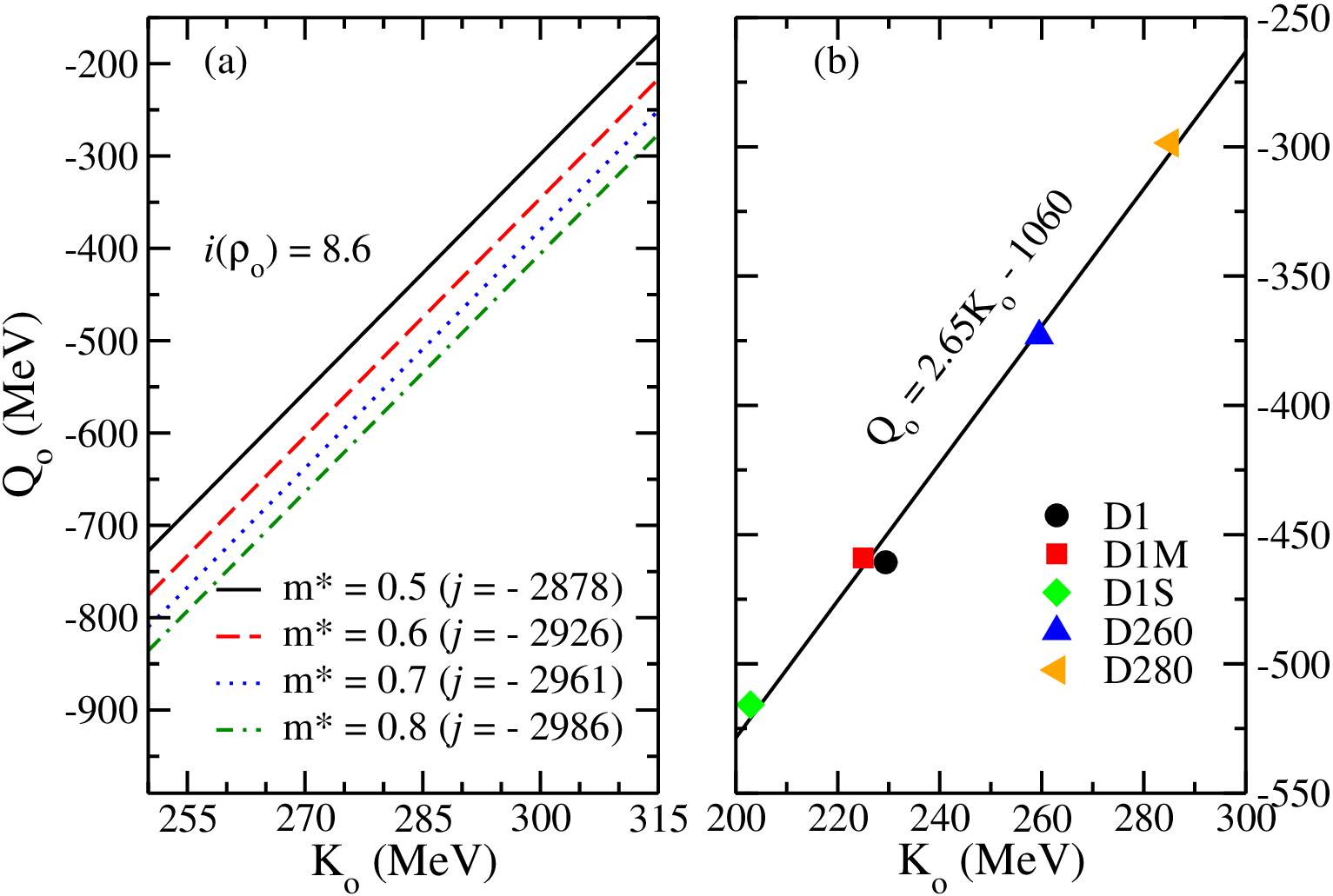}
\vspace{-0.2cm}
\caption{(Color online) $Q_o$ as a function of $K_o$ for (a) the NR limit with different 
$m^*$ values, and for (b) some Gogny parametrizations studied in Ref.~\cite{gogny}. In 
panel (a), $\rho_o=0.15$~fm$^{-3}$ and $B_o=16$~MeV.} 
\label{qo-ko-nr-gogny}
\end{figure}

By using  the coupling constants in terms of the bulk parameters in the 
expressions of $K_o$ and $Q_o$, it is possible to find the following relationship between 
these quantities,
\begin{eqnarray}
Q_o = i(\rho_o)K_o + j(\rho_o, B_o,m^{*}),
\label{qk-nr}
\end{eqnarray}
where
\begin{eqnarray}
i(\rho_o) = \frac{3\left(9M^{2} -73E_{\mbox{\tiny F}}^{o}M +90E_{\mbox{\tiny 
F}}^{o2}\right)}{3M^{2} - 19E_{\mbox{\tiny F}}^{o}M +18E_{\mbox{\tiny F}}^{o2}},
\label{inr}
\end{eqnarray}
and
\begin{multline}
j(\rho_o, B_o,m^{*}) = \frac{1}{3M^2 - 19E_{\mbox{\tiny F}}^{o}M + 18E_{\mbox{\tiny 
F}}^{o2}}\times\\
\times\Bigg\{\frac{24E_{\mbox{\tiny F}}^{o}}{m^{*}}\left(M^{2}-40ME_{\mbox{\tiny F}}^{o}/3 
+60 E_{\mbox{\tiny F}}^{o2}\right)\\
-  162B_{o}\left(3M^{2}-25ME_{\mbox{\tiny F}}^{o} +40E_{\mbox{\tiny F}}^{o2}  \right)\\
-  108E_{\mbox{\tiny F}}^{o}\left( M^{2} -10ME_{\mbox{\tiny F}}^{o} +27E_{\mbox{\tiny 
F}}^{o2} \right)  \Bigg\}.
\end{multline}
From the structure presented in Eq.~(\ref{qk-nr}), we notice that the effective mass 
plays a crucial role for the correlation between $Q_o$ and $K_o$. For fixed values of 
$m^*$, this correlation is a linear one. We remember the reader that $\rho_o$ and $B_o$ 
vary only in a very narrow range around $0.15$~fm$^{-3}$ and $16$~MeV, respectively.
Thus, these bulk parameters can be considered constants for the hadronic mean-field models. 
For different $m^*$ values, Eq.~(\ref{qk-nr}) generates parallel lines as indicated in 
Fig.~\ref{qo-ko-nr-gogny}{\color{purple}a}. The same linear correlation is also observed 
in some Gogny parametrizations, as indicated in 
Fig.~\ref{qo-ko-nr-gogny}{\color{purple}b}.

\begin{figure}[!htb]
\centering
\includegraphics[scale=0.33]{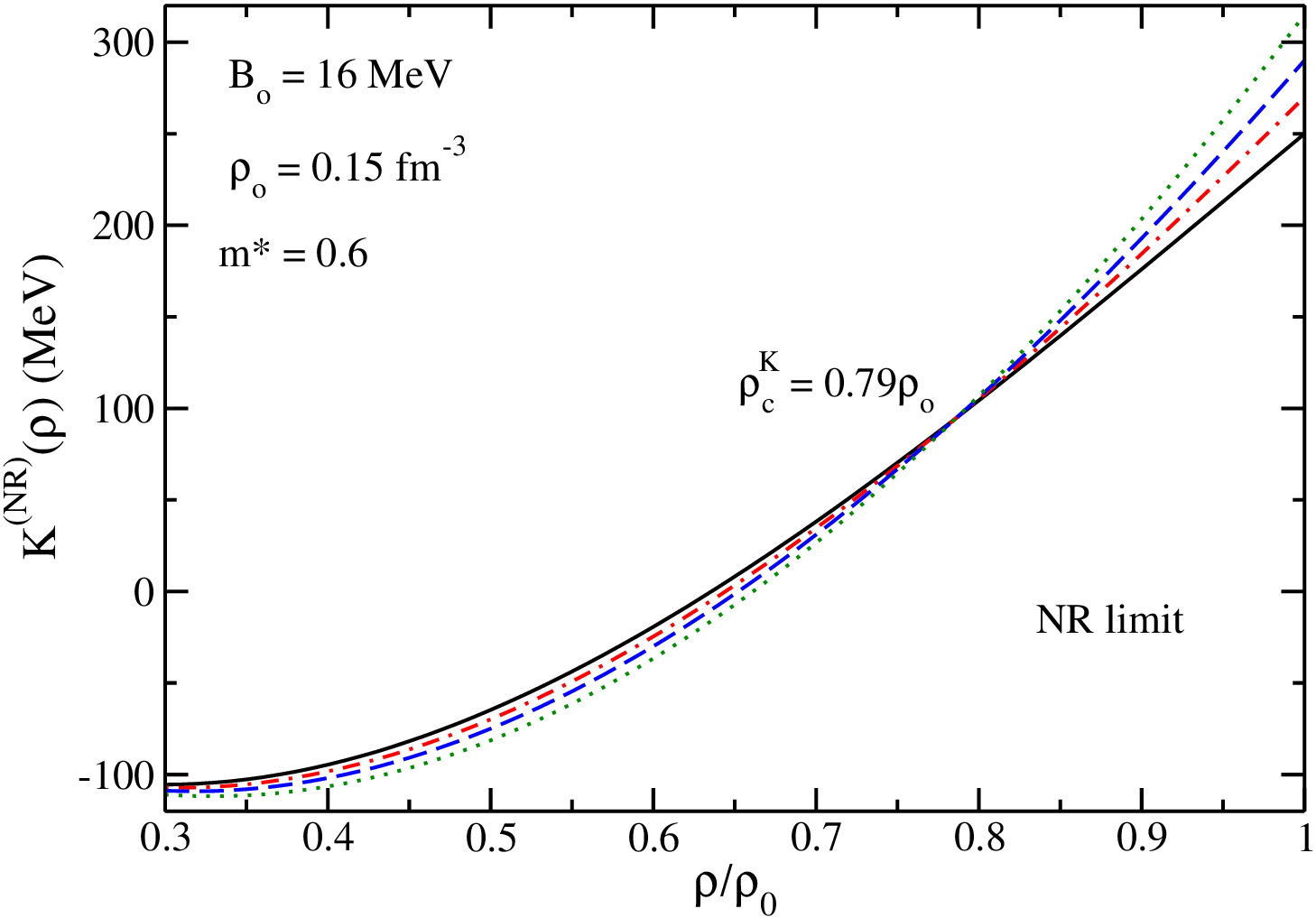}
\vspace{-0.2cm}
\caption{(Color online) Incompressibility as a function of $\rho/\rho_o$ for the NR 
limit, Eq.~(\ref{knr}).} 
\label{crossing-k-nr}
\end{figure}

From the perspective addressed in Sec.~\ref{lincorr}, the linear correlation between 
$K_o$ and $Q_o$ could also be sought, by searching for a possible crossing point in the 
density dependence of the incompressibility. In fact, as pointed out in 
Fig.~\ref{crossing-k-nr}, there are two of them, at $\rho_c^{\mbox{\tiny K}}/\rho_0=0.21$ 
(not shown), and $\rho_c^{\mbox{\tiny K}}/\rho_0=0.79$.
Furthermore, notice that the the second point is quite close to the crossing density 
$\rho_c^{\mbox{\tiny K}}/\rho_0=0.71$ observed also for the Skyrme parametrizations 
studied in Ref.~\cite{margueron}.

Since crossings in the $K^{\mbox{\tiny (NR)}}(\rho)$ function were found, they indicate
a signature of a linear correlation, in this case at least between $K_o$ and $Q_o$, since 
the latter is the bulk parameter associated to the immediately next order derivative of 
$K(\rho)$. Nevertheless, for the NR limit, we can also check analytically if the next 
bulk parameter, $I_o$, correlates with $K_o$. From Eq.~(\ref{enr}) and the definition 
$I=(3\rho)^3\left[\frac{\partial^3(\mathcal{E}/\rho)}{\partial\rho^3}\right]$, one 
obtains,
\begin{multline}
I^{\mbox{\tiny (NR)}}(\rho,y) = \frac{84\lambda 
H_{5/3}}{5M^2}\rho^{\frac{2}{3}}  \times \\
\times\left(-M +\frac{5}{7} G^2_{\mbox{\tiny S}}\rho -\frac{20}{7}A\rho^{2} 
+\frac{330}{7} 
B\rho^{3}\right).
\end{multline}
From this expression, the fourth order derivative of the energy per particle evaluated at 
the saturation density, $I_o=I^{\mbox{\tiny (NR)}}(\rho_o,1/2)$, can 
be written in terms of $\rho_o$, $B_o$, $m^*$, and $K_o$ as
\begin{eqnarray}
I_o = w(\rho_o)K_o + z(\rho_o, B_o,m^{*}),
\label{ik-nr}
\end{eqnarray}
where
\begin{eqnarray}
w(\rho_o)=-\frac{20\left(25E_{\mbox{\tiny F}}^oM-54E_{\mbox{\tiny F}}^{o2}\right)}
{3M^{2} - 19E_{\mbox{\tiny F}}^{o}M +18E_{\mbox{\tiny F}}^{o2}},
\end{eqnarray}
and
\begin{multline}
z(\rho_o,B_o,m^*)=\frac{1}{3M^2-19E_{\mbox{\tiny F}}^oM+18E_{\mbox{\tiny F}}^{o2}}\times\\
\times\Bigg\{\frac{40E_{\mbox{\tiny F}}^o}{m^*}\left(3M^2-56ME_{\mbox{\tiny F}}^o
+324E_{\mbox{\tiny F}}^{o2}\right) \\
+ 1080B_o\left(9ME_{\mbox{\tiny F}}^o -32E_{\mbox{\tiny F}}^{o2}  \right)\\
-72E_{\mbox{\tiny F}}^o\left(4M^2-77ME_{\mbox{\tiny F}}^o+324E_{\mbox{\tiny F}}^{o2}
\right)\Bigg\}.
\end{multline}
Notice that once more, the effective mass needs to be constant for the parametrizations in 
order to ensure a linear dependence between $I_o$ and $K_o$, with angular coefficient 
given by $w(0.15)=-4.16$.

\begin{figure}[!htb]
\centering
\includegraphics[scale=0.33]{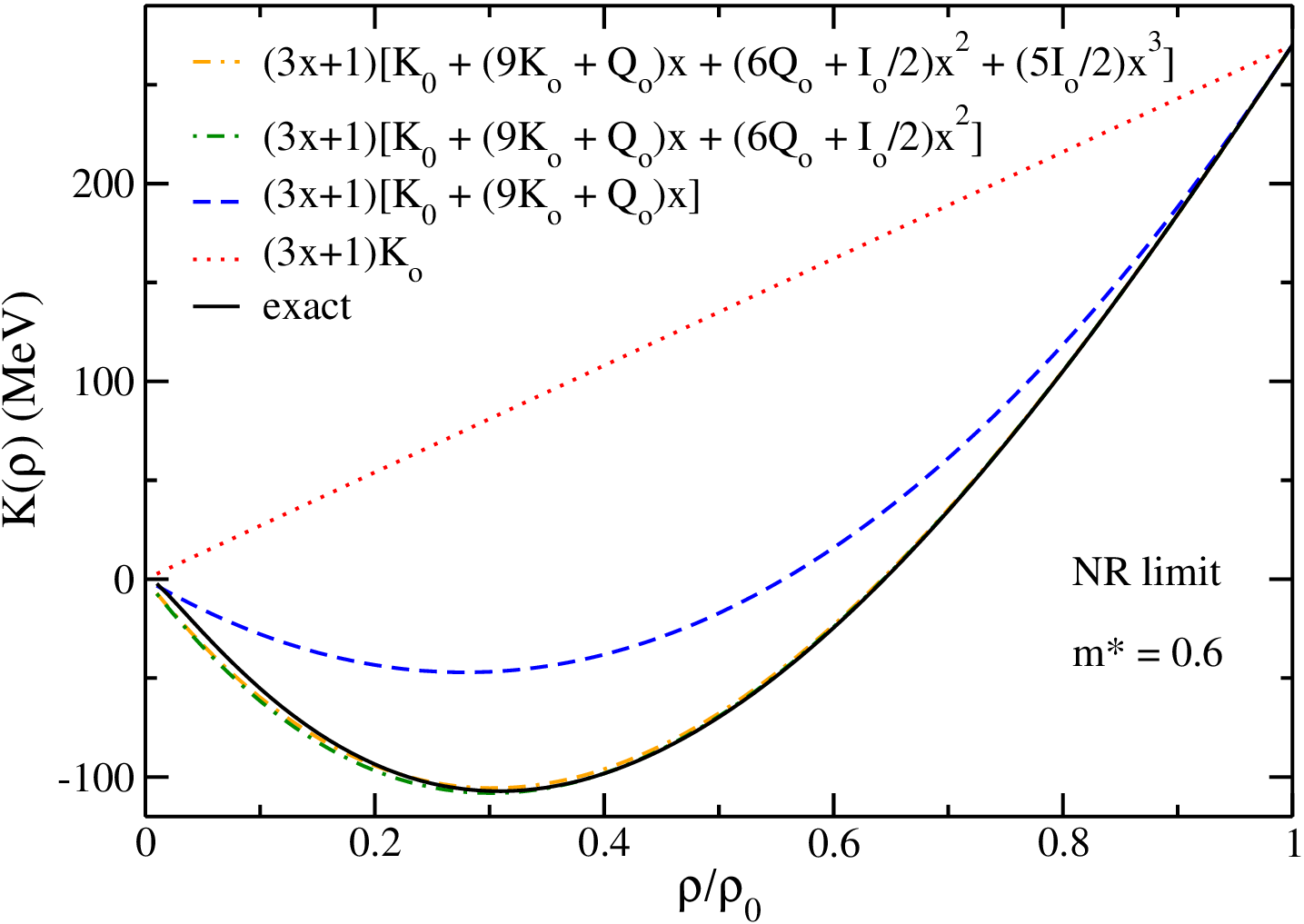}
\vspace{-0.2cm}
\caption{(Color online) Density dependence of the incompressibility compared with its 
expansion for the NR limit.} 
\label{krhoexp-nr}
\end{figure}

For the sake of completeness, we use the angular coefficients $i(\rho_o)$ and $w(\rho_o)$ 
to calculate the crossing density in Fig.~\ref{crossing-k-nr}. First, we consider the 
energy per particle of symmetric nuclear matter as
\begin{eqnarray}
E(\rho)\simeq E_o  + \frac{K_o}{2!}x^2 + \frac{Q_o}{3!}x^3 + \frac{I_o}{4!}x^4,
\label{epp}
\end{eqnarray}
then, the corresponding expansion for $K(\rho)$ reads
\begin{eqnarray}
&K&(\rho) = 18\rho\frac{\partial E}{\partial\rho}+9\rho^2\frac{\partial^2E}{
\partial\rho^2} \nonumber \\
&=& 6(3x+1)\frac{\partial E}{\partial x} + (3x+1)^2\frac{\partial^2E}{\partial x^2} 
\label{kx} \\
&\simeq&(3x+1)\times \nonumber \\
&\times&\left[ K_o + (9K_o+Q_o)x + \left(6Q_o+\frac{I_o}{2}\right)x^2 
+ \frac{5I_o}{2}x^3\right].\quad\,\,\,\,\,
\label{kexp}
\end{eqnarray}
This expansion is observed to be consistent with the exact function, Eq.~(\ref{knr}), 
as showed in Fig.~\ref{krhoexp-nr}.

It is worth noting that despite the extra term in Eq.~(\ref{kx}) compared with 
$\mathcal{F}^{\prime\prime}$ in Eq.~(\ref{fpp}), the final expansion, Eq.~(\ref{kexp}), 
is analogous to the general function $\mathcal{F}^{(m)}(\rho)$ in Eq.~(\ref{fgeral}). 
Therefore, all the procedure developed in Sec.~\ref{lincorr} also applies in the analysis 
of correlations and crossing points for the $K^{\mbox{\tiny (NR)}}(\rho)$ function in 
the isoscalar sector. Indeed, this was done for some Skyrme parametrizations in 
Ref.~\cite{margueron}. From this point of view, it is possible to use Eqs.~(\ref{qk-nr}) 
and (\ref{ik-nr}) to rewrite Eq.~(\ref{kexp}) as
\begin{multline}
K(\rho) = (3x+1)\times \\
\times
\left\lbrace K_o\left[1+(9+i)x+\left(6i+\frac{w}{2}\right)x^2+\frac{5w}{2}x^3\right] 
\right. \\
+\left.\left[j + \left(6j + \frac{z}{2}\right)x + \frac{5z}{2}x^2 \right]x\right\rbrace.
\end{multline}
Thus, the crossings points in Fig.~\ref{crossing-k-nr}{\color{purple}a} are justified if
$1 + (9+i)x_c^{\mbox{\tiny K}} + (6i+w/2){x_c^{{\mbox{\tiny K}}}}^2 + 
(5w/2){x_c^{{\mbox{\tiny K}}}}^3 = 0$. Two solutions of this cubic equation are 
$\rho_c^{\mbox{\tiny K}}/\rho_0=0.21$, and $\rho_c^{\mbox{\tiny K}}/\rho_0=0.79$.

\section{Correlations in \mbox{FR-RMF} models}
\label{rel}

In the context of QHD, protons and neutrons are the fundamental particles interacting each 
other through scalar and vector mesons exchange. In this framework, the fields $\sigma$ 
and $\omega$ represent, respectively, these mesons and mimic the attractive and repulsive 
parts of the nuclear interaction. The main representative of QHD models is the Walecka 
one~\cite{walecka}, in which the only two free parameters are fitted in order to 
reproduce the values of $\rho_o$ and $B_o$.  However, it does not give reasonable values
for $K_o$ ($\sim 500$~MeV), and $M^*_o$ ($\sim 0.54M$). This problem was circumvented by 
Boguta and Bodmer~\cite{boguta}, who added to the Walecka model cubic and quartic 
self-interactions in the scalar field $\sigma$, introducing, consequently, two more free
parameters, which are fitted so as to fix these quantities. All thermodynamic quantities 
of this model are found from its Lagrangian density, given by
\begin{eqnarray}
\mathcal{L} &=& \overline{\psi}(i\gamma^\mu\partial_\mu - M)\psi + 
g_\sigma\sigma\overline{\psi}\psi + \frac{1}{2}(\partial^\mu \sigma \partial_\mu \sigma - 
m^2_\sigma\sigma^2) 
\nonumber \\
&-& \frac{A}{3}\sigma^3 - \frac{B}{4}\sigma^4 
- g_\omega\overline{\psi}\gamma^\mu\omega_\mu\psi -\frac{1}{4}F^{\mu\nu }F_{\mu\nu} + 
\frac{1}{2}m^2_\omega\omega_\mu\omega^\mu
\nonumber\\
&-&\frac{g_\rho}{2}\overline{\psi}\gamma^\mu\vec{\rho}_\mu\vec{\tau}\psi 
-\frac{1}{4}\vec{B}^{\mu\nu}\vec{B}_{\mu\nu} 
+ \frac{1}{2}m^2_\rho\vec{\rho}_\mu\vec{\rho}^\mu,
\label{ld}
\end{eqnarray}
with $F_{\mu\nu}=\partial_\nu\omega_\mu-\partial_\mu\omega_\nu$, and 
$\vec{B}_{\mu\nu}=\partial_\nu\vec{\rho}_\mu-\partial_\mu\vec{\rho}_\nu
- g_\rho (\vec{\rho}_\mu \times \vec{\rho}_\nu)$. The coupling constants are $g_\sigma$, 
$g_\omega$, $g_\rho$, $A$ and $B$. For a complete description of the model, and also 
other kind of \mbox{FR-RMF} ones, such as density dependent, crossed terms and nonlinear 
point-couplings, we address the reader to the recent study of Ref.~\cite{rmf} involving an 
analysis of $263$ relativistic parametrizations under constraints related to symmetric 
nuclear matter, pure neutron matter, symmetry energy, and its derivatives. Here, we mainly 
focus in searching for correlations between bulk parameters for the \mbox{FR-RMF} 
parametrizations described by Eq.~(\ref{ld}).

\subsection{Isovector sector}

\begin{figure}[!htb]
\centering
\includegraphics[scale=0.33]{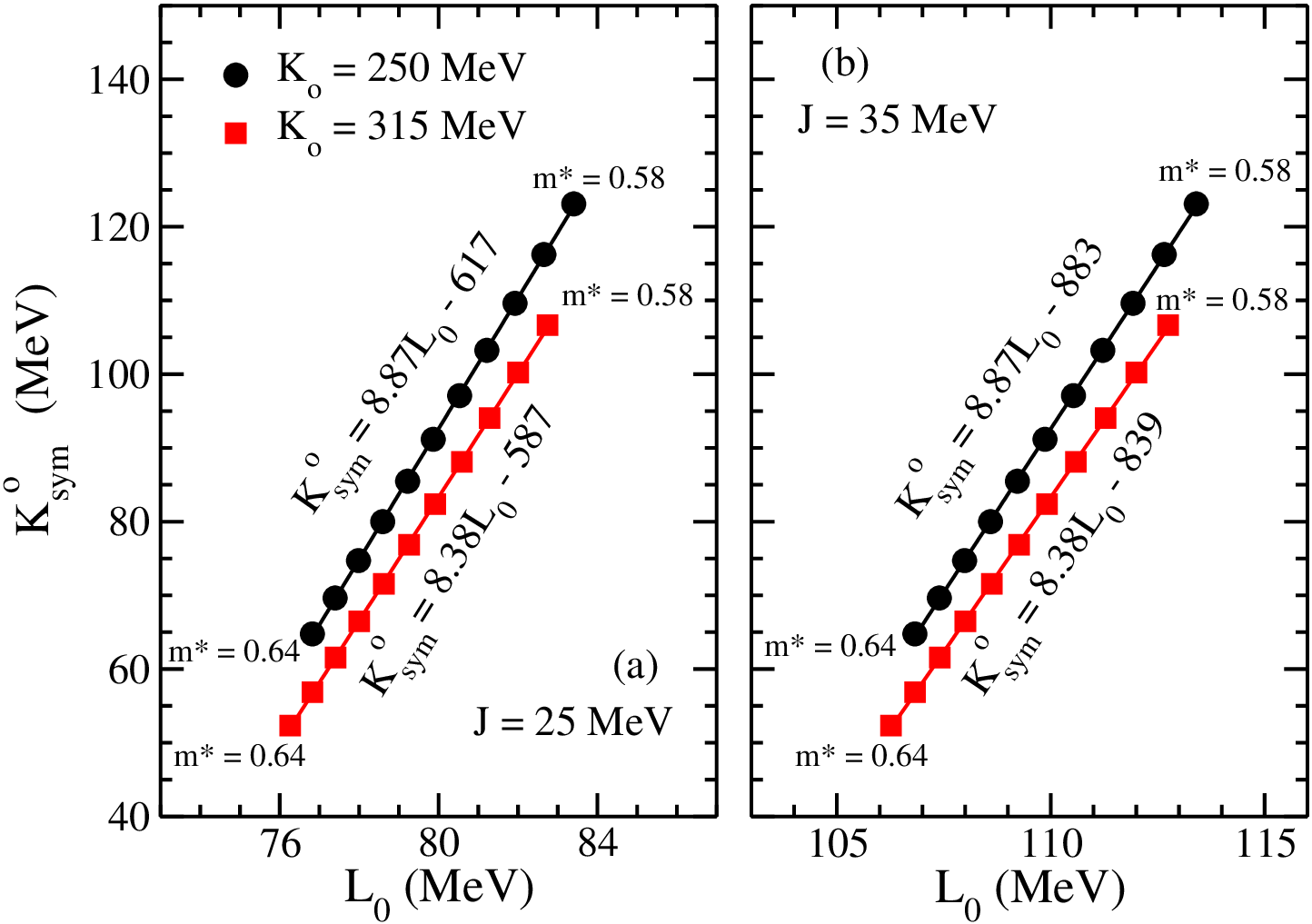}
\vspace{-0.2cm}
\caption{(Color online) $K_{\mbox{\tiny sym}}^o$ versus $L_o$ for Boguta-Bodmer models 
with $\rho_o=0.15$~fm$^{-3}$ and $B_o=16$~MeV. The linear fittings are indicated in both 
panels.} 
\label{ksymo-lo-rel}
\end{figure}

Besides its complete analytical structure, other advantage of the NR limit of NLPC models 
is its usefulness in predictions of correlations in Boguta-Bodmer models, as pointed out 
in Ref.~\cite{bianca}. For instance, the $L_o\times J$ correlation in Eq.~(\ref{lj}) is 
showed to be linear also for these models under the restriction of fixed values of 
effective mass. Like in the NR limit, different values for $K_o$ do not destruct the 
linear dependence, see figure 2b of Ref.~\cite{bianca}. Based on this correspondence, we 
use the framework of the NR limit to confirm other correlations in Boguta-Bodmer models. 
Still at the isovector sector, we showed in Ref.~\cite{bianca} that the linear correlation 
indicated in Eq.~(\ref{ksymo}) holds for the models described by Eq.~(\ref{ld}), if we 
also fix the values of $K_o$ and $J$. Now, we further investigate such a correlation. In 
Fig.~\ref{ksymo-lo-rel}, we show $K_{\mbox{\tiny sym}}^o$ as a function of $L_o$ for 
parametrizations with effective masses submitted to the FRS constraint, Eq.~(\ref{FRS}). 
According to Ref.~\cite{ls-splitting}, this is the range of $m^*$ in which Boguta-Bodmer 
models have to be constrained in order to produce spin-orbit splittings in agreement with 
well established experimental values for the $^{16}\rm{O}$, $^{40}\rm{Ca}$, and 
$^{208}\rm{Pb}$ nuclei.
In this figure, we present curves corresponding to the limiting values of the ranges 
$25\leqslant J \leqslant 35$~MeV~\cite{rmf}, and $250\leqslant K_o \leqslant 
315$~MeV~\cite{stone}. 

\begin{figure}[!htb]
\centering
\includegraphics[scale=0.33]{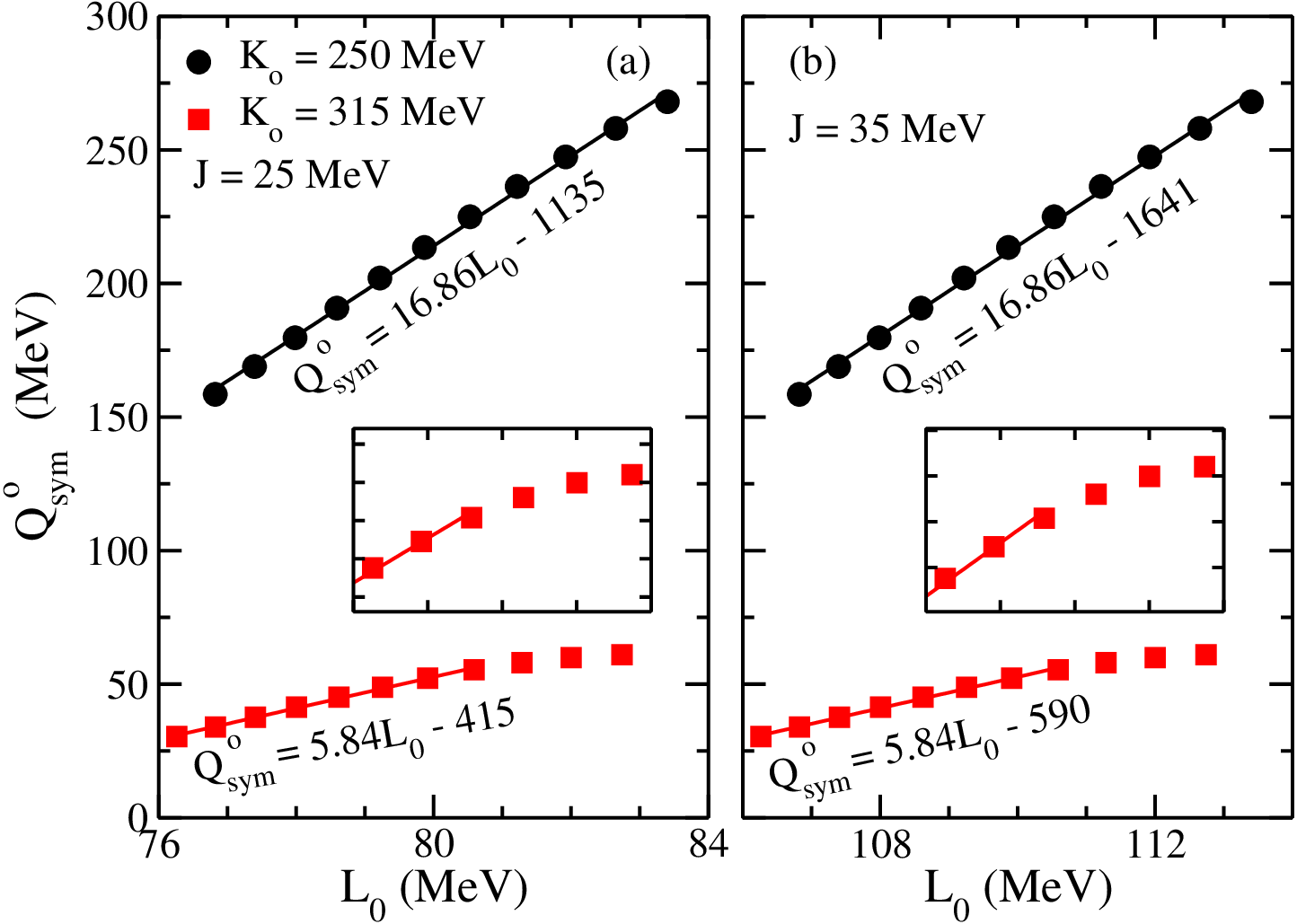}
\vspace{-0.2cm}
\caption{(Color online) $Q_{\mbox{\tiny sym}}^o$ versus $L_o$ for Boguta-Bodmer models 
with $\rho_o=0.15$~fm$^{-3}$, $B_o=16$~MeV, and $0.58\leqslant m^*\leqslant 0.64$. The 
linear fittings are indicated in both panels. In the insets, we show $L_o$ in the ranges 
of (a) $79\leqslant L_o\leqslant 83$~MeV, and (b) $109\leqslant L_o\leqslant 113$~MeV.} 
\label{qsymo-lo-rel}
\end{figure}

From these results, we can conclude that the relation $K_{\mbox{\tiny sym}}^o=p_{rel}L_o + 
q_{rel}$ also works well for the Boguta-Bodmer models submitted to the FRS constraint. 
However, by comparing the angular coefficients $p(\rho_o)$ from Eq.~(\ref{ksymo}) and 
$p_{rel}$, we notice that $p_{rel}$ slightly depends on $K_o$, unlike the nonrelativistic 
case in which the angular coefficient depends only on $\rho_o$. We perform the same 
analysis for the $L_o$ dependence on $Q_{\mbox{\tiny sym}}^o$, motivated by the linear 
correlation presented in Eq.~(\ref{qsymo}). The result is found in 
Fig.~\ref{qsymo-lo-rel}.
As we see, there is a correlation between $Q_{\mbox{\tiny sym}}^o$ and $L_o$. However, the 
linear form $Q_{\mbox{\tiny sym}}^o=u_{rel}L_o + v_{rel}$ is strongly dependent on $K_o$, 
unlike the previous case. For $K_o=315$~MeV, for example, we notice the range of effective 
mass that still ensure a line in the curve fitting is reduced from the 
range of Eq.~(\ref{FRS}) to $0.60\leqslant m^*\leqslant 0.64$. The break of this 
linearity is better depicted in the insets of Fig.~\ref{qsymo-lo-rel}. 

\begin{figure}[!htb]
\centering
\includegraphics[scale=0.33]{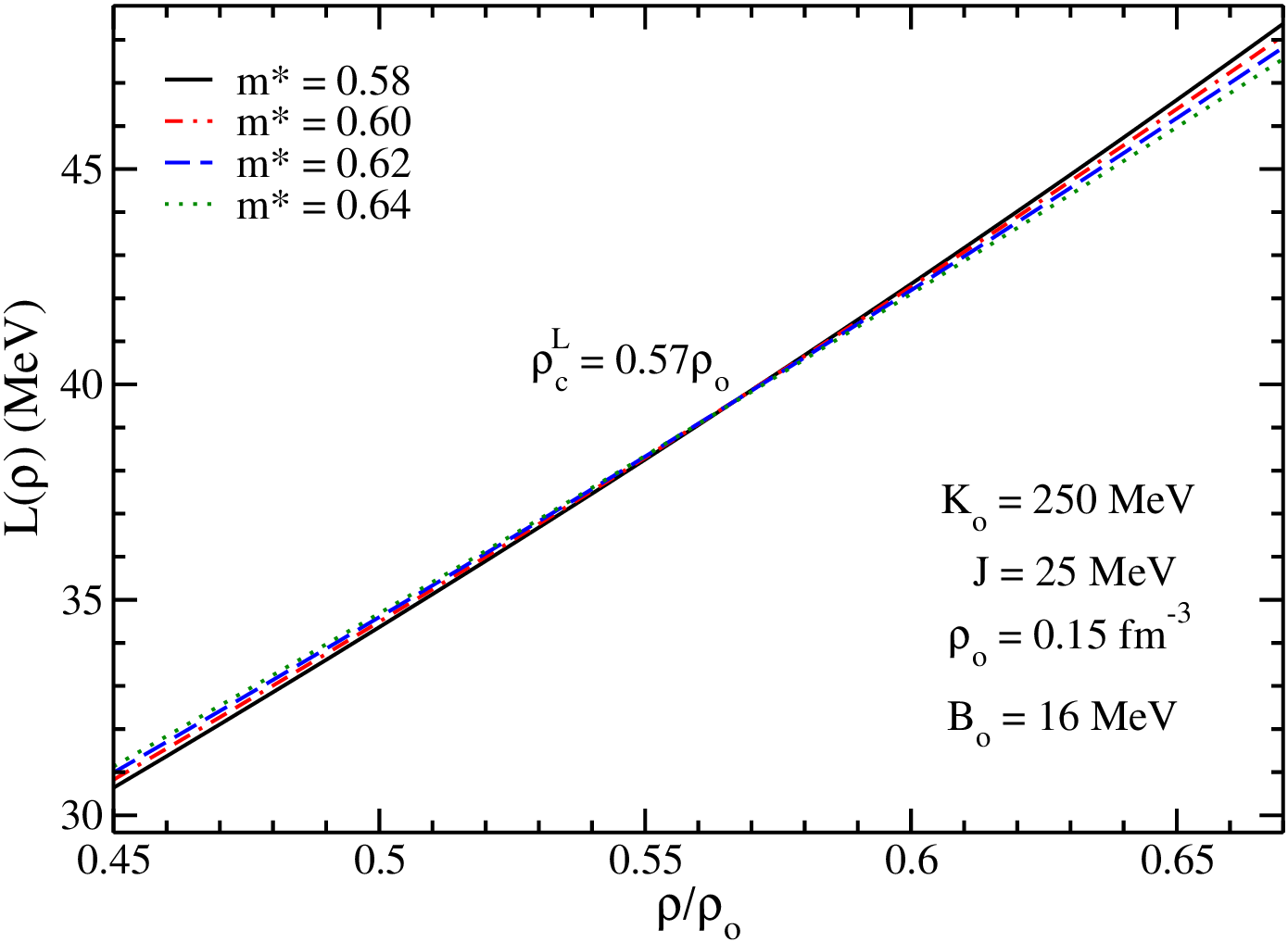}
\vspace{-0.2cm}
\caption{(Color online) Symmetry energy slope as a function of $\rho/\rho_o$ for some 
Boguta-Bodmer parametrizations.} 
\label{crossing-l-rel}
\end{figure}

\begin{figure}[!htb]
\centering
\includegraphics[scale=0.33]{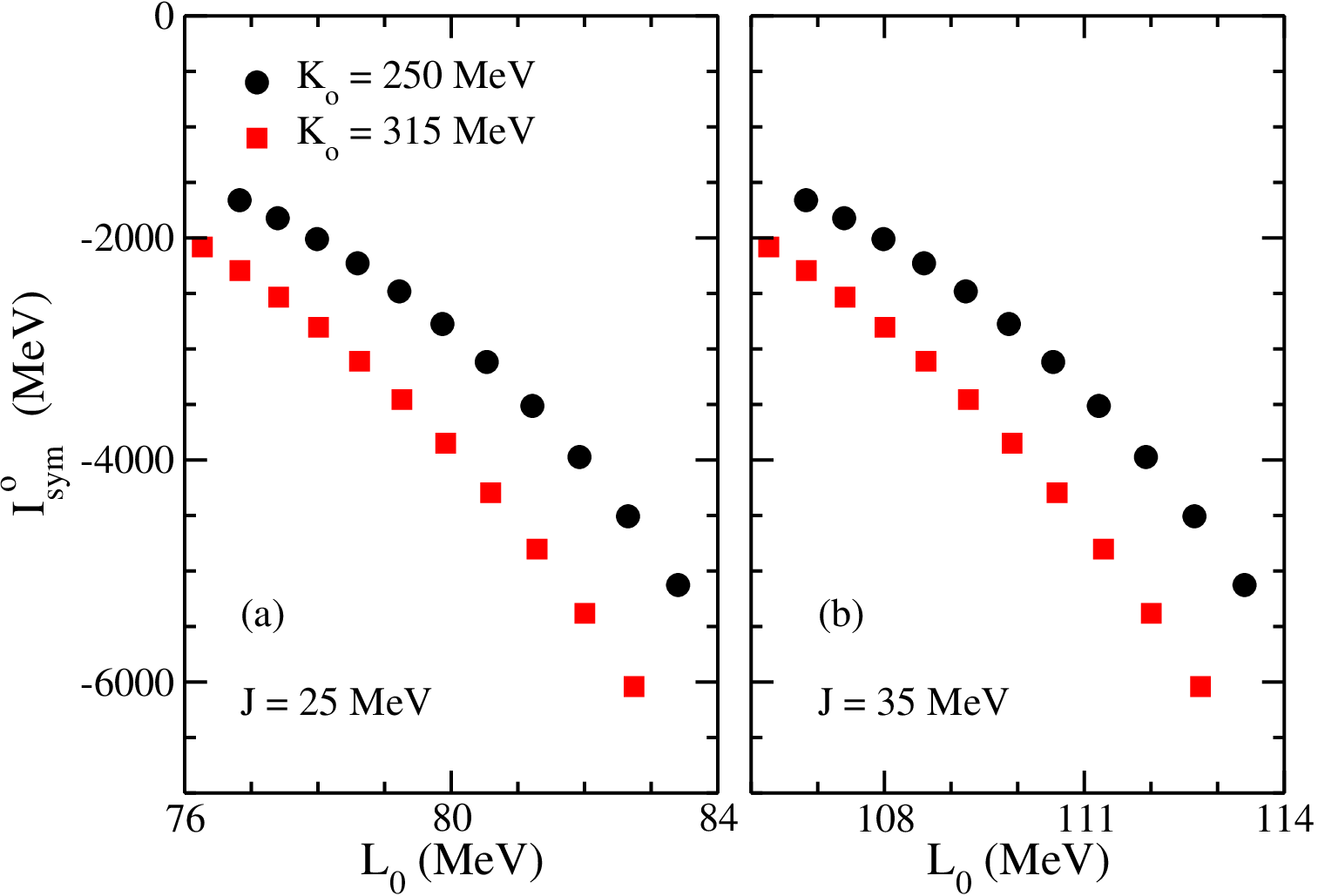}
\vspace{-0.2cm}
\caption{(Color online) $I_{\mbox{\tiny sym}}^o$ versus $L_o$ for Boguta-Bodmer models 
with $\rho_o=0.15$~fm$^{-3}$, $B_o=16$~MeV, and $0.58\leqslant m^*\leqslant 0.64$.} 
\label{isymo-lo-rel}
\end{figure}

One can also use the linear dependences showed in Figs.~\ref{ksymo-lo-rel} 
and~\ref{qsymo-lo-rel}, with the latter guaranteed only for some values of $K_o$ and 
$m^*$, in order to justify a possible crossing point in the density dependence of the 
symmetry energy slope, as we did in the case of nonrelativistic models in 
Sec.~\ref{nonrel}. In fact, there is such a crossing as we can see in 
Fig.~\ref{crossing-l-rel}.
From the angular coefficients of the respective lines of 
Figs.~\ref{ksymo-lo-rel}{\color{purple}a} and \ref{qsymo-lo-rel}{\color{purple}a}, we can 
solve the equation $1 + 8.87x_c^{\mbox{\tiny L}} + \frac{16.86}{2!}{x_c^{\mbox{\tiny 
L}}}^2 = 0$ to find a crossing density at $\rho_c^{\mbox{\tiny L}}/\rho_0=0.61$. This 
value can not be refined since a linear correlation is not found in the next order bulk 
parameter, namely, $I_{\mbox{\tiny sym}}^o$, as we see in Fig.~\ref{isymo-lo-rel}.
We see that a correlation between $L_o$ and $I_{\mbox{\tiny sym}}^o$ holds for the 
relativistic Boguta-Bodmer models, but it is not a linear one, as we verified in the NR 
limit, Eq.~(\ref{isymo}).

\begin{figure}[!htb]
\centering
\includegraphics[scale=0.33]{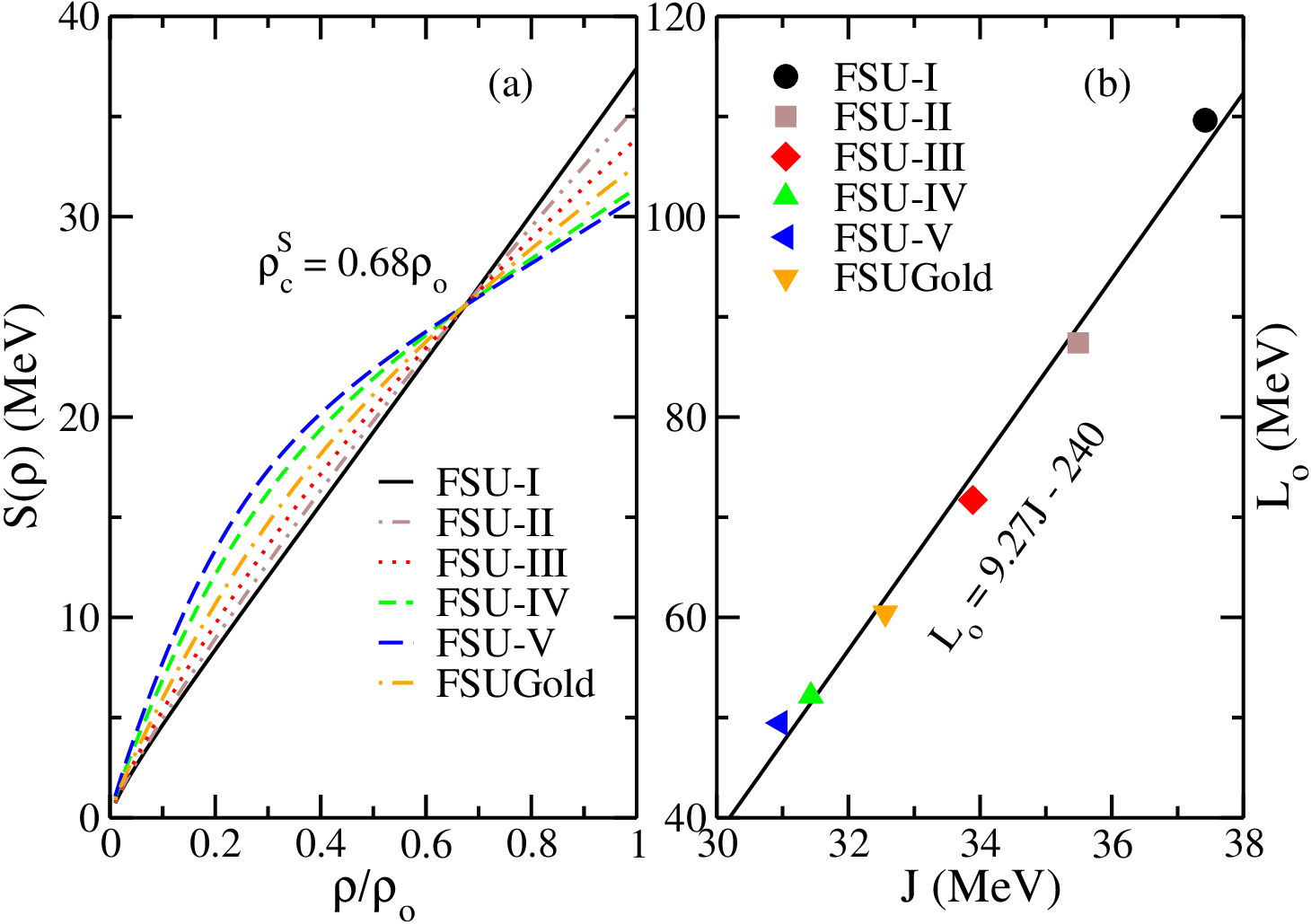}
\vspace{-0.2cm}
\caption{(Color online) (a) Symmetry energy as a function of $\rho/\rho_o$, and (b) $L_o$ 
as a function of $J$ for the parametrizations of Ref.~\cite{cai}.} 
\label{crossing-s-rel}
\end{figure}

Still concerning isovector bulk parameters, we point out to the reader a specif class of 
relativistic models of Ref.~\cite{cai} with mesonic crossed interactions. Following 
notation of Ref.~\cite{rmf}, they are classified as type 4 models ($\sigma^3 + \sigma^4 + 
\omega_0^4+$ cross terms models) and have the terms,
\begin{multline}
\mathcal{L}_{\sigma\omega\rho} =
g_\sigma g_\omega^2\sigma\omega_\mu\omega^\mu
\left(\alpha_1+\frac{1}{2}{\alpha_1}'g_\sigma\sigma\right)
+ \frac{C}{4}(g_\omega^2\omega_\mu\omega^\mu)^2
\\
+\frac{1}{2}{\alpha_3}'g_\omega^2 g_\rho^2\omega_\mu\omega^\mu
\vec{\rho}_\mu\vec{\rho}^\mu + g_\sigma g_\rho^2\sigma\vec{\rho}_\mu\vec{\rho}^\mu
\left(\alpha_2+\frac{1}{2}{\alpha_2}'g_\sigma\sigma\right)
\label{ldcross}
\end{multline}
added to the Lagrangian density of Eq.~(\ref{ld}). The parametrizations of this model 
presented in Ref.~\cite{cai} are constructed in order to fix the symmetry energy not at 
the saturation density, but in a smaller value. They present 
$\mathcal{S}(\rho_c^{\mbox{\tiny S}})=26$~MeV, with $\rho_c^{\mbox{\tiny 
S}}/\rho_o=0.68$. Therefore, the $\mathcal{S}(\rho)$ function presents a crossing point, 
as pointed out in Ref.~\cite{19}, and as one can see in 
Fig.~\ref{crossing-s-rel}{\color{purple}a}.
Therefore, such a crossing indicates a linear behavior between $L_o$ and $J$ for these 
specific parametrizations. This correlation is clearly observed in 
Fig.~\ref{crossing-s-rel}{\color{purple}b}. Furthermore, the crossing density is obtained 
from the angular coefficient, by solving the equation $1 + 9.27x_c^{\mbox{\tiny S}} = 0$. 
The solution of this linear equation leads to $\rho_c^{\mbox{\tiny S}}/\rho_0=0.68$, 
exactly the value verified in Fig.~\ref{crossing-s-rel}{\color{purple}a}.

As a last remark of this subsection, we point out to the reader that the angular and 
linear coefficients of the $L_o\times J$ correlation are not universal, as we can see by 
comparing the linear equation of Fig.~\ref{crossing-s-rel}{\color{purple}b} of the 
relativistic FSU family, with that of Fig.~\ref{lo-ksymo-skyrme}{\color{purple}a} of the 
Skyrme parametrizations. Even among relativistic models, one can not reach such 
universality, as we can see by the comparison of the correlation in 
Fig.~\ref{crossing-s-rel}{\color{purple}b} with that found in Ref.~\cite{ellipses} for 
the relativistic NL3* and IU-FSU families.

\subsection{Isoscalar sector}

\begin{figure}[!htb]
\centering
\includegraphics[scale=0.33]{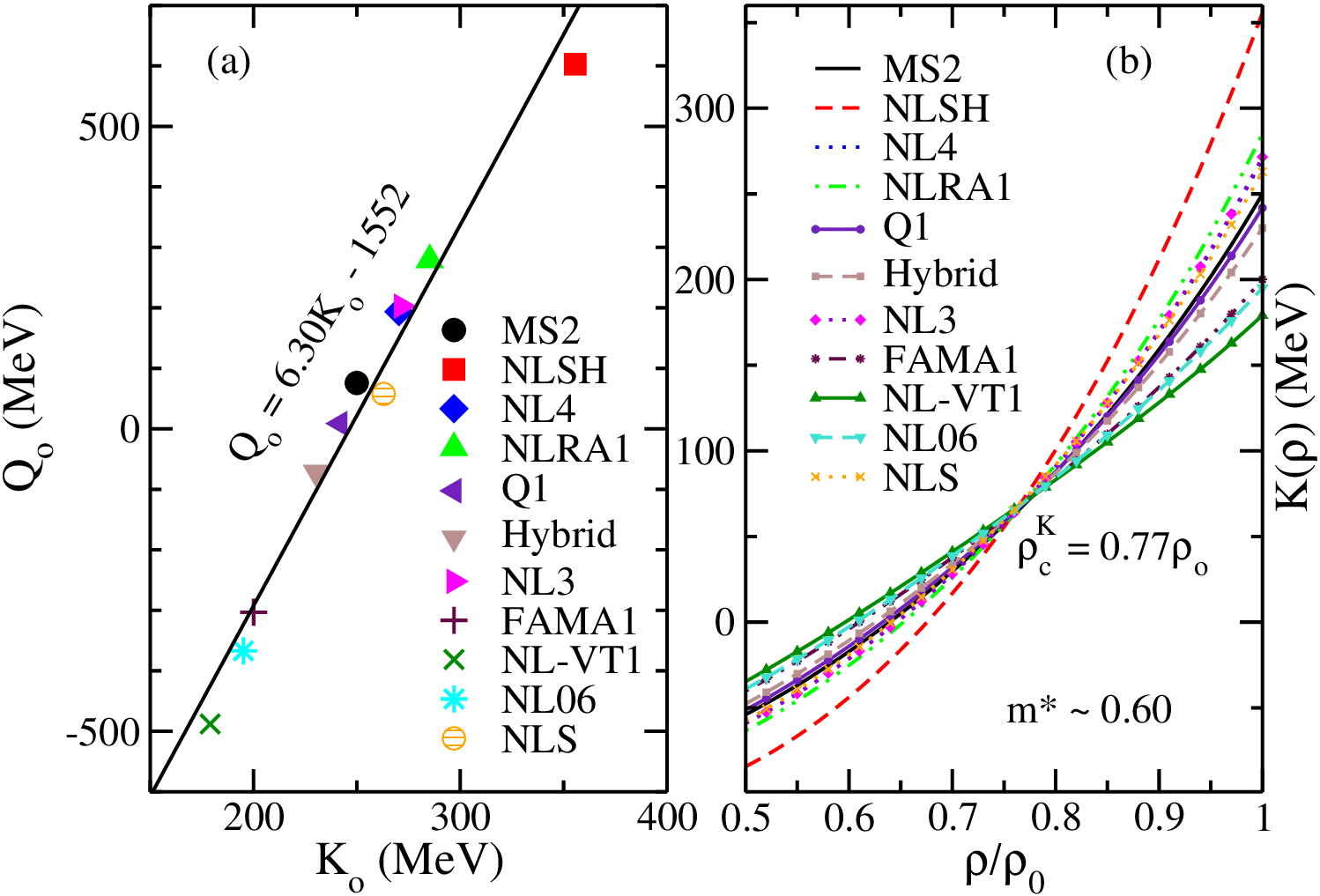}
\vspace{-0.2cm}
\caption{(Color online) (a) $Q_o$ as a function of $K_o$ for some Boguta-Bodmer 
parametrizations with $m^*\simeq0.6$. (b) Incompressibility as a function of $\rho/\rho_o$ 
for the same parametrizations.} 
\label{crossing-k-rel}
\end{figure}

Motivated by the analytical structure relating $K_o$, $Q_o$, and $I_o$, we investigate in 
this section whether the linear dependences presented in the NR limit, see 
Eqs.~(\ref{qk-nr}) and (\ref{ik-nr}), also applies for Boguta-Bodmer models. According to 
the NR limit case, if we keep fixed the effective mass, $K_o$ linearly correlates with 
$Q_o$ as we show in Eq.~(\ref{qk-nr}) and Fig.~\ref{qo-ko-nr-gogny}{\color{purple}a}. For 
the relativistic case of \mbox{FR-RMF} models described by Eq.~(\ref{ld}), we see that 
this condition remains, as one can see in Fig.~\ref{crossing-k-rel}{\color{purple}a} for 
the MS2~\cite{ms2}, NLSH~\cite{nlsh}, NL4~\cite{nl4}, NLRA1~\cite{nlra1}, Q1~\cite{q1}, 
Hybrid~\cite{hybrid}, NL3~\cite{nlsh}, FAMA1~\cite{fama1}, NL-VT1~\cite{nlvt1}, 
NL06~\cite{rmf}, and NLS~\cite{nls} parametrizations presenting $m^*\simeq0.6$.
Such a correlation can be used in order to justify the crossing in the $K(\rho)$ 
function depicted in Fig.~\ref{crossing-k-rel}{\color{purple}b}. Proceeding in that 
direction, we use the expansion of the energy per particle in Eq.~(\ref{epp}) until order 
$x^3$ to calculate the density dependence of the incompressibility. The result of this 
calculation is given by
\begin{eqnarray}
K(\rho)\simeq(3x+1)\left[ K_o + (9K_o+Q_o)x + 6Q_ox^2\right].
\label{kexp-rel}
\end{eqnarray}
Therefore, the linear dependence $Q_o=i_{rel}K_o+j_{rel}$ showed in 
Fig.~\ref{crossing-k-rel}{\color{purple}a} can be used in Eq.~(\ref{kexp-rel}) to furnish
\begin{eqnarray}
K(\rho) &\simeq& (3x+1)\left\lbrace K_o\left[ 1 + ( 9 + i_{rel} )x + 6ix^2 
\right]\right.
\nonumber\\
&+& \left.\left(1 + 6x \right)j_{rel}x\right\rbrace,
\end{eqnarray}
with $i_{rel}=6.30$ and $j_{rel}=1552$~MeV. Thus, one has a crossing point when the 
quadratic equation $1 + (9+i_{rel})x_c^{\mbox{\tiny K}} + 6i_{rel}{x_c^{{\mbox{\tiny 
K}}}}^2=0$ present solution. This is the case for $\rho_c^{\mbox{\tiny K}}/\rho_0=0.78$.

We remark that a crossing point in the $K(\rho)$ function was firstly explained from the 
linear correlation between $K_o$ and $Q_o$ in Ref.~\cite{margueron}. However, the authors 
found such a crossing for some nonrelativistic Skyrme and Gogny parametrizations. For the
relativistic models analysed, they did not found linear correlations or crossing points. 
Indeed, for the nonrelativistic case, they found a crossing at $\rho_c^{\mbox{\tiny 
K}}/\rho_0\simeq 0.7$, a value quite close to ours findings, namely, $\rho_c^{\mbox{\tiny 
K}}/\rho_0=0.77$ and $\rho_c^{\mbox{\tiny K}}/\rho_0=0.79$, for Boguta-Bodmer models and 
the NR limit, respectively, see Figs.~\ref{crossing-k-rel}{\color{purple}b} and 
\ref{crossing-k-nr}.

\begin{figure}[!htb]
\centering
\includegraphics[scale=0.33]{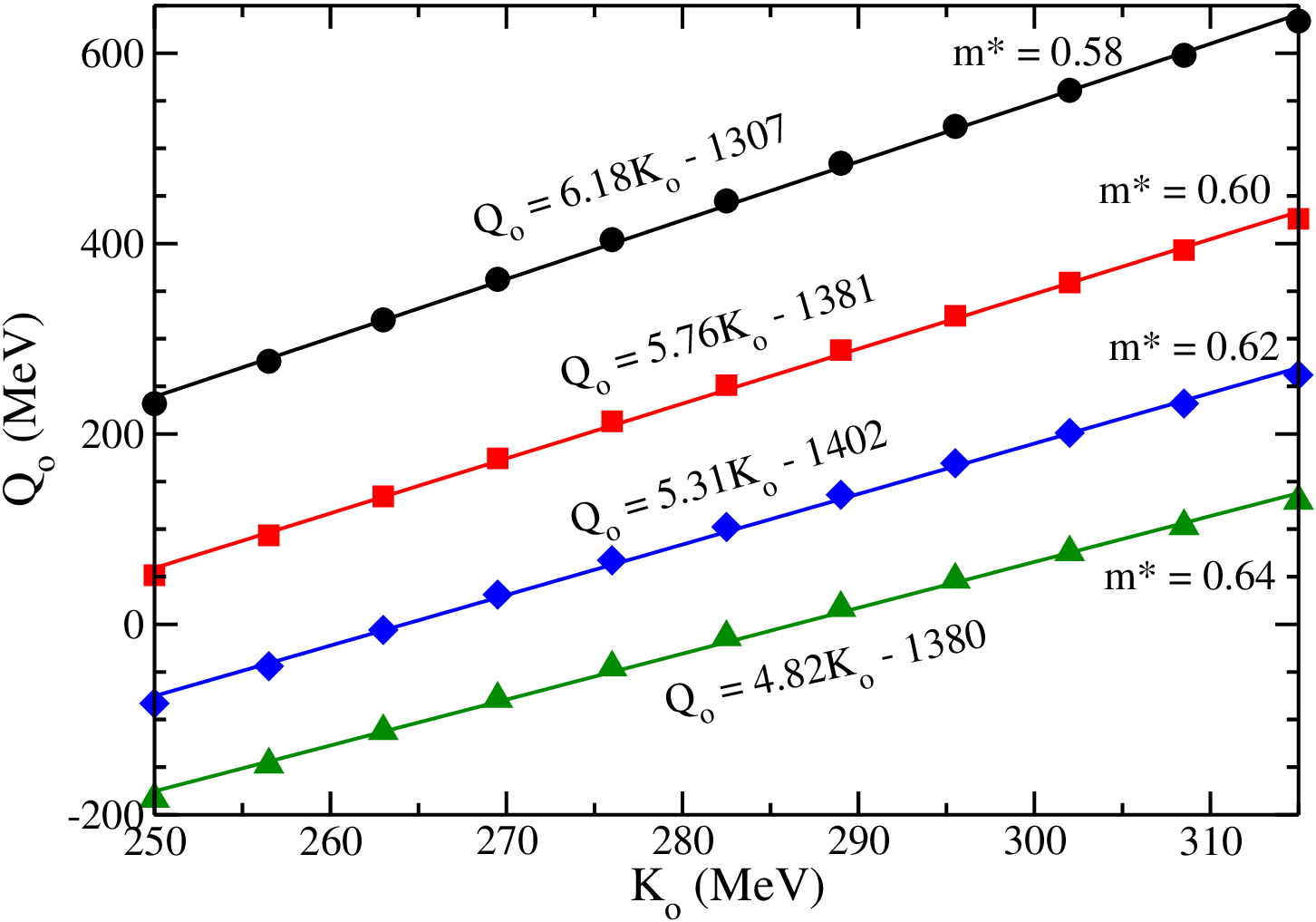}
\vspace{-0.2cm}
\caption{(Color online) $Q_o$ as a function of $K_o$ for Boguta-Bodmer parametrizations 
in which $\rho_o=0.15$~fm$^{-3}$ and $B_o=16$~MeV. The linear fittings are indicated in 
the figure.} 
\label{qo-ko-rel}
\end{figure}

Unlike the linear correlation presented in the NR limit, the angular coefficient 
$i_{rel}$ is slightly dependent on the effective mass. For the former case, we have 
$i=i(\rho_o)$, see Eq.~(\ref{inr}). In Fig.~\ref{qo-ko-rel}, we show this variation 
observing the FRS constraint and the range $250\leqslant K_o \leqslant 315$~MeV.
In particular, notice that for $m^*=0.64$, a value that ensure good values for finite 
nuclei spin-orbit splittings~\cite{ls-splitting}, the range of the skewness coefficient 
is given by $-183\leqslant Q_o \leqslant 130$~MeV. Such a specific constraint for $Q_o$ 
present an overlap of $\simeq 36\%$ with a recent range proposed for this bulk parameter 
in Ref.~\cite{chen-skew}, namely, $-494\leqslant Q_o \leqslant -10$~MeV. In this study, 
the authors analysed models with crossed interactions among the fields, i. e., models 
described by Eq.~(\ref{ld}) added to the terms in Eq.~(\ref{ldcross}). They verified that 
such models, presenting the skewness coefficient in the range of $-494\leqslant Q_o 
\leqslant -10$~MeV, satisfy the suprasaturation constraint for the density 
dependence of the pressure in the symmetric nuclear~\cite{science}, and also the neutron 
star mass constraint, given by $2.01\pm0.04M_\odot$. This latter is due to the recently 
discovered neutron star PSR J0348+0432~\cite{ns}.

\begin{figure}[!htb]
\centering
\includegraphics[scale=0.32]{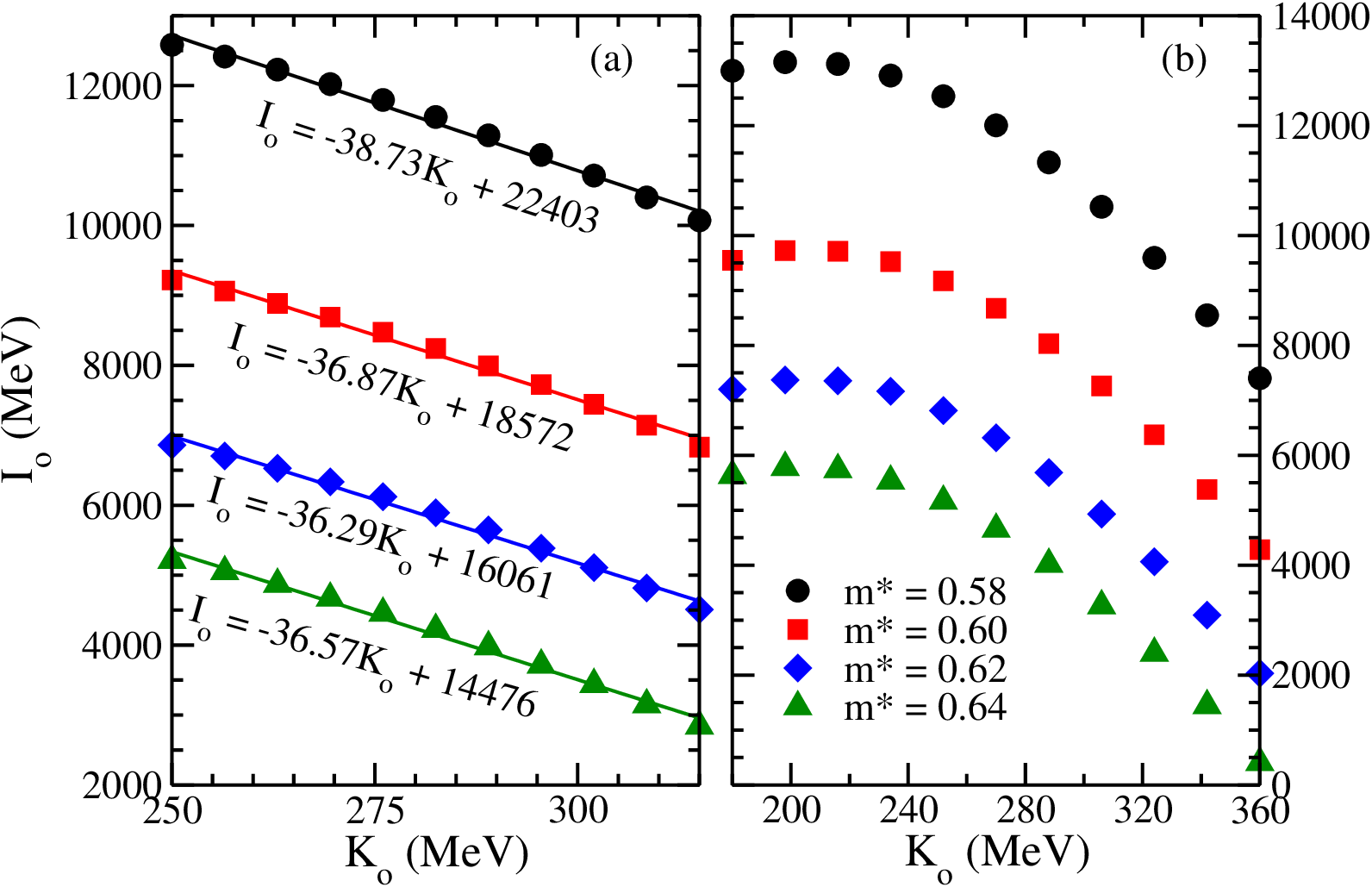}
\vspace{-0.2cm}
\caption{(Color online) $I_o$ as a function of $K_o$ for Boguta-Bodmer parametrizations 
in which $\rho_o=0.15$~fm$^{-3}$ and $B_o=16$~MeV. The curves are constructed in the range 
of (a) $250\leqslant K_o \leqslant 315$~MeV, and (b) $180\leqslant K_o \leqslant 360$~MeV. 
The linear fittings are indicated in panel (a).} 
\label{io-ko-rel}
\end{figure}

Finally, we verify whether the relationship between $K_o$ and $I_o$ presented in 
Eq.~(\ref{ik-nr}) is preserved in the relativistic case. According to the results of 
Fig.~\ref{io-ko-rel}{\color{purple}a}, we see that the linear behavior 
$I_o=w_{rel}K_o+z_{rel}$ still remains in the range of $250\leqslant K_o \leqslant 
315$~MeV, with $w_{rel}$ slightly depending on $m^*$, like in the case of $i_{rel}$.

It is worth to notice that, as showed in Fig.~\ref{io-ko-rel}{\color{purple}b}, for a 
broader range of $K_o$ the linear dependence is blurred, although a correlation between 
the bulk parameters $K_o$ and $I_o$ still remains.

\section{Models with more than one isovector coupling constant} 
\label{twoicc}

\begin{figure}[!htb]
\centering
\includegraphics[scale=0.32]{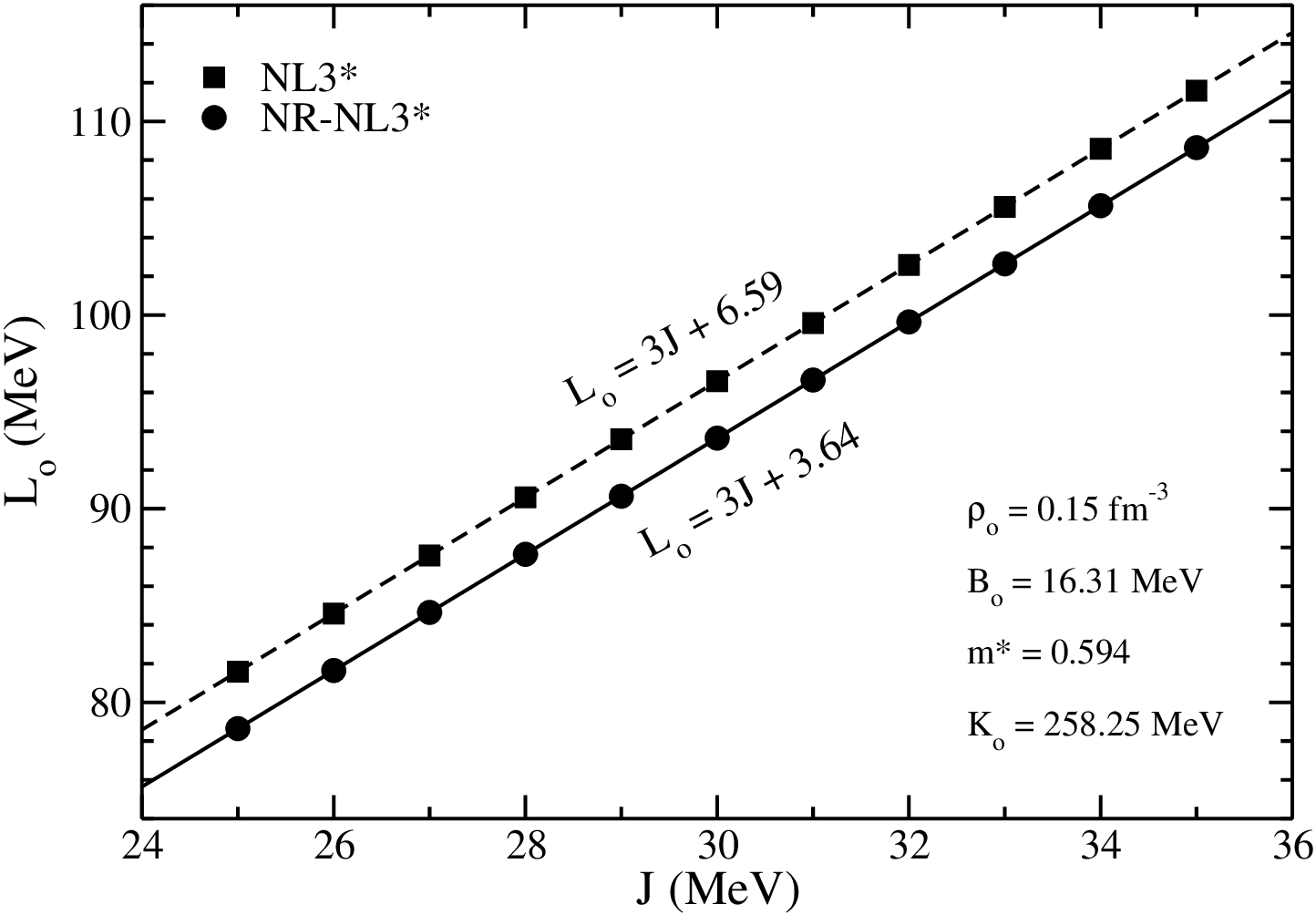}
\vspace{-0.2cm}
\caption{$L_o$ as a function of $J$ for the NL3* Boguta-Bodmer parametrizations and their 
NR limit versions \mbox{NR-NL3*}. These parametrizations were constructed by fixing 
$\rho_o=0.15$~fm$^{-3}$, $B_o=16.31$~MeV, $m^*=0.594$, $K_o=258.25$~MeV, and by running 
$J$.} 
\label{ljnl3-1}
\end{figure}

In previous sections, we have analyzed under what conditions the linear correlations 
presented in the NR limit of point-coupling models are reproduced in the context of 
relativistic Boguta-Bodmer parametrizations. However, our comparisons were restricted to 
relativistic and nonrelativistic models presenting only one isovector coupling constant, 
namely, $g_\rho$ and $G^{2}_{\mbox{\tiny TV}}$, respectively. For the Boguta-Bodmer model, 
$g_\rho$ is related to the interaction strength between the nucleon and the $\rho$ meson. 
For the NR limit model, $G^{2}_{\mbox{\tiny TV}}$ regulates the strength of the term that 
mimics the same kind of interaction (we remind the reader that our NR limit model 
is derived from a relativistic point-coupling model, therefore, a model in which there 
are no meson exchanges). Regarding the specific relationship between $L_o$ and $J$, 
we concluded in Ref.~\cite{bianca} that for the NR limit model, such correlation is 
linear whenever the isoscalar bulk parameters, namely,  $m^*$, $\rho_o$, $B_o$ and $K_o$, 
remain unchanged. We also showed that this same condition also ensures a linear 
correlation between $L_o$ and $J$ for Boguta-Bodmer parametrizations.
Furthermore, the angular coefficients of these correlations are the same, and the absolute 
values of $L_o$ are very close each other, as we can see in Fig.~\ref{ljnl3-1} for the 
NL3* Boguta-Bodmer parametrizations and their respective NR limit versions, namely, the 
NR-NL3* ones.

In order to construct this figure, we fixed the isoscalar parameters values of the NL3* 
model, and varied the $J$ values. We taken such procedure for the exact relativistic NL3* 
parametrization, and for its NR limit version. For the latter, we have used our 
Eq.~(\ref{lj}).

By proceeding one step further in our analysis of the $L_o\times J$ correlation, we now 
study relativistic and nonrelativistic models with more than one isovector parameter in 
order to verify whether the dependence observed in Fig.~\ref{ljnl3-1} still applies. For 
the relativistic model, we use that described by the Lagrangian density of Eq.~(\ref{ld}) 
added to those of Eq.~(\ref{ldcross}) with $\alpha_1=\alpha_2=\alpha'_1=\alpha'_2=C=0$, 
i. e., we choose a model with interaction between the mesons $\omega$ and $\rho$. Thus, 
we are dealing with a model with two isovector parameters, namely, $g_\rho$ and 
$\alpha'_3$.

To take the NR limit of this specific model, we construct the following point-coupling 
Lagrangian density,
\begin{eqnarray}
\mathcal{L}_{\mbox{\tiny NLPC}}&=&
\bar{\psi}(i\gamma^{\mu}\partial_{\mu}-M)\psi
-\frac{1}{2}G^{2}_{\mbox{\tiny V}}(\bar{\psi}\gamma^{\mu}\psi)^{2}
+\frac{1}{2}G^{2}_{\mbox{\tiny S}}(\bar{\psi } \psi)^{2} \nonumber \\
&+&\frac{A}{3}(\bar{\psi}\psi)^{3}
+\frac{B}{4}(\bar{\psi}\psi)^{4}
-\frac{1}{2}G^{2}_{\mbox{\tiny TV}}(\bar{\psi}\gamma^{\mu}\vec{\tau}\psi)^{2} \nonumber\\
&-&\frac{1}{4}G_{\mbox{\tiny 
VTV}}(\bar{\psi}\gamma^{\mu}\vec{\tau}\psi)^{2}(\bar{\psi}\gamma^{\mu}\psi)^{2},
\label{lagd}
\end{eqnarray}
in which the last term mimics the interaction between the mesons $\omega$ and $\rho$. The 
isovector coupling constants of this model are $G^{2}_{\mbox{\tiny TV}}$ and 
$G_{\mbox{\tiny VTV}}$. The symmetry energy and its slope for the NR limit of this model 
are given by,
\begin{eqnarray}
\mathcal{S}^{\mbox{\tiny (NR)}}(\rho)= G^{2}_{\mbox{\tiny TV}}\rho + G_{\mbox{\tiny 
VTV}}\rho^3 + \frac{\lambda\rho^{\frac{2}{3}}}{6M^*(\rho,1/2)},
\label{snr2}
\end{eqnarray}
and
\begin{eqnarray}
L^{\mbox{\tiny (NR)}}(\rho)&=&\frac{\lambda\rho^{\frac{2}{3}}}{3M^{2}} 
\left(M+ \frac{5}{2}G^2_{\mbox{\tiny S}}\rho + 8A\rho^{2} + 
\frac{33}{2}B\rho^3\right)
\nonumber\\
&+& 3G^2_{\mbox{\tiny TV}}\rho + 9G_{\mbox{\tiny VTV}}\rho^3,
\label{lrhonr2}
\end{eqnarray}
respectively. The new isovector coupling constant, $G_{\mbox{\tiny VTV}}$, is found by 
imposing upon the model that the symmetry energy at $\rho_1/\rho_o\equiv r$ is fixed at a 
particular value $\mathcal{S}_1\equiv \mathcal{S}(\rho_1)$. Here, $r$ is a value smaller 
than $1$. Furthermore, we still found $G^{2}_{\mbox{\tiny TV}}$ by requiring that the 
model present a particular value $J$ for the symmetry energy at the saturation density. 
The analytical form of these constants as a function of the bulk parameters can be found 
in the Appendix.

\begin{figure}[!htb]
\centering
\includegraphics[scale=0.32]{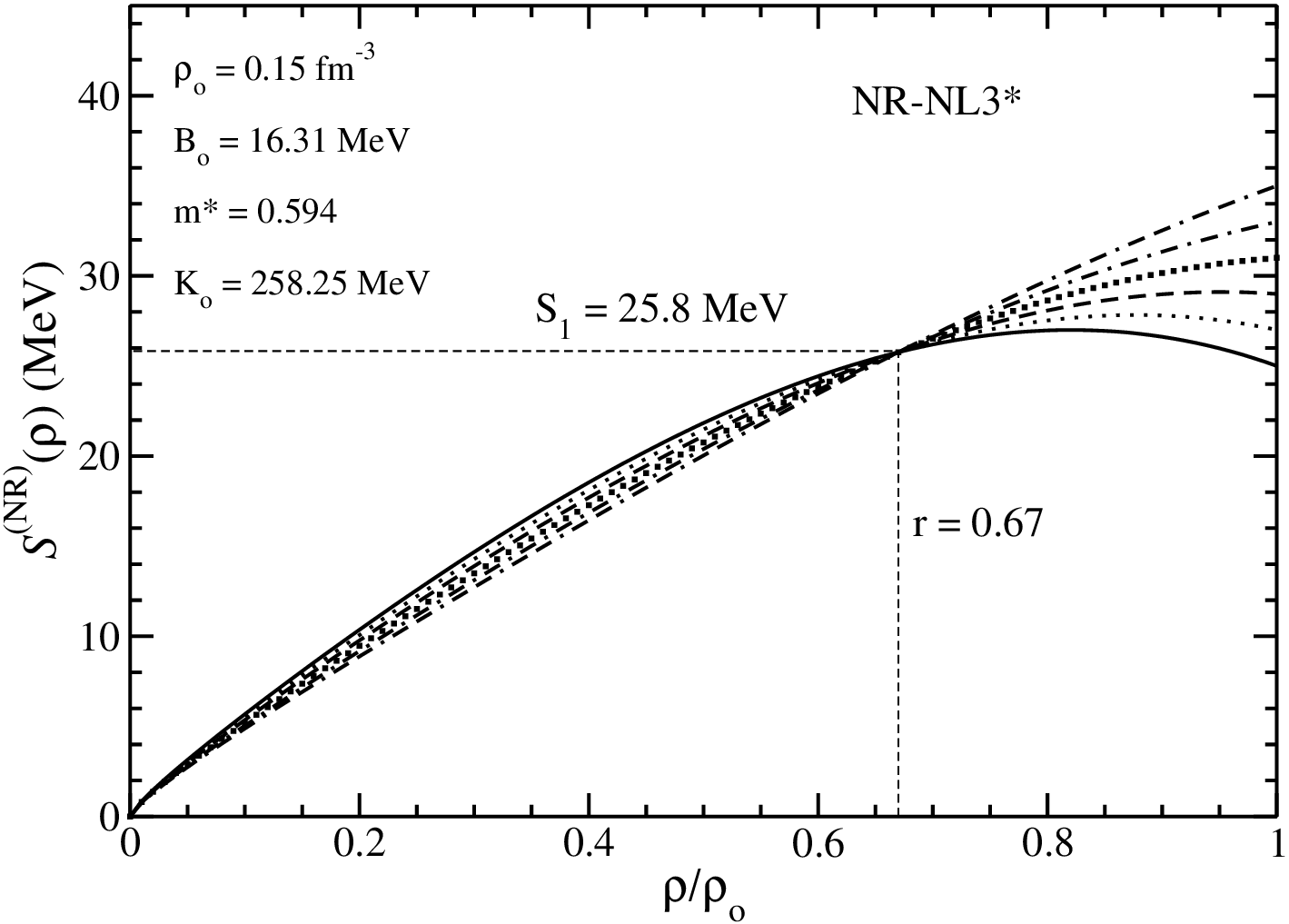}
\vspace{-0.2cm}
\caption{Density dependence of the symmetry energy for the \mbox{NR-NL3*} parametrization 
family with two isovector parameters.} 
\label{crossing-s-nrnl3}
\end{figure}

Such an analytical structure enables us to find the following correlation between $L_o$ 
and $J$,
\begin{eqnarray}
L_o=3\left(\frac{3 - r^2}{1 - r^2}  \right)J + 
b'(m^{*},\rho_{o},B_{o},K_{o},r,\mathcal{S}_1),
\end{eqnarray}
with
\begin{eqnarray}
&b'&(m^{*},\rho_{o},B_{o},K_{o},r,\mathcal{S}_1) = b(m^*,\rho_o,B_o,K_o) \nonumber \\
&+& \frac{10E_{\mbox{\tiny F}}^or^{2/3}}{3(r^2-1)(3M^{2} - 19E_{\mbox{\tiny F}}^oM 
+18E_{F}^{o2})} \times \nonumber\\
&\times&\Bigg\{\frac{E_{\mbox{\tiny F}}^o}{3 m^{*}}\left(13M-66E_{\mbox{\tiny F}}^o 
\right)  
+ \frac{K_{o}}{6} \left(M -6E_{\mbox{\tiny F}}^o\right) \nonumber \\
&-&B_{o}\left(9M-48E_{\mbox{\tiny F}}^o\right) - E_{\mbox{\tiny F}}^o\left( 
7M-36E_{\mbox{\tiny F}}^o \right) 
\nonumber\\
&-& r\bigg[\frac{16E_{\mbox{\tiny F}}^o}{ 3m^{*}}\left(M-6E_{\mbox{\tiny F}}^o 
\right)  + \frac{2}{3} K_{o}\left(M -3E_{\mbox{\tiny F}}^o\right) \nonumber \\
&-& 6B_{o}\left(3M-13E_{\mbox{\tiny F}}^o \right) -2E_{\mbox{\tiny F}}^o\left( 
5M-27E_{\mbox{\tiny F}}^o \right) \bigg]
\nonumber\\
&-& r^2\bigg[ - \frac{E_{\mbox{\tiny F}}^o}{m^{*}}\left(M-10E_{\mbox{\tiny F}}^o 
\right) -\frac{K_{o}}{2} \left(M-2E_{\mbox{\tiny F}}^o \right) \nonumber \\
&+& 3B_{o}\left(3M-10E_{\mbox{\tiny F}}^o\right) +3E_{\mbox{\tiny 
F}}^o\left(M-6E_{\mbox{\tiny F}}^o\right) \bigg]\Bigg\} \nonumber\\
&+& \frac{6}{(r^2-1)} \Bigg[\frac{\mathcal{S}_1}{r} + \frac{5E_{\mbox{\tiny 
F}}^o}{9m^{*}}(1-r^{2/3}) 
+\frac{5E_{\mbox{\tiny F}}^o}{9}(r^{2/3} - r^{-1/3})\Bigg].\nonumber\\
\end{eqnarray}
Notice that now, a linear correlation between $L_o$ and $J$ is established if the 
function $b'(m^{*},\rho_{o},B_{o},K_{o},r,\mathcal{S}_1)$ is a constant, i. e., if the 
quantities $m^{*}$, $\rho_{o}$, $B_{o}$, $K_{o}$, $r$, and $\mathcal{S}_1$ are kept 
fixed. Moreover, if we now look at the $\mathcal{S}(\rho)$ function for a particular 
parametrization family, namely, that in which the set $m^{*}$, $\rho_{o}$, $B_{o}$, 
$K_{o}$, $r$, and $\mathcal{S}_1$ is fixed and $J$ runs a certain range, we see a 
crossing point, differently from the NR limit case presenting only one isovector coupling 
constant. We show this finding in Fig.~\ref{crossing-s-nrnl3} for the \mbox{NR-NL3*} 
family.

\begin{figure}[!htb]
\centering
\includegraphics[scale=0.32]{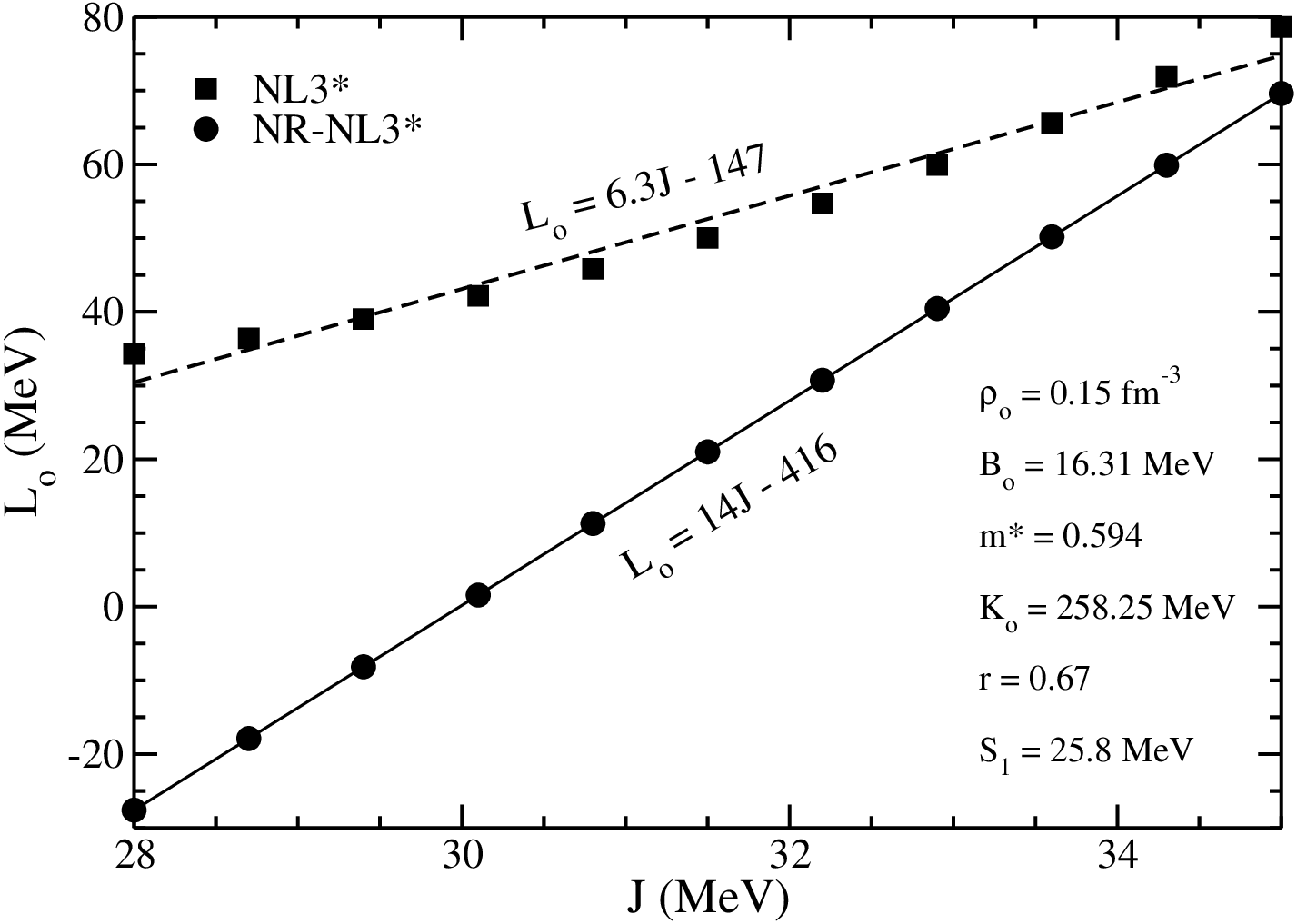}
\vspace{-0.2cm}
\caption{$L_o$ as a function of $J$ for the relativistic NL3* parametrization family
and their NR limit versions \mbox{NR-NL3*}. These parametrizations were constructed by 
fixing $\rho_o=0.15$~fm$^{-3}$, $B_o=16.31$~MeV, $m^*=0.594$, $K_o=258.25$~MeV, 
$r=0.67$, $\mathcal{S}_1=25.8$~MeV, and by running $J$. Both models present two isovector 
coupling constants.} 
\label{ljnl3-2}
\end{figure}

By looking at the finite range relativistic model with the $\omega$ and $\rho$ mesons 
interaction, we verified that a linear correlation between $L_o$ and $J$ also holds if we 
apply the same conditions observed in the NR limit case, i. e., fixed values of $m^{*}$, 
$\rho_{o}$, $B_{o}$, $K_{o}$, $r$, and $\mathcal{S}_1$. A direct comparison between these 
two correlations, analogous to that presented in Fig.~\ref{ljnl3-1}, is displayed in 
Fig.~\ref{ljnl3-2}.
Here, we restricted our analysis for $J$ in a range of values greater than 
$\mathcal{S}_1$.

\begin{figure}[!htb]
\centering
\includegraphics[scale=0.32]{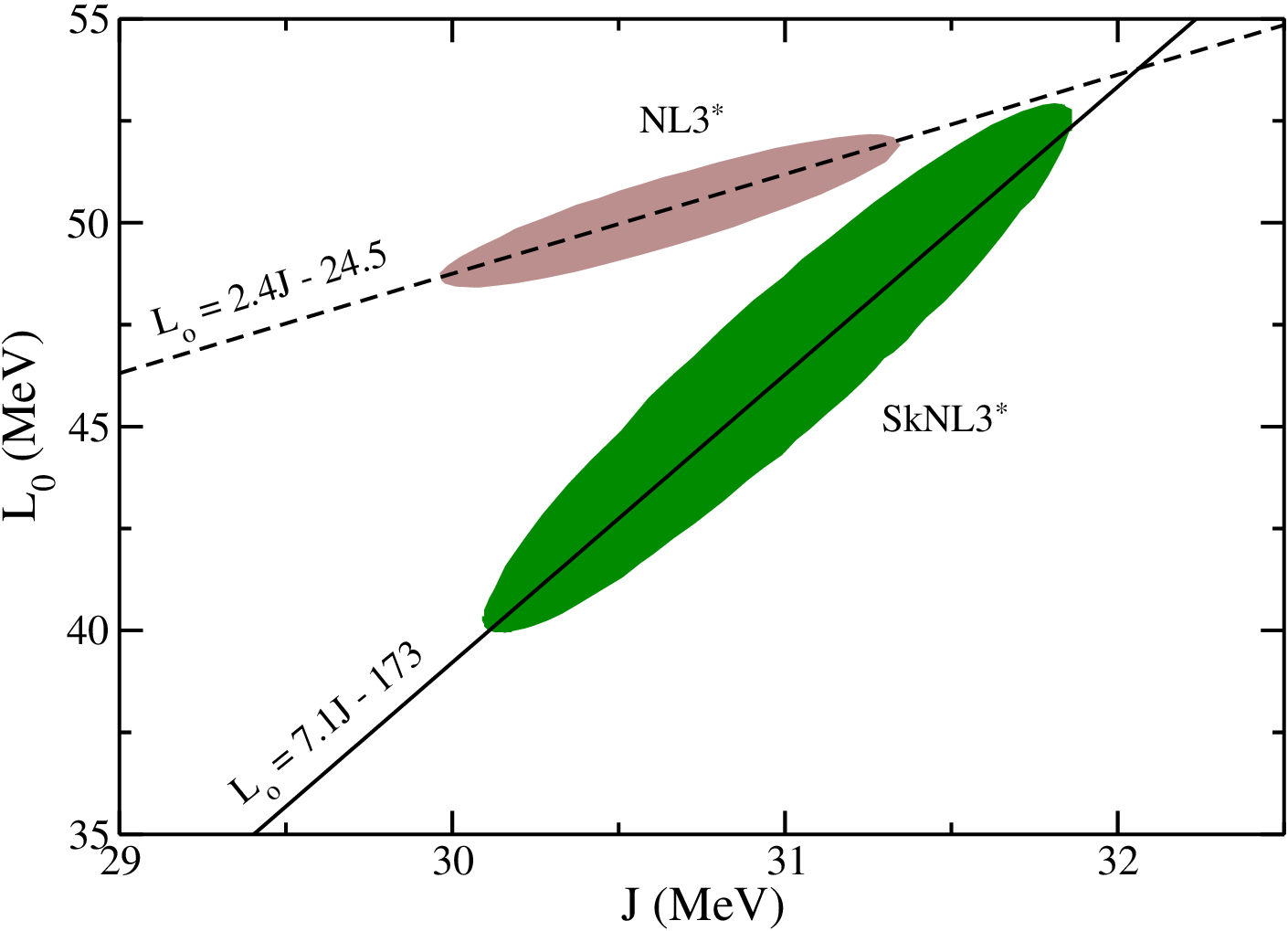}
\vspace{-0.2cm}
\caption{(Color online) Correlation bands for the $L_o$ and $J$ bulk parameters of the 
NL3* parametrization family and SkNL3* one, extracted from Fig.~3a of 
Ref.~\cite{ellipses}. The full lines are linear fits.} 
\label{ellipses}
\end{figure}

From Fig.~\ref{ljnl3-2}, we can notice that the NR limit version of NL3* parametrizations 
with two isovector coupling constants, presents a different slope for the $L_o\times J$ 
linear correlation, differently from the case showed in Fig.~\ref{ljnl3-1}, where we 
tested models with only one isovector parameter. This result is in qualitative agreement 
with the findings obtained in Ref.~\cite{ellipses}, where the authors compared the same 
NL3* parametrization family (two isovector parameters) with a nonrelativistic Skyrme 
parametrization family named as SkNL3*. For this family, the isoscalar bulk parameters 
present the same values as in the relativistic NL3* model. The authors also imposed 
that the energy per neutron predictions, at subsaturation densities, of the SkNL3* and 
NL3* models were compatible with the band constraint depicted in Fig.~2 of 
Ref.~\cite{ellipses}. As a consequence, they found correlation bands (ellipses) for the 
$L_o$ and $J$ bulk parameters, as one can see in Fig.~\ref{ellipses}.

These ellipses were constructed by the authors of Ref.~\cite{ellipses} from the covariance 
analysis method. In Fig.~\ref{ellipses}, we extracted such bands and constructed the 
linear fits. Notice the nonrelativistic line presenting a greater slope in comparison with 
the relativistic one, exactly the same qualitative behavior observed in 
Fig.~\ref{ljnl3-2}, where we have constructed the correlations only observing the 
conditions under which they are linear ones. Furthermore, our ratio for the NL3* slope to 
the NR-NL3* one is not much different to the same ratio of Fig.~\ref{ellipses}, namely, 
$2.22$ for ours (Fig.~\ref{ljnl3-2}), and $2.96$ for Ref.~\cite{ellipses}, or 
Fig.~\ref{ellipses}, obtained from the covariance analysis method.

\section{Summary and conclusions} 
\label{sum-con}

In this work, we analysed the arising of correlations between isovector and isoscalar 
bulk parameters of hadronic nonrelativistic, and relativistic mean-field models. In 
particular, we discussed the connection of the crossing point in the density dependence of 
a particular bulk quantity, with the specific linear correlation between this quantity 
with its immediately next order bulk parameter. In the isovector sector, for instance, if 
there is a crossing point in the density dependence of the symmetry energy, then, it can 
be explained by the linear correlation between the symmetry energy, $\mathcal{S}$, and 
its slope, $L=3\rho(\partial\mathcal{S}/\partial\rho)$, both evaluated at the saturation 
density, i. e., there will be a linear correlation between $J=\mathcal{S}(\rho_o)$ and 
$L_o=L(\rho_o)$. In summary, the crossing points can be seen as a signature, or a route, 
in the searching of linear correlations among bulk parameters, as discussed in 
Sec.~\ref{lincorr}.

In the nonrelativistic framework, we presented correlations in some Skyrme~\cite{skyrme} 
and Gogny parametrizations, see Figs.~\ref{ksymo-qsymo-gogny}, 
\ref{ksymo-qsymo-isymo-skyrme}, \ref{lo-ksymo-skyrme}, \ref{qsymo-isymo-skyrme} and 
\ref{qo-ko-nr-gogny}{\color{purple}b}, as well as in parametrizations generated from the 
NR limit of NLPC models. By using the analytical structure of the latter model, we could 
write its five coupling constants in terms of the bulk parameters $\rho_o$, $B_o$, $K_o$, 
$m^*$, and $J$ in order to investigate the conditions in which the linear correlations of 
$L_o$ with $K_{\mbox{\tiny sym}}^o$, $Q_{\mbox{\tiny sym}}^o$ and $I_{\mbox{\tiny 
sym}}^o$, in the isovector sector, and of $K_o$ with $Q_o$ and $I_o$, in the isoscalar 
one holds. For these parametrizations, we showed that $L_o$ linearly correlates with 
$K_{\mbox{\tiny sym}}^o$, $Q_{\mbox{\tiny sym}}^o$ and $I_{\mbox{\tiny sym}}^o$ if we keep 
fixed the values of $J$ and $K_o$, see Eqs.~(\ref{ksymo}), (\ref{qsymo}) and 
(\ref{isymo}). Following analogous procedure, we found that parametrizations with fixed 
effective mass lead to linear correlations of $K_o$ with $Q_o$ and $I_o$, according to 
Eqs.~(\ref{qk-nr}), (\ref{ik-nr}), and the respective subsequent discussions. For some of 
these linear correlations, we discussed how they could have been found from the searching 
of crossing points in the bulk parameter as a function of the density. We pointed out 
that the crossing at $\rho_c^{\mbox{\tiny L}}/\rho_0=0.47$ ($\rho_c^{\mbox{\tiny 
K}}/\rho_0=0.79$) exhibited in Fig.~\ref{crossing-l-nr} (Fig.~\ref{crossing-k-nr}) for the 
$L^{\mbox{\tiny (NR)}}$ ($K^{\mbox{\tiny (NR)}}$) function, for instance, is a signature 
of the linear correlations between $L_o$ and $K_{\mbox{\tiny sym}}^o$ ($K_o$ and $Q_o$), 
at least.

Regarding the relativistic mean-field models~\cite{rmf}, we mainly studied that presenting 
cubic and quartic self-interaction in the scalar field $\sigma$, namely, the 
Boguta-Bodmer model~\cite{boguta}. The reason for this choice was based on our previous 
work of Ref.~\cite{bianca}. In that work, we showed that some correlations among bulk 
parameters presented in the NR limit of NLPC models, are also valid for this particular 
relativistic model. We further studied the correlations for bulk parameters of 
isovector and isoscalar sectors, mainly in the ranges of effective mass, symmetry energy, 
and incompressibility given by $0.58\leqslant m^*\leqslant 0.64$, $25\leqslant J \leqslant 
35$~MeV, and $250\leqslant K_o \leqslant 315$~MeV, respectively. The first range was 
proved to be experimentally consistent with finite nuclei spin-orbit splittings, 
according to Ref.~\cite{ls-splitting}. The second is compatible with experimental values 
from analyses of different terrestrial nuclear experiments and astrophysical 
observations~\cite{rmf,bali}, and the latter was based on the recent reanalysis of data 
on isoscalar giant monopole resonance energies~\cite{stone}.

In the isovector sector, we showed that $L_o$ also correlates with $K_{\mbox{\tiny 
sym}}^o$, $Q_{\mbox{\tiny sym}}^o$ and $I_{\mbox{\tiny sym}}^o$ like in the NR limit. 
However, the linear behavior of $L_o$ with $I_{\mbox{\tiny sym}}^o$ is broken in the 
range of $m^*$ and $K_o$ analysed, as pointed out in Fig.~\ref{isymo-lo-rel}. For the 
correlation between $L_o$ and $Q_{\mbox{\tiny sym}}^o$, we verified that the linear 
behavior is blurred only at higher values of $K_o$, see Fig.~\ref{qsymo-lo-rel}. We still 
concluded that the linear dependence of $L_o$ on $K_{\mbox{\tiny sym}}^o$ is preserved 
for fixed values of $J$, as in the case of the NR limit, according to the results 
presented in Fig.~\ref{ksymo-lo-rel}. This specific linear correlation was used to 
justify the crossing point exhibited in Fig.~\ref{crossing-l-rel} for the $L(\rho)$ 
function.

By comparing the behavior of $K_o$ and $Q_o$ in the isoscalar sector, we verified that 
these quantities are linearly correlated if the effective mass is kept fixed, exactly as 
deduced in the NR limit case. This correlation was displayed in 
Figs.~\ref{crossing-k-rel}{\color{purple}a} and \ref{qo-ko-rel}. We also used the angular 
coefficient presented in the former figure in order to justify the crossing point in 
the incompressibility function of the parametrizations showed  in 
Fig.~\ref{crossing-k-rel}{\color{purple}b}. Furthermore, we notice that for $m^*=0.64$, 
and $250\leqslant K_o \leqslant 315$~MeV, $Q_o$ varies in the range of $-183\leqslant Q_o 
\leqslant 130$~MeV, and present an overlap of about $36\%$ with the range of 
$-494\leqslant Q_o \leqslant -10$~MeV recently proposed in Ref.~\cite{chen-skew}. 

Lastly, we verified that the linear behavior between $K_o$ and $I_o$ is still valid for 
the Boguta-Bodmer models with fixed effective mass and for the range of $250\leqslant K_o 
\leqslant 315$~MeV, see Fig.~\ref{io-ko-rel}{\color{purple}a}. Nevertheless, the linearity 
is broken for a broader range of $K_o$ but with a correlation $I_o=I_o(K_o)$ still 
applying, see Fig.~\ref{io-ko-rel}{\color{purple}b}.

\section*{Acknowledgements}

We thank the support from Coordena\c c\~ao de Aperfei\c coamento de 
Pessoal de N\'ivel Superior (CAPES), and Conselho Nacional de Desenvolvimento
Cient\'ifico e Tecnol\'ogico (CNPq) of Brazil. O.~L. also acknowledges the support
of the grant $\#$2013/26258-4 from S\~ao Paulo Research Foundation (FAPESP). 
M.~D. acknowledges support from Funda\c{c}\~ao de Amparo \`a Pesquisa do Estado 
do Rio de Janeiro (FAPERJ), grant $\#$111.659/2014.

\appendix
\section{Coupling constants of the NR limit}
\label{appendix}

In the NR limit of the NLPC models, the coupling constants can be written in terms of the 
bulk parameter, namely, $m^*$, $\rho_o$, $B_o$, $K_o$ and $J$, as

\begin{eqnarray}
G^{2}_{\mbox{\tiny TV}} &=& \frac{J}{\rho_{o}} 
-\frac{\lambda}{6M}\frac{1}{m^{*}} \rho_{o}^{-\frac{1}{3}},
\label{ctegtv2}
\end{eqnarray}
\begin{eqnarray}
A &=& \frac{M/3\rho_{o}^{2}}{\left( 3M^{2} - 19E_{\mbox{\tiny F}}^{o}M +18E_{\mbox{\tiny 
F}}^{o2} \right)} \Bigg\{
\frac{8E_{\mbox{\tiny F}}^{o}}{ m^{*}}\left(M-6E_{\mbox{\tiny F}}^{o} \right)  \nonumber\\
&+&  K_{o}\left(M -3E_{\mbox{\tiny F}}^{o}\right)-9B_{o}\left(3M-13E_{\mbox{\tiny 
F}}^{o}\right) \nonumber\\
&-&3E_{\mbox{\tiny F}}^{o}\left( 5M-27E_{\mbox{\tiny F}}^{o} \right) \Bigg\},
\label{ctea}
\end{eqnarray}
\begin{eqnarray}
B &=& \frac{M/3\rho_{o}^{3}}{\left( 3M^{2} - 19E_{\mbox{\tiny F}}^{o}M +18E_{\mbox{\tiny 
F}}^{o2} \right)} \Bigg\{
 - \frac{E_{\mbox{\tiny F}}^{o}}{m^{*}}\left(M-10E_{\mbox{\tiny F}}^{o} \right) 
\nonumber\\
&-&\frac{K_{o}}{2} \left(M-2E_{\mbox{\tiny F}}^{o} 
\right)+3B_{o}\left(3M-10E_{\mbox{\tiny F}}^{o}\right)\nonumber\\
&+&3E_{\mbox{\tiny F}}^{o}\left(M-6E_{\mbox{\tiny F}}^{o} \right) \Bigg\},
\label{cteb} 
\end{eqnarray}
\begin{eqnarray}
G^{2}_{\mbox{\tiny S}}&=& \frac{M}{\rho_{o}} \left(  \frac{1}{m^{*}}-1\right) -2A\rho_{o} 
-3B\rho_{o}^{2},
\label{ctegs2}
\end{eqnarray}
and
\begin{eqnarray}
G^{2}_{\mbox{\tiny V}} &=& \frac{M}{\rho_{o}} \left(  \frac{1}{m^{*}}-1\right) 
-\frac{E_{\mbox{\tiny F}}^{o}}{m^{*}\rho_{o}}  
 -\frac{B_{o}}{\rho_{o}}  -A\rho_{o} -2B\rho^{2}_{o}.\qquad
\label{ctegv2}
\end{eqnarray}

In the case of the NR limit obtained from the Lagrangian density of Eq.~(\ref{lagd}), the 
two isovector coupling constants are written as
\begin{eqnarray}
G_{\mbox{\tiny VTV}}&=&
\Bigg\{J-\frac{\mathcal{S}_1}{r}-\frac{5E_{\mbox{\tiny F}}^o}{9m^{*}}(1-r^{2/3})
-\frac{5E_{\mbox{\tiny F}}^o}{9M} \Bigg[M\left(1 -\frac{1}{r}\right)   
\nonumber\\
&+& 2A\rho_o^2(1 - r) + 3B\rho_o^3(1 - r^2)\Bigg] 
r^{2/3} \Bigg\}\frac{1}{\rho_o^3(1-r^2)},\nonumber\\
\end{eqnarray}
and
\begin{eqnarray}
G^{2}_{\mbox{\tiny TV}} &=& \frac{J}{\rho_{o}} 
-\frac{\lambda}{6M}\frac{1}{m^{*}} \rho_{o}^{-\frac{1}{3}} - G_{\mbox{\tiny VTV}}\rho_o^2,
\end{eqnarray}
with $r=\rho_1/\rho_o$ and $\mathcal{S}_1=\mathcal{S}(\rho_1)$.

\end{document}